\documentclass[
  aps,
  pra,
  reprint,
  superscriptaddress,
  longbibliography,
  nofootinbib
]{revtex4-2}

\usepackage[utf8]{inputenc}
\usepackage[T1]{fontenc}
\usepackage{graphicx}
\usepackage{amsmath,amssymb,amsfonts,amsthm,bm}
\usepackage{physics}
\usepackage{dsfont}
\usepackage{upgreek}
\usepackage{xcolor}
\usepackage{hyperref}
\hypersetup{
  colorlinks=true,
  linkcolor=blue,
  citecolor=blue,
  urlcolor=blue
}
\DeclareSymbolFont{usualmathcal}{OMS}{cmsy}{m}{n}
\DeclareSymbolFontAlphabet{\mathcal}{usualmathcal}

\newcommand{\range}[2]{\in \left[#1,\,#2\right]}

\begin{document}

\title{From single-particle to many-body chaos in Yukawa--SYK: theory and a cavity-QED proposal}

\author{David Pascual Solis}
\email{david.pascualsolis@unitn.it}
\affiliation{Pitaevskii BEC Center, CNR-INO and Department of Physics, University of Trento, Via Sommarive 14, Trento, I-38123, Italy}
\affiliation{INFN-TIFPA, Trento Institute for Fundamental Physics and Applications, Via Sommarive 14, Trento, I-38123, Italy}

\author{Alex Windey}
\email{alex.windey@unitn.it}
\affiliation{Pitaevskii BEC Center, CNR-INO and Department of Physics, University of Trento, Via Sommarive 14, Trento, I-38123, Italy}
\affiliation{INFN-TIFPA, Trento Institute for Fundamental Physics and Applications, Via Sommarive 14, Trento, I-38123, Italy}

\author{Soumik Bandyopadhyay}
\email{soumik.bandyopadhyay@unitn.it}
\affiliation{Pitaevskii BEC Center, CNR-INO and Department of Physics, University of Trento, Via Sommarive 14, Trento, I-38123, Italy}
\affiliation{INFN-TIFPA, Trento Institute for Fundamental Physics and Applications, Via Sommarive 14, Trento, I-38123, Italy}

\author{Andrea Legramandi}
\email{andrea.legramandi@unitn.it}
\affiliation{Pitaevskii BEC Center, CNR-INO and Department of Physics, University of Trento, Via Sommarive 14, Trento, I-38123, Italy}
\affiliation{INFN-TIFPA, Trento Institute for Fundamental Physics and Applications, Via Sommarive 14, Trento, I-38123, Italy}

\author{Philipp Hauke}
\email{philipp.hauke@unitn.it}
\affiliation{Pitaevskii BEC Center, CNR-INO and Department of Physics, University of Trento, Via Sommarive 14, Trento, I-38123, Italy}
\affiliation{INFN-TIFPA, Trento Institute for Fundamental Physics and Applications, Via Sommarive 14, Trento, I-38123, Italy}

\begin{abstract}
Understanding how quantum systems evolve from integrable to fully chaotic behavior remains a central open problem in physics. 
In this work, we show that the Yukawa--Sachdev--Ye--Kitaev (YSYK) model is a particularly rich testbed to address this question.  
This model extends the paradigmatic SYK model of many-body chaos and holography by introducing boson-mediated random all-to-all fermionic interactions. 
Using spectral and dynamical chaos markers, we perform a comprehensive numerical characterization tailored to the mesoscopic regime relevant for quantum simulation. 
We show that, at finite size, the YSYK model interpolates between single-particle and many-body chaos, with the interaction strength acting as a tunable control parameter interpolating between SYK$_2$ and SYK$_4$ behavior. 
We introduce a framework enabling direct and quantitative comparison with these benchmark models, and we uncover prethermalization plateaus as well as delayed and incomplete scrambling appearing in an intermediate regime that weakly breaks integrability. 
We further propose a feasible optical-cavity realization of the YSYK model using ultracold atoms, opening the door to studies beyond the capacities of numerical investigations. 
Our results establish the YSYK model as a benchmark platform connecting single-particle and many-body chaos, and they provide a quantitative reference point for future quantum simulations. 
\end{abstract}

\maketitle

\section{Introduction}

Understanding the nature of quantum chaos across different physical regimes remains a central challenge in modern theoretical physics~\cite{Berry1977, BGS1984, haake_quantum_1991, Mehta2004, DAlessio2016}. Solvable models such as the Sachdev–Ye–Kitaev (SYK)~\cite{SY93,Kitaev2015} model play a pivotal role in this endeavor, as they provide analytical tractability of complex phenomena such as thermalization~\cite{Deutsch1991,Srednicki1994,Rigol2008,Sonner:2017hxc, Larzul_2022, Bandyopadhyay_2023,Louw_2022,Paviglianiti_2023,Jaramillo_2025,Perugu_2025} and information scrambling~\cite{Maldacena:2016hyu,Polchinski:2016xgd,Maldacena_2016}, while also connecting concepts across multiple areas of physics. For instance, the SYK model serves as a minimal framework for studying the strange-metal phase of unconventional superconductors~\cite{Chowdhury2022} and, in the context of quantum gravity, offers an insightful realization of the AdS/CFT correspondence~\cite{Cotler2016,Maldacena:2016upp} through its holographic duality with Jackiw–Teitelboim (JT) gravity~\cite{Jackiw:1984je,Teitelboim:1983ux}. Despite these exceptional features, the SYK model is a maximally chaotic system~\cite{Maldacena_2016}. In its standard formulation, it is therefore unable to capture more intricate phenomena such as the onset of ergodicity breaking~\cite{Turner:2017fxc,Suntajs2020,Sels2021,Schiulaz2019} and the crossover from chaotic to integrable or many-body localized dynamics~\cite{Basko_2006,Kravtsov_2015,Abanin:2018yrt}, which remain central challenges in developing a comprehensive understanding of quantum chaos.
In this work, we study the Yukawa--SYK (YSYK) model~\cite{Schmalian_YSYK, Wan2020,wang_quantum_2020,pan_yukawa-syk_2021}, which we argue naturally interpolates between many-body chaotic behavior and regimes with either slowed, constrained, or non-ergodic dynamics.

\begin{figure}[t!]
    \centering \includegraphics[width=1.0\linewidth]{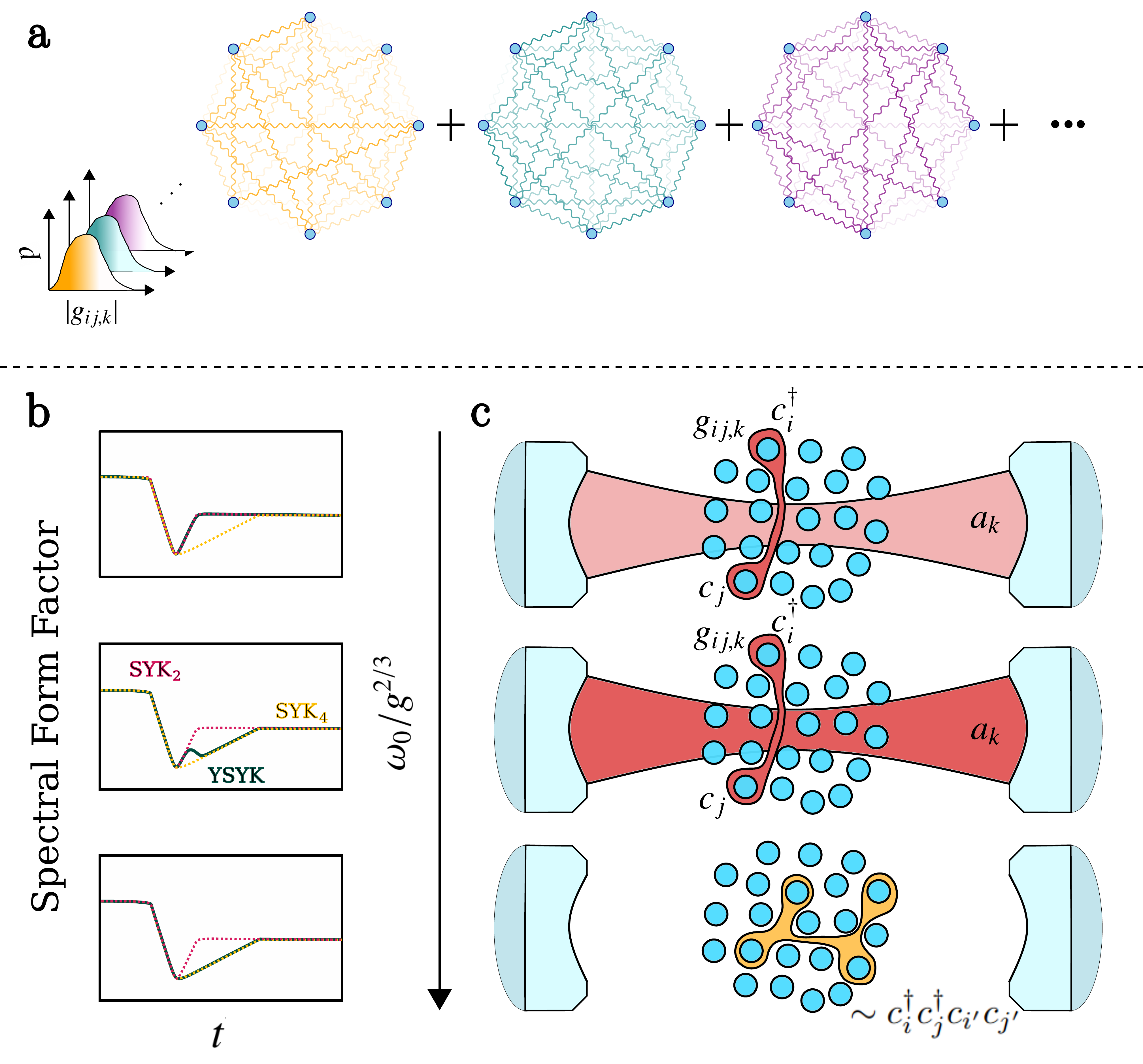}
    \caption{\textbf{(a)} Schematic illustration of the interaction term in the spinless Yukawa--SYK model. A set of $N$ spinless fermionic modes (blue dots) interact pairwise via $M$ independent bosons (wiggly lines), through random couplings $g_{ij,k}$. Each bosonic mode is shown in a different color, and the line shading represents the coupling strengths. The couplings are drawn independently from a Gaussian Unitary Ensemble (GUE). 
    \textbf{(b)} As illustrated by the spectral form factor as a chaos diagnostic, tuning the boson mass $\omega_0$ relative to the coupling strength $g$ drives a finite-size crossover from a single-particle–chaotic to a many-body–chaotic regime, with an intermediate region where features of both coexist. \textbf{(c)} Proposed optical-cavity realization of the YSYK model. Cavity photons act as the bosonic modes, while fermions are represented by cold atoms trapped inside the cavity. An engineered optical speckle pattern introduces disorder in the light–matter coupling strengths. At very small photon detuning (boson mass), the photon only dresses hopping between fermionic modes and the model behaves as quadratic (SYK$_2$-like) single-particle chaotic, while at large detuning the photon can be adiabatically eliminated, yielding an  effective quartic (SYK$_4$-like) many-body-chaotic theory.}
    \label{fig:first_schematic}
\end{figure}

The YSYK model generalizes the SYK framework by introducing bosonic fields that mediate the fermionic interactions (Fig.~\ref{fig:first_schematic}\textbf{a}), leading to a richer and more natural physical structure~\cite{patel_universal_2023,Li:2024kxr}. Like the SYK model, it is one of the few strongly coupled systems that are exactly solvable in the large-$N$ limit, yet its dynamics are substantially more diverse. The maximally chaotic~\cite{Marcus_2019, davis_quantum_2023} SYK-like phase emerges only at low temperatures, whereas at higher energies the system exhibits a crossover to a Fermi-liquid regime. Unlike in mass-deformed SYK~\cite{Lunkin_2020}, the SYK phase remains stable in the thermodynamic limit.
The presence of boson-mediated interactions, commonly found in real materials, refines the phenomenology of non-Fermi liquids and enables the unified description of both normal and superconducting phases~\cite{HAUCK_YSYK}, making the YSYK model an effective description of strongly correlated electronic systems. From the holographic viewpoint, the additional bosonic sector enriches the gravitational dual beyond the minimal JT setup~\cite{inkof_quantum_2022,Schmalian:2022web,esterlis2025quantum}, potentially providing a route toward more complex holographic phases and toward a connection with holographic superconductors~\cite{Hartnoll:2008kx}.
Yet, despite its theoretical appeal and the considerable insights obtained in the large-$N$ limit, the finite-size behavior and systematic chaos characterization of the YSYK model remain largely uncharted. Exploring this regime is essential for uncovering finite-size effects, studying chaos diagnostics that are difficult to access analytically---such as level statistics---and for characterizing realistic implementations of the YSYK model in controllable experimental platforms at mesoscopic sizes, where the large-$N$ techniques applicable to the idealized model no longer hold.

In this work, we provide a comprehensive analysis of the chaos properties of the YSYK model at finite system sizes. By analyzing spectral statistics~\cite{Wigner_1951, Brody81, Oganesyan2007, Atas2013}, the spectral form factor (SFF)~\cite{haake_quantum_1991,Mehta2004, Gharibyan2018,Saad:2018bqo}, and out-of-time-ordered correlators (OTOCs)~\cite{Larkin1969,Shenker_2014,Shenker:2014cwa,Maldacena_2016}, we show that the boson mass $\omega_0$ in the YSYK model serves as a tunable parameter that smoothly interpolates between single-particle and many-body chaos\footnote{This is different from what happens in low-rank SYK  models, where this role is played by the ratio of the dynamic fermionic degrees of freedom to static bosonic fields~\cite{Bi:2017yvx, Kim_2020}.}. To quantify this crossover, we introduce effective perturbative descriptions at small and large boson masses, together with normalization factors derived from random matrix theory,
which enable a direct and quantitative comparison between the YSYK and the complex SYK$_q$ model~\cite{Sachdev:2015efa,Davison:2016ngz,Gu_2020}.
We show that the model continuously bridges two regimes: at small boson mass, it approaches SYK$_2$, which is characterized by single-particle chaos~\cite{Liao2020,Winer2020,Legramandi2024}, while at large boson mass it exhibits SYK$_4$-like behavior associated with many-body quantum chaos (see Fig.~\ref{fig:first_schematic}\textbf{b}). 
In the intermediate regime, the system displays rich dynamical features whose finite-size phenomenology is reminiscent of slowed ergodic dynamics, including localization crossovers~\cite{Garcia-Garcia2018, Finite_size_OTOC}, partial scrambling~\cite{weak_quantum_chaos,He_MBL_OTOC,Swingle_slow_scrambling,OTOC_quantum_Ising_chain}, and prethermalization plateaus~\cite{Rakovszky2018,Nandy22,Dieplinger2023,Baumgartner:2024orz}. We show analytically and numerically that this integrable phenomenology persists up to late times $t\sim \omega_0^{-5/2}$, after which the non-integrable behavior starts emerging.

Moreover, the Yukawa interaction makes this model particularly promising for experimental realization, as it corresponds to the standard QED vertex that naturally arises in light–matter coupled systems~\cite{Mivehvar02012021}. Building on previous proposals for analog SYK simulators~\cite{cavity_proposal,Baumgartner:2024ysk}, we present a feasible implementation of the YSYK model using ultracold atoms in an optical cavity (Fig.~\ref{fig:first_schematic}\textbf{c}), leveraging long-range photon-mediated interactions and disorder engineering available in current platforms \cite{Baghdad_2023,sauerwein_engineering_2023, Forsi_cavity_control24}. 
Compared to earlier proposals for realizing the complex SYK$_4$ model, the characteristic dynamical timescales are enhanced by up to three orders of magnitude.
The proposed setup provides a tunable platform to investigate the finite-$N$ crossover from single-particle to many-body chaos and to explore emergent phenomena of fundamental importance ranging from unconventional superconductivity~\cite{Schmalian_YSYK,HAUCK_YSYK} to aspects of quantum gravity via the holographic principle~\cite{Maldacena:2016upp,Cotler2016,Jackiw:1984je,Teitelboim:1983ux}.
Our work thus follows a two-fold aim: First, to provide finite-size observables of the YSYK model, also beyond those accessible in large-$N$ path-integral formalism and able to serve as benchmarks for future quantum simulations. Second, to demonstrate a realistic route for cavity-QED realizations to reach regimes beyond the capacities of numerics.

The rest of the paper is organized as follows. Section~\ref{sec:models} introduces a spinless variant of the YSYK Hamiltonian and the complex SYK$_q$ models ($q=2,4$) that we use as benchmarks throughout. Section~\ref{sec:markers} summarizes the considered chaos markers: density of states, gap ratio statistics, spectral form factor, and out-of-time-ordered correlators. Section~\ref{Sec:Results} reports our numerical results across the interaction control ratio, beginning with an overview and then detailing the strong-coupling (SYK$_2$-like) and weak-coupling (emergent SYK$_4$-band) regimes. Section~\ref{sec:experiment} outlines an optical-cavity implementation and discusses estimates of achievable parameters and dissipation strengths, showing that all regimes of YSYK from weak to strong fermion--boson coupling are accessible with state-of-the-art quantum simulators. Section~\ref{sec:conclusion} presents our conclusions and potential directions for future work. 
Several Appendices contain technical details as well as supplementary results. 

\section{Yukawa--SYK and SYK$_q$ $(q=2,4)$ as Limiting Models}
\label{sec:models}
This section introduces the spinless-fermion variant of the YSYK model used throughout this work. 
For context, we also briefly review the complex SYK$_2$ and SYK$_4$ models, between which YSYK interpolates as shown in Sec.~\ref{Sec:Results}. 

\subsection{Yukawa--SYK}
The YSYK model couples $N$ complex fermions to $M$ dynamical bosons through random, all-to-all Yukawa interactions.
For simplicity, in contrast to previously studied spinful variants~\cite{Schmalian_YSYK, Wan2020}, we take the fermions to be spinless, halving the local Hilbert space dimension.
The model is governed by the Hamiltonian
\begin{align}
    H_{\rm YSYK} = & -\mu \sum_{i=1}^N c_i^\dagger c_i + \frac{1}{2}\sum_{k=1}^M \left( \pi_k^2 + \omega_0^2 \phi_k^2 \right) \nonumber \\
    & + \frac{1}{\sqrt{MN}} \sum_{i,j=1}^N \sum_{k=1}^M g_{ij,k}\, c_i^\dagger c_j\, \phi_k.
\label{eq:hamiltonian}
\end{align}
The scalar bosonic fields $\phi_k$ have a bare mass $\omega_0$, with conjugate momenta $\pi_k$, and $c_i$ ($c_i^\dagger$) are annihilation (creation) operators for spinless fermions in mode $i$. The fermionic chemical potential $\mu$ can be set to zero as we work in the fixed sector of half-filling. For each $i,j$, and $k$, the couplings $g_{ij,k}$ are independent random numbers drawn from the Gaussian unitary ensemble (GUE) with zero mean and variance $\overline{|g_{ij,k}|^2}=g^2$. 
One of the main challenges in numerically studying the YSYK model is the presence of bosonic fields, which renders the Hilbert space infinite-dimensional. To make the problem tractable, we impose an occupancy cutoff $N_b$ for each oscillator. The Hamiltonian expressed in terms of standard creation and annihilation operators reads
\begin{align}
    H =& \sum_{k=1}^M \omega_0 \left( a_k^\dagger a_k + \tfrac{1}{2} \right) \nonumber \\ 
    & + \frac{1}{\sqrt{2\omega_0 MN}} \sum_{i,j=1}^N \sum_{k=1}^M g_{ij,k}\, c_i^\dagger c_j\, \left( a_k + a_k^\dagger \right).
\label{eq:hamiltonian_explicit}
\end{align}
This spinless YSYK model has two competing energy scales: the bare boson frequency $\omega_0$ and the disorder scale set by $g$. 

In Sec.~\ref{Sec:Results}, we explore the regimes resulting from tuning the dimensionless ratio $\omega_0/g^{2/3}$. As we show qualitatively and quantitatively, tuning this control parameter interpolates between an SYK$_2$-like regime for $\omega_0/g^{2/3}\!\ll\!1$ and an SYK$_4$-like regime for $\omega_0/g^{2/3}\!\gg\!1$. Before introducing the employed chaos markers in Sec.~\ref{sec:markers}, we briefly review the complex SYK$_q$ model.

\subsection{Complex SYK$_q$}
\label{sec:CSYK}
The complex SYK$_q$ model~\cite{Sachdev:2015efa,Davison:2016ngz, Gu_2020} features $N$ spinless complex fermions with $q$-body random interactions. The Hamiltonian is
\begin{equation}
\label{eq:SYKq_H}
    H_{\text{SYK}_q} = \hspace*{-0.5cm} \sum_{1 \le i_1 < \cdots < i_{q/2} \le N \atop 1 \le j_1 < \cdots < j_{q/2} \le N}\hspace*{-0.5cm} 
    J_{i_1 \ldots i_{q/2},\,j_1 \ldots j_{q/2}}
    \,c_{i_1}^\dagger \cdots c_{i_{q/2}}^\dagger\,
    c_{j_1} \cdots c_{j_{q/2}} \,,
\end{equation}
with couplings drawn from a complex Gaussian distribution~\cite{Gu_2020},
\begin{equation}
    \overline{|J_{i_1 \ldots i_{q/2},\,j_1 \ldots j_{q/2}}|^2}
= \frac{J^2 (q/2)! (q/2-1)!}{N^{q-1}}.
\label{eq:SYK_q_variance}
\end{equation}
Unlike the Majorana version of the SYK model, its complex variant possesses a conserved U(1) charge, which---similarly to YSYK---allows one to work with a fixed particle number; in this work, we always focus on the half-filling sector.

The model with $q=2$ (SYK$_2$) corresponds to an integrable theory of non-interacting fermions with random hopping. Although the couplings are drawn from the GUE and thus the model exhibits quantum chaos at the single-particle level, the full many-body Hamiltonian is quadratic and hence remains integrable.  
As a result, while the single-particle spectrum follows random matrix statistics, the many-body spectrum exhibits uncorrelated energy levels, reflecting the absence of many-body chaos. Notably, the chaotic properties of the single-particle sector imprint themselves on certain characteristics of the many-body spectrum~\cite{Liao2020, Winer2020,Legramandi2024}. 

In contrast, the SYK$_q$ model with $q \geq 4$ exhibits genuine many-body chaos~\cite{Garcia-Garcia2016, Garcia-Garcia2017}. It features a holographic low-temperature phase dual to near-extremal black holes, whose electric charge corresponds to the conserved U(1) quantum number of the complex SYK$_4$ model~\cite{Sachdev:2015efa,Bhattacharya:2017vaz}, and has thus become a paradigmatic model for holographic quantum matter. 

\section{Chaos markers}  
\label{sec:markers}

Quantum chaos manifests itself in both the energy spectrum and the dynamics of many-body systems on a wide range of energy and time scales~\cite{BGS1984, haake_quantum_1991, Mehta2004, Berry1977, DAlessio2016}. These features, evident in spectral correlations and dynamical behavior, contrast sharply with those of non-chaotic systems, such as integrable and localized models, with significant implications for thermalization~\cite{Rigol2008, Rigol2006}, entanglement growth~\cite{Calabrese_2005, Kim_2013, Nandkishore_2015, Dima_2019, Lukin_2019}, and transport~\cite{Thouless1974,Altshuler1986, Zotos_1997, H-Meisner_2003, Bertini_2021}.
In this Section, we present the chaos markers that we use in our analysis of the YSYK model. In this work, we will always consider markers averaged over different independent disorder realizations of the model. 

\subsection{Spectral probes}
\label{sec:SpecProb}

The main idea behind spectral probes is that integrable or localized systems are characterized by the presence of an extensive number of symmetries, which lead to an uncorrelated energy spectrum~\cite{Berry1977}. 
In contrast, chaotic quantum systems exhibit correlated spectral properties, typical of a random matrix theory (RMT) ensemble~\cite{BGS1984, Mehta2004, haake_quantum_1991}, and these correlations are expected to be universal, i.e., independent of the microscopic details of the system under consideration~\cite{Wigner_1951, Wigner1958,Dyson1962}. 

\subsubsection{Density of States}
\label{subsec:dos}
The \emph{density of states} (DOS) $\rho(E)$ provides a coarse-grained view of the spectrum.
For a Hamiltonian $H$ with $D$ eigenenergies $\{E_n\}_{n=1}^D$, we define
\begin{equation}
\rho(E) = \frac{1}{D} \sum_{n=1}^{D} \delta\bigl(E - E_n\bigr),
\end{equation}
with $\delta(\cdot)$ the Dirac delta. Averaging $\rho(E)$ over disorder realizations suppresses sample-to-sample fluctuations and yields a smooth envelope for comparisons across parameters. While edge-resolved features of $\rho(E)$ can, in principle, carry signatures relevant to chaos, resolving them with sufficient fidelity is typically impractical in our setting. Accordingly, we use $\rho(E)$ primarily as a baseline for interpreting finer spectral and dynamical probes. In our case, the emergence of well-separated features in $\rho(E)$ will signal crossovers between regimes of the YSYK model.

\subsubsection{Gap Ratio Distribution}
\label{sec:GapRatio}

The disorder-averaged DOS does not resolve correlations between neighboring energy levels, which are essential for distinguishing between chaotic and non-chaotic quantum systems.
Finer information comes from the distribution of adjacent energy-level spacings.
In this context, a particularly robust 
probe is the statistics of \emph{gap ratios}~\cite{Oganesyan2007, Atas2013}.
For energy levels $\{E_n\}$ ordered in ascending energy, the gap ratio is defined as the minimum of the ratios between consecutive level spacings, $s_n = E_{n+1} - E_n$,
\begin{equation}
r_n = \frac{\min(s_n,s_{n-1})}{\max(s_n,s_{n-1})}, \quad n = 2, \dots, D - 1.
\end{equation}

The ensemble-averaged distribution $P$ of the gap ratios $r_n$ provides valuable information on level correlations, while its average value $\langle r \rangle$ serves as a compact diagnostic of spectral statistics and allows one to easily distinguish integrable from chaotic behavior~\cite{Luitz2015,  Garcia-Garcia2018}.
In integrable systems or those in the localized phase, level spacings are uncorrelated, resulting in a Poissonian distribution of $r_n$. The probability distribution and mean value in this case are known analytically~\cite{Oganesyan2007},
\begin{equation}
P_{\rm P}(r) =  2 (1 + r)^{-2}, \qquad \langle r\rangle_{\rm P} = 2 \ln 2 - 1 \approx 0.386.
\end{equation}

In contrast, quantum chaotic systems exhibit level repulsion~\cite{Mehta2004, haake_quantum_1991}. Atas \emph{et al.}~\cite{Atas2013} derived a closed-form approximation for the gap ratio distribution for the three Dyson symmetry classes ($\upbeta = 1, 2, 4$), corresponding to GOE, GUE, and GSE, respectively,
\begin{equation}
P_{\rm WD}^{(\upbeta)}(r)
= \frac{1}{\mathcal{N}_\upbeta}
\frac{(r + r^2)^\upbeta}
{(1 + r + r^2)^{1 + \tfrac{3\upbeta}{2}}},
\end{equation}
with $\mathcal{N}_\upbeta$ a normalization constant. For the unitary class ($\upbeta = 2$), typically relevant for systems with broken time-reversal symmetry, the mean gap ratio is
$\langle r\rangle_{\rm GUE} \approx 0.599.$ In our setting with complex $g_{ij,k}$, time-reversal is generically broken, making GUE statistics the appropriate benchmark for chaotic behavior.

\subsubsection{Spectral Form Factor}
\label{sec:SFF}

The \emph{spectral form factor} (SFF)~\cite{haake_quantum_1991, Mehta2004, Cotler2016, Gharibyan2018,Saad:2018bqo} is sensitive to correlations between eigenvalues on energy scales beyond nearest neighbors. It captures long-range rigidity in the spectrum and is closely tied to the temporal evolution of quantum observables~\cite{Tavora2016,Herrera2017,Herrera2018,Santos2020}.
It is defined as the disorder-averaged modulus squared of the partition function analytically continued to real time,
\begin{align}
    K(t, \beta) & = \overline{\Bigl|\frac{Z(\beta+it)}{Z(\beta)}\Bigr|^2} \nonumber \\
    & = \overline{\left[\frac{\sum_{m,n =1}^{D} e^{-\beta(E_{m}+E_{n})}e^{-it(E_m-E_{n})}}{(\sum_{n=1}^D e^{-\beta E_n})^2}\right]}.
\label{Eq:SFF}
\end{align} 
In chaotic systems, the SFF exhibits the well-known \emph{dip–ramp–plateau} structure~\cite{Cotler2016, Mehta2004,haake_quantum_1991}, a hallmark of RMT. At early times the SFF shows a rapid decay—referred to as the \emph{slope}—which is non-universal and depends on the DOS. This decay reaches a minimum at the \emph{dip}, before starting to grow linearly at intermediate times, forming the so-called \emph{ramp}. The linearity of the ramp is due to long-range spectral rigidity and universal level repulsion—key signatures of quantum chaos. The time scale at which the ramp emerges, often called the \emph{ramp time} ($ t_{\rm ramp}$), is generally close to the Thouless time $t_{\rm Th}$,  which marks the crossover between non-universal system-specific dynamics and universal late-time RMT behavior and which has been connected to the system’s transport characteristics~\cite{Thouless1974, Altshuler1986, Altshuler1988,Evers_2008}~\footnote{In spatially extended chaotic systems, the Thouless energy $E_{\text{Th}} \sim \hbar \mathcal{D} / L^2$, where $\mathcal{D}$ is the diffusion constant related to DC conductivity~\cite{Altshuler1986}.This leads to a Thouless time $t_{\text{Th}} = 1 / E_{\mathrm Th} \sim L^2 / \mathcal{D}$, associated with typical diffusive transport in chaotic systems.}. 
In practice, $t_{\rm ramp}$ may appear slightly later than $t_{\rm Th}$ due to masking by the non-universal slope and finite-size fluctuations~\cite{Altshuler1986, Altshuler1988, Evers_2008, Liu_2018,Chan2018}. However, since these effects are typically small compared to other time scales, we will use $t_{\rm ramp}$ and $t_{\rm Th}$ interchangeably. 
At late times, the SFF saturates to a constant value, forming the \emph{plateau}. 
This saturation reflects the finite Hilbert-space dimension, where the discreteness of the spectrum becomes manifest and the dynamics are sensitive to individual energy levels. The characteristic time scale for resolving the discreteness of the spectrum is set by the Heisenberg time $t_{\rm H}= 2 \pi / \Delta (E)  \sim D$, where $\Delta(E)$ is the mean level spacing.
In chaotic many-body systems, the plateau typically emerges at times $t_{\rm plateau} \sim t_{\rm H}$, whereas it occurs earlier in integrable or localized systems.

A particularly insightful diagnostic emerges from the \emph{ratio} of the Heisenberg to the Thouless time, $t_{\rm H} / t_{\text{Th}}$: in fully chaotic systems, $t_{\rm H} \gg t_{\text{Th}}$, indicating that universal spectral correlations develop well before individual energy levels are resolved. As the system becomes less chaotic, $t_{\text{Th}}$ increases and may eventually exceed $t_{\rm H}$, i.e., $ t_{\text{Th}} \gtrsim t_{\rm H}$, signaling the breakdown of RMT behavior~\cite{Suntajs2020,Sels2021, Prakash2021}. Thus, the crossover $t_{\text{Th}} \sim t_{\rm H}$ can be interpreted as a boundary between chaotic and non-chaotic regimes~\cite{Schiulaz2019,Suntajs2020,Gharibyan2018}~\footnote{For example, near the many-body localization (MBL) transition, transport slows down (e.g., subdiffusively with $t_{\text{Th}} \sim L^z$, $ z > 2$) and ultimately halts in the MBL phase, where $t_{\text{Th}} \sim e^{L/\xi}$ grows exponentially with system size~\cite{Gopalakrishnan2016,Serbyn2016,Serbyn2017}.
Interestingly, integrable systems also exhibit a large separation of timescales, with $t_{\text{Th}} \sim L$ due to ballistic quasiparticle propagation. 
This illustrates that, while a large $t_{\rm H}/t_{\text{Th}}$ is a necessary condition for quantum chaos, it is not sufficient---spectral statistics and other dynamical correlations must be examined jointly~\cite{Prosen2002}.}. 

In the SYK$_4$ model, which exhibits genuine many-body chaos, $t_{\rm ramp}$ is estimated to scale as $\sqrt{N}\log{N}$~\cite{Altland2018} or even $\log{N}$~\cite{Gharibyan2018}, and hence is parametrically shorter than $t_{\rm H}$,  which scales exponentially with $N$. This separation of timescales reflects the chaotic and complex nature of the many-body spectrum. In contrast, the SYK$_2$ model lacks a true linear ramp, consistent with its integrable nature. However, due to the GUE structure of the single-particle Hamiltonian, signatures of chaos persist in the form of an \emph{exponential ramp} in the SFF~\cite{Liao2020, Winer2020, Legramandi2024}. Single-particle level spacing also governs the plateau time in SYK$_2$, which is $t_{\rm plateau} = 2N$~\cite{Liao2020,Winer2020}, in stark contrast to the exponentially large plateau time of SYK$_4$ and underscoring the difference between single-particle and many-body chaos. 
The SFF therefore provides a powerful dynamical diagnostic of ergodicity and chaos, complementing short-range spectral probes such as the gap ratio.

\subsection{Dynamical probe: Out-of-time-ordered correlators}

To complement the spectral probes, which primarily characterize long-time (or small energy) properties, one can also consider dynamical chaos markers that capture the early-time behavior of quantum systems. Chief among these are the \emph{out-of-time-ordered correlators} (OTOCs)~\cite{Larkin1969,Shenker_2014,Shenker:2014cwa,Maldacena_2016}, which track the growth of initially local perturbations, probing operator spreading and information scrambling~\cite{Roberts:2018mnp}.
A way of quantifying how initially local perturbations evolve into nonlocal operators under Heisenberg evolution is by evaluating the squared commutator
\begin{equation}
  C(t)
  = \bigl\langle
      [A(t),B(0)]^\dagger\, [A(t),B(0)]
    \bigr\rangle_\beta,
\label{Eq: commutator_squared}
\end{equation}
where $\langle \ldots \rangle_\beta$ denotes the thermal average relative to the Gibbs state $\rho_\beta = \frac{e^{-\beta H}}{\mathrm{Tr}\,(e^{-\beta H})}$ and 
$A(t) =$ $ e^{i H t} A e^{-i H t}$. 
For unitary operators $A$ and $B$, Eq.~\eqref{Eq: commutator_squared} can be written as
\begin{equation}
    C(t) = 2 - 2 F(t),
\label{Eq:F_1_minus_2C}
\end{equation} 
where we introduced the OTOC $F(t)$, defined as
\begin{equation}
    F(t) = \Re\left[\mathrm{Tr}\,(\rho_\beta A^\dag(t) B^\dag A(t) B)\right].
\label{Eq:OTOC}
\end{equation}

A system is said to scramble quantum information if the commutator in Eq.~\eqref{Eq: commutator_squared} grows exponentially at early times for generic, sufficiently local operators $A$ and $B$. The scrambling time is then defined as the time at which the corresponding OTOC saturates to a value close to zero.
The characteristic exponent governing the commutator growth is the quantum Lyapunov exponent, which quantifies the rate at which perturbations spread through the system. This mirrors classical chaos and provides a direct quantum analog of sensitive dependence on initial conditions~\cite{Cotler_butterfly_effect, Varikuti_2024}. Maldacena, Shenker, and Stanford proposed a fundamental bound on chaos and the growth of the commutator~\cite{Maldacena_2016}, which is saturated by systems believed to be the fastest scramblers in nature—most notably black holes and the SYK$_4$ model~\cite{Kitaev2015, Fast_scramblers}. 

It is important to distinguish scrambling from quantum chaos in a broader sense~\cite{Dowling_2023}. While exponential decay of the OTOC is a strong indicator of information scrambling, it can also occur in systems with unstable semiclassical dynamics, even when they are integrable~\cite{Scrambling_and_chaos}. In such cases, the OTOC may exhibit transient exponential decay but fails to fully decay at late times, often displaying persistent oscillations—hallmarks of non-chaotic or integrable dynamics~\cite{OTOC_and_quantumchaos, Hummel_2019, Kidd_2021}.
In contrast, systems near to integrability often show either no exponential decay or saturation of OTOCs, which instead saturate to a nonzero value depending on the symmetries and the finite system size~\cite{Finite_size_OTOC, He_MBL_OTOC, Swingle_slow_scrambling, OTOC_quantum_Ising_chain}. For the SYK$_2$ model with Majorana fermions, $F(t)$ shows a power-law decay towards the saturation value with superimposed oscillations, which is believed to be a generic feature of quantum many-body integrable systems. Moreover, around $t \sim \sqrt{N}$ the decrease of the OTOC is halted and increases linearly before saturating at the plateau time $t \sim 2N$~\cite{garcíagarcía2025}.
Conversely, some non-integrable systems with clear random matrix spectral statistics do not show an exponential growth of $C(t)$~\cite{weak_quantum_chaos, Dora_2017}. 
As these considerations suggest, no single diagnostic fully captures the multifaceted nature of quantum chaos, and complementary tools—static and dynamical—must be used together for a reliable diagnosis.

\subsection{Chaos markers across different models} \label{sec:time_comparison}

Together, these spectral and dynamical probes provide a coherent set of chaos diagnostics that span a complementary range of temporal and energy scales. In models like YSYK---where SYK$_2$-like single-particle and SYK$_4$-like many-body features coexist---this composite view resolves competing effects and delineates the underlying dynamical regimes.

One aim of our work is to permit for a quantitative comparison of the chaos markers in the YSYK and complex SYK$_q$ models. This requires---in particular for the SFF and OTOC---rescaling the coupling constants to suitably match the characteristic energy scales of both models. To this end, as an operational procedure we impose that the time evolution of two models coincide at early times. As we will see, this rescaling makes the dynamics of the models match quantitatively even at much later times.  

For the SFF, a short-time expansion of Eq.~\eqref{Eq:SFF} yields
\begin{align}
    K(t) & = 1 - t^2 \frac{\rm{Tr} (H^2)}{\rm{Tr} \mathds{1}} + t^2\frac{\rm{Tr}(H)^2}{(\rm{Tr} \mathds{1})^2} + \mathcal{O}(t^3) \nonumber \\ 
    & = 1 - \sigma_H^2 t^2 + \mathcal{O}(t^4) \, .
\end{align}
Imposing that the early-time evolution of two different Hamiltonians, $H$ and $H^\prime$, possibly acting on distinct Hilbert spaces, be identical requires rescaling the time coordinate as 
\begin{equation}
    t^\prime = \alpha_{\text{SFF}} t \, , \qquad \alpha_{\text{SFF}} = \sigma_{H}/\sigma_{H^\prime}
\end{equation}

For rescaling the OTOC, instead of considering its direct short-time expansion, we equivalently evaluate the commutator squared $C(t)$ defined in Eq.~\eqref{Eq: commutator_squared}. For the simplified case in which $A$ and $B$ are Hermitian unitary operators and $[A,B]=0$, we obtain
\begin{equation}
    C(t) = \frac{t^2}{\mathrm{Tr}\, \mathds{1}} \mathrm{Tr}\, \left( \Big[[H,A],B\Big]^2 \right) + \mathcal{O}(t^3),
\end{equation}
and therefore the OTOC time rescaling is given by
\begin{equation}
\label{eq:C_OTOC_def}
    t^\prime = \alpha_{\text{OTOC}} t \, , \quad \alpha_{\text{OTOC}} = \sqrt{\frac{\mathrm{Tr}\, \left( [[H,A],B]^2 \right) / \mathrm{Tr}\, \mathds{1}}{\mathrm{Tr}\, \left( [[H^\prime,A],B]^2 \right) / \mathrm{Tr}\, \mathds{1}^\prime}} \, .
\end{equation}
As we will see in the next section, these rescalings permit quantitative matching of YSYK observables to those of SYK$_2$ and SYK$_4$, respectively. 

\section{Results}
\label{Sec:Results}

\begin{figure*}[ht!]
    \centering \includegraphics[width=1.0\textwidth]{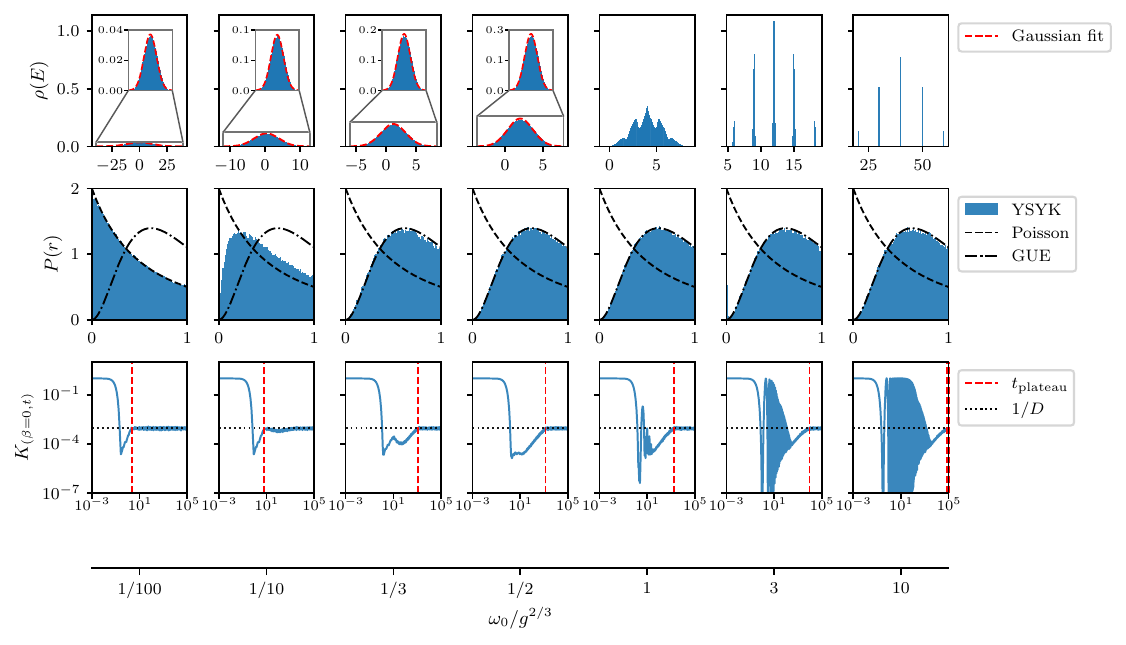}
    \caption{Spectral diagnostics of the YSYK model with $N=8$ fermionic and $M=4$ bosonic modes (occupation cutoff $N_b$=1) across the coupling range of the YSYK model.  
    \textbf{Top row:} Density of states $\rho(E)$. 
    At small $\omega_0/g^{2/3}$, the distributions closely follow a Gaussian fit (red dashed curves).  At
    $\omega_0/g^{2/3}\sim 1$ the DOS develops bands separated by the boson mass scale $\omega_0$. 
    \textbf{Middle row:} The gap-ratio distribution $P(r)$ is close to Poisson at
    $\omega_0/g^{2/3}=1/100$ and shifts toward GUE (dash-dotted)  as $\omega_0/g^{2/3}$ increases. 
    \textbf{Bottom row:} Spectral form factor $K(t)$ at $\beta=0$.  For small
    $\omega_0/g^{2/3}$ there is a steep superlinear ramp, typical of single-particle chaotic systems. 
    Around $\omega_0/g^{2/3}\sim 1/2$ a clear linear ramp emerges, a feature of many-body spectral rigidity. 
    For larger $\omega_0/g^{2/3}$, the linear ramp persists, but with $(2\pi/\omega_0)$-periodic modulations due to DOS clustering.  
    The dotted black line
    marks the plateau value $1/D$, and the red dashed vertical line indicates the
    extracted \emph{plateau time}. All the data are averaged over $500$ disorder realizations.}
 \label{fig:master_diagram_M4_N8}
\end{figure*}

In this section, we examine how the chaos markers introduced above evolve as the competing energy scales of the YSYK model are tuned. Our analysis combines exact diagonalization of the Hamiltonian in Eq.~\eqref{eq:hamiltonian_explicit}, which resolves the crossover phenomenology in the mesoscopic regime, with perturbative treatments of the asymptotic small- and large-$\omega_0$ limits, which anchor the emergence of SYK$_2$- and SYK$_4$-like behavior, respectively. Because the bosonic sector is infinite-dimensional, the numerics are performed with a local bosonic occupation cutoff $N_b$; unless stated otherwise, we focus on the hard-core case $N_b=1$, which already captures the relevant crossover phenomenology and the limiting behaviors discussed below. Details of the numerical implementation are given in Appendix~\ref{Appendix: numerical_implementation}, while the dependence on the bosonic cutoff is discussed in Appendix~\ref{Appendix:finite_size_effects_OTOCs}.
Section~\ref{sec:phase_diagram} provides an overview showing that varying the fermion--boson coupling (or equivalently tuning the boson mass) provides a tunable interpolation between single-particle-dominated behavior and strongly interacting, many-body chaos. Sections~\ref{sec:strong} and~\ref{sec:weak} then analyze in detail the limits of strong and weak fermion–boson coupling, respectively, demonstrating that the YSYK model quantitatively reproduces SYK$_2$-like behavior in the strong-coupling regime and SYK$_4$-like behavior in the weak-coupling regime.

\subsection{Mapping Chaos Across Coupling Strengths}
\label{sec:phase_diagram}

We start by presenting a synoptic characterization of the YSYK model as we tune the dimensionless ratio $\omega_0/g^{2/3}$, which controls the relative strength of the interactions with respect to the bare boson mass.
Figure~\ref{fig:master_diagram_M4_N8} assembles three complementary indicators: the DOS $\rho(E)$, the gap-ratio distribution $P(r)$, and the infinite temperature SFF $K(t)$.

In the strong-coupling limit, $\omega_0/g^{2/3}\to 0$, the boson mass term of the Hamiltonian becomes negligible compared to the interaction contribution to the Hamiltonian. The system is then dominated by a quadratic fermion theory with hopping strengths renormalized by the bosons. Consistent with this picture, at very small $\omega_0/g^{2/3}$ our numerics show a Gaussian DOS envelope with Poissonian gap-ratio distribution and a SFF which follows a superlinear ramp, as expected from a system that is chaotic only at the single-particle level. 
As $\omega_0/g^{2/3}$ is increased, the DOS remains smooth and close to Gaussian, but the local level statistics change rapidly: $P(r)$ moves from Poisson to GUE, indicating significant level repulsion. The SFF mirrors this evolution: the initial superlinear growth at $\omega_0/g^{2/3}=0.1$ is progressively replaced by an emerging linear ramp whose onset shifts to earlier times, until the superlinear behavior vanishes entirely around $\omega_0/g^{2/3}=0.5$, signaling the build-up of long-range spectral rigidity characteristic of many-body quantum chaos. 

The signatures of strong many-body chaos—GUE-like level statistics and a clear linear ramp in the SFF—remain robust up to around $\omega_0/g^{2/3}=1$. In this regime, the boson mass term in the Hamiltonian starts to shape the DOS, which develops $N_b (M+1)$ distinct peaks (one for each boson occupation sector) separated by the boson frequency scale $\omega_0$. Upon increasing $\omega_0/g^{2/3}$ further, boson-occupancy sectors split the DOS into well-separated clusters; $P(r)$ stays GUE-like, and the SFF retains its linear ramp but acquires early-time oscillations with period $2\pi/\omega_0$---an imprint of the $\omega_0$-spaced clustering that shifts the ramp onset to later times.
Taken together, the DOS, gap-ratio distributions, and SFFs in Fig.~\ref{fig:master_diagram_M4_N8} support a unified interpretation based on the competition of energy scales. At small $\omega_0/g^{2/3}$, the system is governed by effectively quadratic, single-particle-chaotic dynamics, as reflected in the Gaussian DOS envelope, Poissonian many-body level statistics, and superlinear SFF ramp. As $\omega_0/g^{2/3}$ increases, many-body spectral rigidity builds up and is strongest in the broad intermediate window where a clear linear ramp emerges. For still larger $\omega_0/g^{2/3}$, the spectrum splits into boson-occupancy bands: the global harmonic structure weakens spectral rigidity across bands, while chaotic correlations persist within each band. The sensitivity of this overall picture to the accessible system sizes and to the ratio $M/N$ is discussed in Appendix~\ref{Appendix:gap_ratio_finite_size}, where we use the mean gap ratio as a compact diagnostic. The next subsections refine these observations into quantitative diagnostics in the strong- and weak-coupling regimes, respectively.

\subsection{Strong fermion--boson coupling: SYK$_2$-like regime}
\label{sec:strong}

In the strong fermion–boson coupling regime, the interaction term dominates the Hamiltonian in Eq.~\eqref{eq:hamiltonian_explicit}. 
Neglecting the bosonic mass term allows one to choose a basis for the bosons that diagonalizes their contribution to the interaction term (given by the operator $a_k+a_k^\dagger$). 
Within each sector of bosonic eigenvalues, the resulting Hamiltonian becomes a quadratic fermionic theory 
whose spectrum lacks the long-range rigidity characteristic of fully chaotic systems, and the bosons simply rescale the fermion interaction strength. The resulting late-time dynamics is also consistent with a perturbative picture of weakly broken integrability in the small-$\omega_0$ regime (see Eq.~\eqref{Eq: SYK2_time_rescaling}). 
In this section, we aim (i) to make these statements more robust through a quantitative comparison with the complex SYK$_2$ model, and
(ii) to track how small but finite boson mass restores genuine many-body chaos. 
To this end, we focus on two complementary diagnostics: the SFF $K(t)$, tracking the scaling of the ramp and plateau times, and  the infinite-temperature OTOC $F(t)$.

\subsubsection{Spectral Form Factor}
\begin{figure}[t!]
    \label{fig:syk2_sff_inset}
    \centering
    \includegraphics[width=1.0\linewidth]{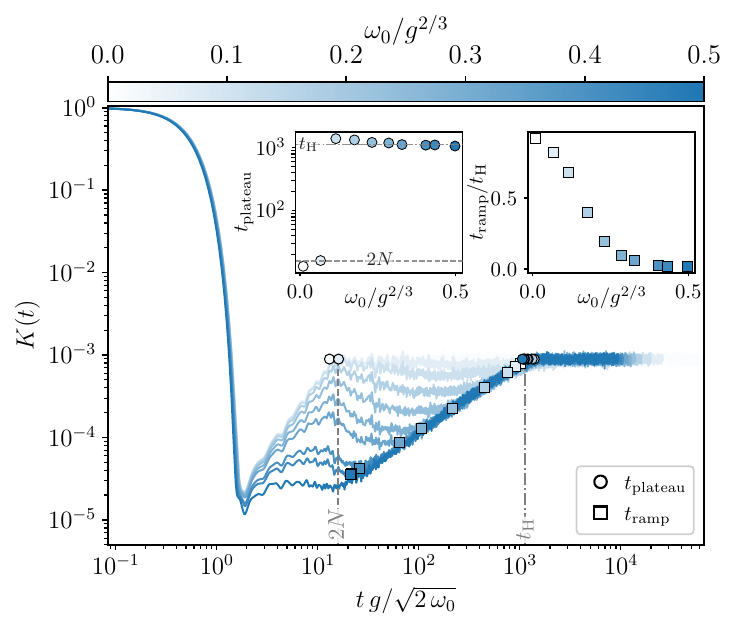}
    \caption{\textbf{Main panel:} Log-log plot of the SFF $K(t)$ versus the rescaled time $tg/\sqrt{2\omega_0}$ for $N=8$ fermionic modes and $M=4$ bosonic modes with cutoff $N_b=1$.
    Curves are color–coded by $\omega_0$ (color bar at top) in the range $\omega_0 \in \left[0.01,0.5\right]$. Dashed vertical lines mark the SYK$_2$ plateau time $2N$ and the Heisenberg time $t_\mathrm{H}$. Solid circles indicate the numerically extracted plateau–onset times $t_{\rm plateau}(\omega_0)$, while squares denote 
    the fitted ramp–onset times $t_{\rm ramp}(\omega_0)$. 
    \textbf{Insets:} (top middle) Linear–scale plot of $t_{\rm plateau}$ versus $\omega_0$, with horizontal dashed lines at 
    $t_\mathrm{H}$ and at the SYK$_2$ plateau time $2N$. For very small $\omega_0$, the numerics cannot resolve deviations from SYK$_2$–dominated behavior. From around $\omega_0=0.1$ onwards, the plateau time is close to the Heisenberg time, indicative of fully many–body chaotic behavior. 
    (top right) Ratio $t_{\rm ramp}(\omega_0)/t_H$ as a function of $\omega_0$. The ramp time becomes parametrically smaller than the Heisenberg time as $\omega_0$ increases, indicating the appearance of many-body chaos. }
    \label{fig:syk2_sff_full}
\end{figure}

Throughout our analysis of the strong–coupling regime, we adopt the time rescaling introduced in Sec.~\ref{sec:time_comparison} to compare YSYK dynamics with those of the complex SYK$_2$ model.
As shown in Appendix~\ref{Appendix: comparison coupling constants_small omega}, for the case of hard-core bosons ($N_b=1$), the rescaled time for the YSYK model,
\begin{equation}
\label{eq:SYK2_time_rescaling}
    \alpha \, t = \frac{g t}{\sqrt{2 \omega_0}} \, , \qquad \alpha = \alpha_\text{SFF} \approx \alpha_{\rm OTOC} \, ,
\end{equation}
is to be compared with the SYK$_2$ time evolution in units of $Jt$. This rescaling also allows for a quantitative comparison between YSYK realizations with different values of $\omega_0$ in regimes dominated by single-particle dynamics.

Figure~\ref{fig:syk2_sff_full} shows how the SFF interpolates from single-body to many-body chaos as $\omega_0$ is increased. At very small $\omega_0$, the SFF reaches the plateau at time $\alpha t_{\rm plateau} =2N$, which exactly matches the plateau time for the SYK$_2$ model~\cite{Winer2020,Liao2020}. For slightly larger $\omega_0$, the SFF still reaches close to the plateau height around $\alpha t_{\rm plateau} =2N$, but afterwards bends downwards and develops a secondary dip, before it rises again to settle into the plateau at around $t_\mathrm{H}$. 
As $\omega_0$ increases, the height of the first peak decreases and the behavior after the secondary dip evolves into a linear ramp, signaling the emergence of many-body level repulsion.
A similar behavior also emerges in mass-deformed SYK~\cite{Nosaka:2018iat}, with a superlinear ramp at early times, followed by an intermediate decrease and a linear ramp that leads into the final plateau. 
The presence of these intermediate features is typical of systems that are not yet fully chaotic, reflecting dynamical regimes such as delayed thermalization~\cite{Nandy22} or Hilbert-space diffusion~\cite{Baumgartner:2024orz}.

The crossover from single-particle chaos to many-body chaos can be tracked through the evolution of the plateau time, indicated by the circles in Fig.~\ref{fig:syk2_sff_full} and in the top-middle inset.
At around $\omega_0 = 0.1$, $\alpha t_{\rm plateau}$ sharply departs from the SYK$_2$ value of $2N$ and approaches the Heisenberg time $\alpha t_{\rm H} \sim D$. 
A further indicator for the increasing many-body chaotic influence is given by the ratio $t_{\rm ramp}/t_{\rm H}$ (top-right inset), which decreases from unity toward zero, signaling that the Thouless time becomes parametrically shorter than the Heisenberg time and many-body chaos has emerged.

\subsubsection{Out-of-time-order Correlator}

\begin{figure}
  \centering
  \includegraphics[width=1.0\linewidth]{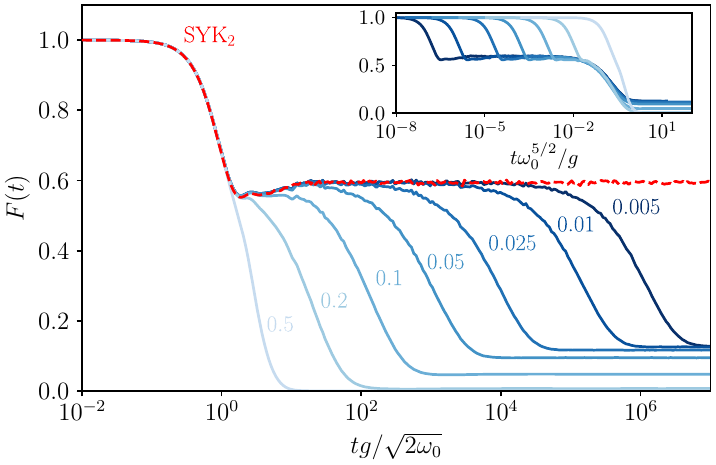}
  \caption{Disorder averaged OTOCs of the YSYK model for small values of $\omega_0$ (blue lines) with $N=8$ fermions and $M=4$ bosonic modes with $N_b = 1$, averaged over $1000$ samples, alongside the target complex SYK$_2$ model (red-dashed) with same number of fermions, averaged over $10 000$ samples. For small enough boson mass $\omega_0$, $F(t)$ shows a two-step scrambling process: it first reaches a prethermal plateau that coincides with the non-zero saturation value of SYK$_2$ before decaying further and becoming fully scrambling. Larger values of $\omega_0$ do not show a prethermal plateau, indicating the non-perturbative deviation from the SYK$_2$ regime. Inset: The OTOCs for sufficiently small $\omega_0$ collapse when time is rescaled by the characteristic scale $\omega_0^{5/2} g$ derived from a perturbative analysis (see Appendix~\ref{Appendix: Magnus_expansion_small_omega0}).}
  \label{Fig: OTOC_cSYK2_regime_latetimes}
\end{figure}

We now turn our discussion to the OTOC $F(t)$ as defined in Eq.~\eqref{Eq:OTOC}. In what follows, we consider the infinite temperature case $\beta=0$ and take the unitary Hermitian operators $A = 2c^\dag_i c_i - 1$ and $B= 2c^\dag_j c_j - 1$, for arbitrary $i,j$. In Fig.~\ref{Fig: OTOC_cSYK2_regime_latetimes}, we plot the OTOCs for several small values of $\omega_0 = 0.005, \ldots, 0.5$ alongside that of complex SYK$_2$ with the same number of complex fermions $N$. 
After the initial decay up to the scrambling time, the time evolution of the complex SYK$_2$ OTOC shows a superlinear ramp starting at $\alpha t \sim \sqrt{N}$ before saturating to a non-zero value---characteristic of integrable systems that are not fully scrambling---at the complex SYK$_2$ SFF plateau time, $\alpha t = 2N$~\cite{garcíagarcía2025}. As shown in Ref.~\cite{garcíagarcía2025} for SYK$_2$, the non-decaying behavior of the OTOC at late times arises because this regime is dominated by spectral statistics, exhibiting strong similarities with the SFF discussed above.

The YSYK OTOCs closely follow the early-time decay and ramp behavior of the complex SYK$_2$ model for small $\omega_0$, and saturate to the same intermediate plateau value. 
However, at sufficiently large times, we observe an $\omega_0$-dependent deviation from the SYK$_2$ saturation value. 
This behavior can be explained as follows. 
In the limit where the boson mass term is negligible, $\omega_0\to 0$, the YSYK Hamiltonian can be diagonalized over the different bosonic sectors. Each sector is governed by a quadratic fermionic Hamiltonian, with conserved charges given by the occupation number of the fermionic normal modes. 
When $\omega_0$ is small but finite, these symmetries are weakly broken. 
The early-time dynamics are governed by an effective quadratic fermionic theory, with a large number of quasi-symmetries that constrain scrambling to occur within fixed sectors of the Hilbert space. 
As the system evolves, the effects of a finite boson mass become significant at an $\omega_0$-dependent time scale, at which transitions between different symmetry sectors are enabled that allow the system to explore the full Hilbert space. Consequently, the system becomes fully scrambling at late times and $F(t)$ further decays to zero. A similar two-step scrambling process has been observed in~\cite{Rakovszky2018, Dieplinger2023} and has been linked to the emergence of a prethermal plateau.

The time scale for this secondary scrambling process can be estimated by treating the boson mass term in the Hamiltonian as a small perturbation. As detailed in Appendix~\ref{Appendix: Magnus_expansion_small_omega0}, this analysis reveals the emergence of a new time scale, $t \sim g/\omega_0^{5/2}$, beyond which many-body interactions start to kick in causing the OTOCs to deviate from the intermediate SYK$_2$ plateau and to decay further. As the inset of Fig.~\ref{Fig: OTOC_cSYK2_regime_latetimes} shows, the late time decay in $F(t)$ for different values of $\omega_0$ collapses onto a single curve when time is rescaled accordingly. 
The data for $\omega_0 = 0.5$ does not follow this collapse, since this boson mass no longer constitute a perturbative deviation from the single-body chaotic regime. Indeed, for larger values of $\omega_0 = 0.2, 0.5$, the intermediate plateau disappears and the OTOCs directly become fully scrambling, as one would expect from a many-body chaotic system.

The final late-time saturation value of the OTOCs for the YSYK model is not exactly zero due to finite-size effects, but is expected to go to zero in the thermodynamic limit~\cite{Finite_size_OTOC}. In Appendix~\ref{Appendix:finite_size_effects_OTOCs}, we demonstrate that the late-time saturation value of the OTOCs indeed decreases with increasing $N$. We also show that a finite boson cutoff does not fundamentally alter the behavior of the OTOC, once the time has been properly rescaled.
\subsection{Weak-coupling regime: Emergent SYK$_4$ bands}
\label{sec:weak}
We now consider the weak-coupling regime, $\omega_0 / g^{2/3} \gg 1$, where the energy spectrum of the YSYK Hamiltonian fragments into well‐separated and approximately equally spaced clusters due to the dominating boson mass term. 
Thanks to the Yukawa coupling, each subspace of fixed boson occupancy becomes dressed by the random fermion–boson interactions, which causes level repulsion inside each cluster. As we will see, the finite bandwidth of the clusters is governed by $\Delta E_{\rm cl}\propto{g^2}/{\omega_0^2}$.

Just as for the SYK$_2$-limit, it is important to match the corresponding time scales between the YSYK and the complex SYK$_4$ models. As detailed in Appendix~\ref{Appendix:large_omega_rescaling}, in this regime the adequate rescaling factor for both the SFF and the OTOC is given by
\begin{equation}
\label{eq:small_coupling_rescaling}
    \alpha_\# \, t = C_{\#} \, \frac{g^2}{\omega_0^2} t \, ,
\end{equation}
where $C_{\#}$ is an $N,M$-dependent numerical constant which differs between the OTOC ($C_{\rm OTOC}$) and the SFF ($C_{\rm SFF}$). In this regime, the bosonic cutoff becomes progressively irrelevant, since the low-energy dynamics are dominated by the lowest bosonic sector.

\begin{figure}[t!]
  \centering
  \includegraphics[width=1.0\linewidth]{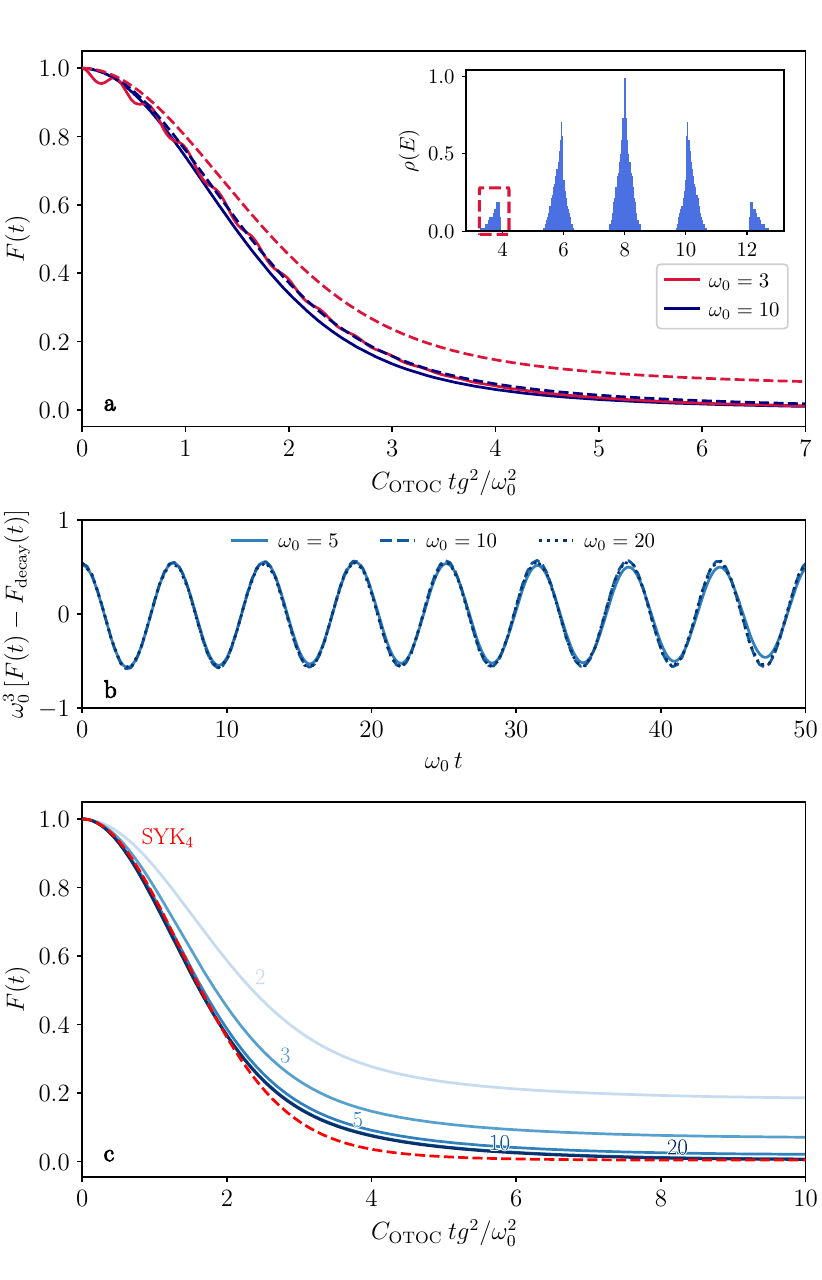}
  \caption{Disorder averaged YSYK OTOCs $F(t)$ for $N=8, M=4$ and $N_b=1$ in the large $\omega_0$ limit averaged over $1000$ samples. 
  \textbf{(a)} 
  $F(t)$ for $\omega_0 = 3$ and $10$ versus the rescaled time in Eq.~\eqref{eq:small_coupling_rescaling}. The full OTOC (solid lines) shows oscillations that decrease with $\omega_0$. These are absent when restricting the OTOCs to the lowest energy peak in the DOS (dashed lines). The difference between full and restricted OTOC decreases as $\omega_0$ increases.  
  Inset: A representative DOS $\rho(E)$ for $\omega_0=2$, showing distinct energy peaks, with the lowest energy peak highlighted. 
  \textbf{(b)} The oscillating part of $F(t)$,  obtained using a numerical low-pass (Savitzky–Golay) filter. Rescaling the amplitude by $\omega_0^3$ and plotting with respect to time $\omega_0 t$ lets curves for different values of $\omega_0$ collapse. 
  \textbf{(c)} $F(t)$ restricted to the low energy peaks. As the boson mass $\omega_0$ increases, the data show improved agreement with the SYK$_4$ OTOC with same number of complex fermions. 
  Curves for $\omega_0=10$ and $20$ essentially coincide, signaling the convergence of $F(t)$ for large $\omega_0$.}
  \label{Fig: OTOC_cSYK4_regime_combined}
\end{figure}
\subsubsection{Out-of-time-ordered Correlators}
In Fig.~\ref{Fig: OTOC_cSYK4_regime_combined}{\bf a}, we plot the disorder averaged YSYK OTOCs in the weak coupling limit for $\omega_0 = 3$ and $10$. The solid curves correspond to the full $F(t)$, whereas the dashed curves are OTOCs with contributions coming only from the lowest-energy peak in the DOS (inset). Numerical details on the OTOC calculations are provided in Appendix~\ref{Appendix: OTOC_numerics}. The clustering of the spectrum is reflected in the OTOCs: the full $F(t)$ shows oscillatory behavior overlaid on the decay, originating from the different boson-filling sectors in the energy spectrum; OTOCs restricted to the low-energy peak do not show such oscillations.

As the boson mass is increased, the low-energy-peak OTOC better approximates the full-spectrum one. 
Quantitatively, the oscillation amplitude decreases as $\omega_0^{-3}$, while the period scales as $\omega_0^{-1}$. 
This behavior is typical of off-resonant Rabi oscillations~\footnote{The unusual $\omega_0^{-3}$ behavior in the damping of the amplitude is due to the presence of an additional factor $1/ \sqrt{\omega_0}$ in the interaction term in Eq.~\eqref{eq:hamiltonian_explicit}.} and micromotion in systems subject to fast periodic driving~\cite{Goldman:2014xja,Eckardt:2015mtt}. 
To illustrate this behavior, in Fig.~\ref{Fig: OTOC_cSYK4_regime_combined}{\bf b} we subtract the decaying part of $F(t)$ from the full OTOC to isolate the oscillations. Upon rescaling the amplitude and time axes accordingly, the data for different $\omega_0$ collapses, confirming the abovementioned scaling for both quantities. 

Lastly, Fig.~\ref{Fig: OTOC_cSYK4_regime_combined}{\bf c} compares the OTOC of the low-energy peak of YSYK with the one of complex SYK$_4$ with the same number of complex fermions. With increasing values of $\omega_0$, the YSYK OTOCs more closely approach the early-time decay of the SYK$_4$ model and, at late times, saturate to zero---as expected for a chaotic model with fully scrambling behavior. If $\omega_0/g^{2/3}$ is not sufficiently large, as is the case, e.g., for $\omega_0=2$, then the OTOC clearly deviates from the SYK$_4$ result, signaling a departure from the fast-scrambling regime. 

\subsubsection{Effective Schrieffer–Wolff Hamiltonian}
The tendency of the YSYK OTOCs to approach the SYK$_4$ behavior with increasing boson mass $\omega_0$ can be understood in degenerate perturbation theory through a Schrieffer--Wolff (SW) transformation, which systematically integrates out the fast bosonic excitations to produce a purely fermionic Hamiltonian (see Appendix~\ref{Appendix: SWT} for details). To second order in $g/\omega_0$, one gets
\begin{align}
   H_{\rm eff}
   &= P_0\,H\,P_0 + \frac{1}{2}\,P_0\,[S,\,V]\,P_0 
   \nonumber\\[4pt]
   &=     -\frac{1}{2\omega_0^2N}
     \sum_{i,j,i',j'} 
       \left(\frac{1}{M}\sum_{k}g_{ij,k}g_{i'j',k}\right)
       c_i^\dagger c_j
       c_{i'}^\dagger c_{j'}.
   \label{eq:Heff_SYK4_like}
\end{align}
Here $P_0$ denotes the projector onto the zero-boson subspace, $S$ solves $[S,H_0]=-V$, and $H_0$ and $V$ are the boson mass term and Yukawa-interaction parts of the Hamiltonian in Eq.~\eqref{eq:hamiltonian_explicit}, respectively. 

The effective Hamiltonian has a quartic, SYK$_4$-like interaction, with a coupling strength that scales as $g^2/\omega_0^2$, consistent with the rescaling anticipated in Eq.~\eqref{eq:small_coupling_rescaling}. 
This perturbative analysis elucidates the behavior of the OTOCs: for large $\omega_0$-values, the SW transformation becomes increasingly accurate, and the effective Hamiltonian acting within the subspace $P_0$ (corresponding to the lowest-energy peak) provides an increasingly better approximation to the full OTOC.

Moreover, in the large-$\omega_0$ limit, the bosonic degrees of freedom become effectively frozen, such that each energy peak in the DOS is dominated by purely fermionic interactions with no oscillatory behavior.
In Appendix~\ref{Appendix:finite_size_effects_OTOCs}, we explore some of the finite-size effects on the OTOCs and show that, for a smaller number of boson modes $M$, the agreement with complex SYK$_4$ worsens. This is not surprising, as the rank of the tensor coupling the four fermions, $\frac{1}{M}\sum_{k}g_{ij,k}g_{i'j',k}$, directly depends on the number of bosonic modes, and an insufficient rank fails to reproduce exact SYK$_4$ dynamics~\cite{Bi:2017yvx,Kim_2020}.

\begin{figure}[t!]
    \centering
    \includegraphics[width=1.0\linewidth]{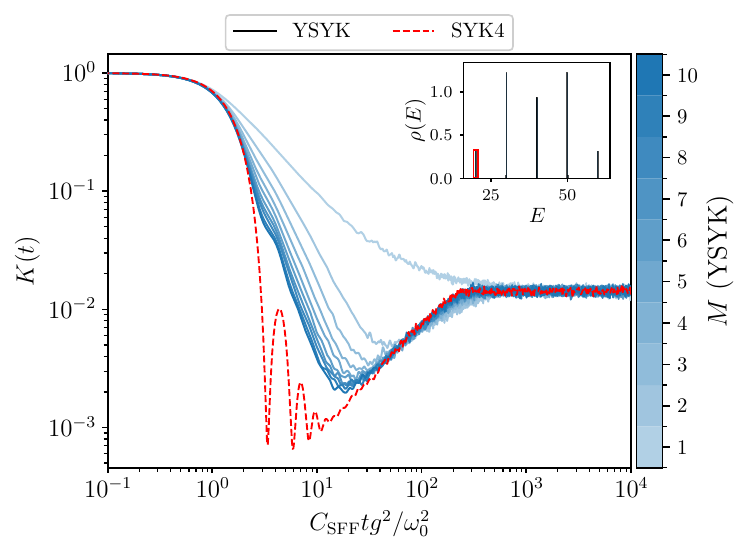}
    \caption{Disorder averaged SFF of the YSYK model in the weak fermion-boson coupling regime, where the DOS decomposes into distinct clusters. 
    When choosing the correct rescaling of time, SFFs for increasing number of bosonic modes $M$ (darker shades of blue) better approximate the complex SYK$_4$ benchmark (red dashed). 
    Data correspond to $N=8$ spinless fermions at half filling, $N_b=1$, and averages over 500 disorder realizations.
    \textbf{Inset:} Disorder-averaged DOS, exemplified for $\omega_0=10$. The red box marks the lowest-energy peak, whose levels are used in the SFF. 
    }
    \label{fig:syk4_limit}
\end{figure}

\subsubsection{Spectral Form Factor}
As anticipated in Fig.~\ref{fig:master_diagram_M4_N8} (bottom row, rightmost panels) and consistent with the discussion above, the early-time dynamics of the full SFF in the small $\omega_0$ regime is dominated by fast oscillations governed by the bosonic mass scale. 
Restricting our analysis to the lowest-energy peak allows us to eliminate these fast oscillations, 
providing a clearer view of the emergence of many-body correlations and enabling a direct comparison with the complex SYK$_4$ model.
In Fig.~\ref{fig:syk4_limit}, we compare the SFF for the energy levels within the lowest-energy cluster (highlighted by the red box in the inset) with that of complex SYK$_4$  (red dashed line) at different numbers of bosonic modes $M$. After rescaling the time axis according to Eq.~\eqref{eq:small_coupling_rescaling}, with $C_{\rm SFF}$ defined in Appendix~\ref{Appendix:large_omega_rescaling}, the YSYK SFF exhibits the characteristic ramp–plateau structure, with plateau height and Heisenberg time consistent with a chaotic system of $N$ fermions at half-filling. 
As the rank of the effective four-fermion interaction given in Eq.~\eqref{eq:Heff_SYK4_like} increases with the number of bosonic modes $M$, the ramp of the YSYK model progressively approaches the one of the SYK$_4$ SFF, signaling proper spectral rigidity.

Nonetheless, the correspondence with the SYK$_4$ model is not exact, particularly in the non-universal slope region. This discrepancy arises because normal-ordering the effective Hamiltonian in Eq.~\eqref{eq:Heff_SYK4_like} generates a disordered quadratic term that is absent in the SYK Hamiltonian in Eq.~\eqref{eq:SYKq_H}. Although this additional term is subextensive, and thus becomes negligible in the large-$N,M$ limit~\footnote{Differently from the SYK$_4$ model, some couplings in the effective Hamiltonian in Eq.~\eqref{eq:Heff_SYK4_like} do not have vanishing mean. One can then wonder if these can produce significant quadratic contributions. Fortunately, the only term that is extensive, and which may thus be significant in the thermodynamic limit, is an irrelevant constant shift when we work at fixed filling.}, it can still produce noticeable effects for the relatively small system sizes considered here: 
(i) it alters the non-universal slope--dip part of the SFF, (ii) it bends the nominally linear ramp, yielding a mild sub-linear growth, and (iii) it smoothens the kink at the ramp--plateau junction characteristic of chaotic systems in the GUE universality class. In Appendix~\ref{Appendix: toy_model}, we study the behavior of a normal-ordered interaction term mimicking the structure of the effective Hamiltonian, which corresponds to a low-rank version of the complex SYK$_4$ model~\cite{Bi:2017yvx, Kim_2020}. This clean interaction term exactly reproduces the SFF of the complex SYK model when the number of bosonic modes is sufficiently large.

The analyses of the OTOC and SFF consistently indicate that the YSYK model reproduces fully chaotic, SYK$_4$-like behavior in the limit of large boson mass, or, equivalently, in the limit of weak fermion–boson coupling. This mechanism underlies recent proposals for realizing SYK$_4$ physics using cold atoms in cavity-QED platforms~\cite{cavity_proposal, Baumgartner:2024ysk}.
However, the YSYK model spans a broader range of physical regimes, further enhancing its appeal as a target for experimental exploration. 
In the next section, we show how an optical cavity platform can implement the YSYK Hamiltonian, achieving an interaction rate that significantly improves over earlier implementation proposals.

\section{Experimental Realization}
\label{sec:experiment}
While analytic path-integral approaches can tackle the YSYK model in the limit of large $N,M$~\cite{Schmalian_YSYK, Wan2020,wang_quantum_2020,pan_yukawa-syk_2021,davis_quantum_2023,HAUCK_YSYK,inkof_quantum_2022,Schmalian:2022web,esterlis2025quantum}, they have difficulties 
when one is interested in quantum corrections appearing in orders of $1/N$. On the other side, numerics are limited to small system sizes. 
This leaves an equally challenging as relevant intermediate regime open. Quantum simulators open the prospect to proceed into this regime.
The finite-size analysis developed in this work is a necessary step towards developing such quantum simulator experiments, as it provides quantitative benchmark predictions. 
In this section, we propose a quantum-simulation setup for YSYK in cavity-QED. 
Such an experimental realization is particularly compelling, as it naturally includes bosons (the cavity photons) without the need for any cutoff. 
More broadly, the ability of YSYK to interpolate between a disordered free-fermion regime and a strongly correlated, chaotic one makes it a versatile platform for exploring a wide range of quantum many-body phenomena beyond the original SYK model. These include the emergence of superconductivity from incoherent metallic states~\cite{Schmalian_YSYK, HAUCK_YSYK}, the evolution of quantum chaos across interaction strengths, and aspects of holography in table-top systems.

\subsection{Cavity setup}

Platforms like cold atoms in optical cavities are particularly well suited for this task. They naturally provide long-range interactions mediated by cavity photons~\cite{Mivehvar02012021}, which can play the role of the Yukawa field, see Fig.~\ref{fig:first_schematic}\textbf{c}. 
Moreover, these platforms offer excellent control over system size as well as tunable disorder~\cite{Baghdad_2023, sauerwein_engineering_2023, Forsi_cavity_control24}.
In this section, we show how the setup presented in Ref.~\cite{cavity_proposal,Baumgartner:2024ysk} can be adapted to simulate the YSYK model, leading to significantly faster timescales.

The starting point is the same light–matter Hamiltonian as in  Refs.~\cite{cavity_proposal, Baumgartner:2024ysk}, where the atomic gas inside the cavity is modeled as a fermionic spinor with two internal levels and mass $m_\mathrm{at}$, confined by a 2D trapping potential $V_{\mathrm{t}}(\mathbf{r})$. The matter is coupled to $M$ cavity modes and the transition between the ground ($\ket{g}$) and the excited $(\ket{e}$) electronic level is driven by a pump laser with Rabi frequency $\Omega_{\mathrm{d}}$.

Working in the frame rotating at the drive frequency $\omega_{\mathrm{d}}$ and applying the rotating‐wave approximation, the total many-body light-matter Hamiltonian reads
\begin{equation}
H_{\mathrm{mb}}
=H_{\mathrm{kt}}
+H_{\mathrm{a}}
+H_{\mathrm{c}}
+H_{\mathrm{ac}}
+H_{\mathrm{ad}},
\label{eq:micro_hamiltonian}
\end{equation}
with
\begin{align}
H_{\mathrm{kt}} &= \sum_{s=e,g}\int d^2r\,\psi_s^\dagger(\mathbf{r})\Bigl[-\frac{\nabla^2}{2m_\mathrm{at}}+V_{\mathrm{t}}(\mathbf{r})\Bigr]\psi_s(\mathbf{r}),\\
H_{\mathrm{a}} &= -\int d^2r\,\Delta_{\mathrm{da}}(\mathbf{r})\,\psi_e^\dagger(\mathbf{r})\psi_e(\mathbf{r}),\\
H_{\mathrm{c}} &= \sum_m\Delta_m\,a_m^\dagger a_m, \label{eq:H_c_def}\\
H_{\mathrm{ac}} &= \tfrac12\sum_m\int d^2r\,[\,\Omega_m\,g_m(\mathbf{r})\,a_m\,\psi_e^\dagger(\mathbf{r})\psi_g(\mathbf{r})+\mathrm{h.c.}],\\
H_{\mathrm{ad}} &= \int d^2r\,[\,\Omega_{\mathrm{d}}\,g_{\mathrm{d}}(\mathbf{r})\,\psi_e^\dagger(\mathbf{r})\psi_g(\mathbf{r})+\mathrm{h.c.}].
\end{align}

Here, $\psi_{g}(\mathbf{r})$ and $\psi_{e}(\mathbf{r})$  are the fermionic field operators for the ground and excited electronic state respectively, and $a_m$ annihilates a boson in the cavity mode $m$. The kinetic Hamiltonian $H_{\mathrm{kt}}$ governs the atomic dynamics, while $H_{\mathrm{a}}$ is the excited state energy in the rotating frame, as $\Delta_{\mathrm{da}}(\mathbf{r})=\omega_{\mathrm{d}}-\omega_{\mathrm{a}}(\mathbf{r})$ measures the laser detuning from the excited state transition frequency $\omega_{\mathrm{a}}(\mathbf{r})$. 

Further, $H_{\mathrm{c}}$ describes the free‐photon Hamiltonian in the rotating frame, with $\Delta_m=\omega_{\mathrm{c}}^{(m)}-\omega_{\mathrm{d}}$ the detuning between the bare cavity resonance $\omega_{\mathrm{c}}^{(m)}$ and the drive frequency.
The term $H_{\mathrm{ac}}$ describes the quantized atom–cavity coupling: each mode $m$ drives the $\ket{g}\leftrightarrow\ket{e}$ transition with single‐photon Rabi frequency $\Omega_m$ and spatial profile $g_m(\mathbf{r})$.  Finally, $H_{\mathrm{ad}}$ is the classical (laser) drive of the same atomic transition, with mode envelope $g_{\mathrm{d}}(\mathbf{r})$. 

One fundamental feature of the above model is that $\omega_{\mathrm{a}}(\mathbf{r})$
can be made spatially dependent by dressing the excited atomic state with an auxiliary laser whose frequency is near-resonant with a higher excited-state transition.
This scheme enables the introduction of engineered disorder into the system: by making the intensity profile of the auxiliary laser spatially disordered, e.g., by a speckle pattern \cite{sauerwein_engineering_2023,cavity_proposal}, the excited atomic level experiences a disordered AC stark shift, which propagates into the effective light--matter couplings in the fermionic ground state, randomizing them similarly to the targeted YSYK model.


To see this, we consider the dispersive regime, $|\Delta_{\mathrm{da}}(\mathbf{r})|\gg\abs{\Omega_{\mathrm{d}}},\abs{\Omega_m}$. In this parameter range, the excited‐state is far-off resonant and can be adiabatically eliminated. The associated field operator reaches a quasi‐steady solution, $\partial_t\psi_e\approx0$, and the Heisenberg equation imposes
\begin{equation}
\psi_e(\mathbf{r})\;\approx\;\frac{\bigl(\Omega_{\mathrm{d}}\,g_{\mathrm{d}}(\mathbf{r})+\tfrac12\sum_m\Omega_m\,g_m(\mathbf{r})\,a_m\bigr)}{\Delta_{\mathrm{da}}(\mathbf{r})}\;\psi_g(\mathbf{r}),
\end{equation}
where we neglected terms subleading in $\abs{\Omega_{\mathrm{d}}}/|\Delta_{\mathrm{da}}(\mathbf{r})|$, $\abs{\Omega_m}/|\Delta_{\mathrm{da}}(\mathbf{r})|$. 
Substituting this expression back into the atom–light terms of $H_{\mathrm{mb}}$ and keeping only terms up to $\mathcal{O}(\Omega_{\mathrm{d},m}^2/\Delta_{\mathrm{da}})$ yields
\begin{equation}
\begin{split}
H_{\mathrm{mb}} \simeq{}& H_{\mathrm{kt}} + \sum_m\Delta_m\,a_m^\dagger a_m \\
&+ \int d^2r\;\frac{\bigl|\Omega_{\mathrm{d}}\,g_{\mathrm{d}}(\mathbf{r})
+\tfrac12\sum_m\Omega_m\,g_m(\mathbf{r})\,a_m\bigr|^2}
{\Delta_{\mathrm{da}}(\mathbf{r})}\;\psi_g^\dagger(\mathbf{r})\psi_g(\mathbf{r}) \, .
\end{split}
\label{Eq:H_mb_AE_fermion}
\end{equation}
which provides the effective photon‐mediated interaction among ground‐state atoms.


Expanding the squared absolute value in Eq.~\eqref{Eq:H_mb_AE_fermion} and expressing
the ground-state field operator in terms of the lowest $N$ trap-level eigenmodes $\psi_g(\mathbf{r}) = \sum_{\ell} \phi_{\ell}(\mathbf{r})\,c_{\ell},$ yields three distinct contributions: (i) a spatially dependent AC–Stark shift, which can be compensated by a two-tones drive scheme and is therefore set to zero; (ii) a Yukawa-type coupling; and (iii) a photon-number–dependent term that can be made parametrically small by choosing $\abs{\Omega_{\mathrm{d}}} \gg \abs{\Omega_m}$. The detailed derivation is given in Appendix~\ref{app:light_matter_Hamiltonian}.
After these simplifications, the resulting many-body light–matter Hamiltonian reads
\begin{align}
H_{\rm eff}
= & \sum_i \epsilon_i \,c_i^\dagger c_i
+\sum_m \Delta_m \,a_m^\dagger a_m \nonumber \\
& +\frac12\sum_{m,i,j}(\mathcal{M}^{(1)}_{m,ij}\,a_m^\dagger+\mathrm{h.c.})\,c_{i}^\dagger c_{j},
\label{eq:experimental_YSYK}
\end{align}
where $\epsilon_i$ are the trap energy levels and 
\begin{equation}
    \mathcal{M}^{(1)}_{m,ij}
=\int d^2r\;\frac{\Omega_{\mathrm{d}}^*\,\Omega_m\,g_{\mathrm{d}}(\mathbf{r})\,g_m(\mathbf{r})}{\Delta_{\mathrm{da}}(\mathbf{r})}\;\phi_i^*(\mathbf{r})\phi_j(\mathbf{r}).
\label{eq:linear_int}
\end{equation}
The Hamiltonian in Eq.~\eqref{eq:experimental_YSYK} coincides with the one of the YSYK model in Eq.~\eqref{eq:hamiltonian_explicit} if we choose a mode family with the same frequency $\Delta_m = \Delta_{\mathrm{cd}}$~\cite{Johansen2022}
and in the case of small trap frequency, where we can neglect $\sum_i \epsilon_i \,c_i^\dagger c_i$. We then identify the Yukawa coupling of our model to scale as
\begin{equation}
\label{Eq:three-leg}
    \frac{g_{ij,k}}{\sqrt{2\omega_0 }}\sim \frac{\abs{\Omega_{\mathrm{d}}}\,\abs{\Omega_m}}{\abs{\Delta_{\mathrm{da}}}}.
\end{equation}
Notice that the same result can be obtained by targeting a single cavity mode and using a Trotterized cycle made of $M$ steps, each realizing a different speckle configuration~\cite{Baumgartner:2024ysk}.

To summarize, an effective YSYK model can be obtained when the following hierarchy of energy scales is satisfied:
\begin{equation}
    \text{YSYK:} \quad |\Delta_{\mathrm{da}}|\gg\abs{\Omega_{\mathrm{d}}} \gg \abs{\Omega_m}. 
    \label{eq:YSYK_scales}
\end{equation}

By imposing the further scale separation $\abs{\Delta_{\mathrm{cd}}}\gg\abs{\Omega_{\mathrm{d}}\,\Omega_m/\Delta_{\mathrm{da}}}$ [corresponding to large boson mass in Eq.~\eqref{eq:hamiltonian_explicit}], it is possible to perform a Schrieffer–Wolff transformation to project the Hamiltonian in Eq.~\eqref{eq:experimental_YSYK} onto the subspace of vanishing photon population,  and thus to recover the SYK$_4$ regime of Ref.~\cite{cavity_proposal}. 
As detailed in Appendix \ref{Appendix: SWT}, 
this procedure results in a purely fermionic quartic Hamiltonian 
\begin{equation}
     H_{\mathrm{eff}}^{(4)} =-\sum_{i,i',j,j'}J_{ii';\,jj'}\,c_i^\dagger c_{i'}\,c_j^\dagger c_{j'},
\label{eq:Heff_4fint}
\end{equation} 
with SYK$_4$‐like couplings given by
\begin{equation}
\label{Eq:four-leg}
    J_{ii';\,jj'}
=\sum_{k=1}^M\frac{\mathcal{M}^{(1)\,*}_{k,ii'}\,\mathcal{M}^{(1)}_{k,jj'}}{\Delta_{\mathrm{cd}}}
\;\sim\;
\frac{\abs{\Omega_{\mathrm{d}}^2}\,\abs{\Omega_m^2}}{\abs{\Delta_{\mathrm{da}}}^2\,\abs{\Delta_{\mathrm{cd}}}}.
\end{equation}
The required energy hierarchy now reads:
\begin{equation}
\label{eq:SYK4_scales}
    \text{SYK}_4:  |\Delta_{\mathrm{da}}|\gg\abs{\Omega_{\mathrm{d}}} \gg \abs{\Omega_m} \, , \quad \abs{\Delta_{\mathrm{cd}}}\;\gg\;\frac{\abs{\Omega_{\mathrm{d}}}\,\abs{\Omega_m}}{|\Delta_{\mathrm{da}}|}\,.
\end{equation}

\subsection{Experimental feasibility and parameter estimates}

A suitable implementation of Yukawa–SYK should satisfy the hierarchy in Eq.~\eqref{eq:YSYK_scales} and at the same time ensure sufficient disorder in the boson--fermion interactions to approach the random couplings of the ideal Yukawa-SYK model. To illustrate the accessibility of this regime, we consider a concrete example \cite{sauerwein_engineering_2023}, using $^6$Li atoms confined in a two-dimensional pancake by a harmonic trap at frequency $\omega_{\rm trap}/2\pi\range{20}{50}\,\mathrm{Hz}$, which corresponds to a harmonic oscillator length of $x_0=\sqrt{\hbar/(m_\mathrm{at}\,\omega_{\rm trap})}\approx6\text{–}9\,\mu\mathrm{m}$,
where $m_\mathrm{at}$ is the atom mass.
If we consider a near‑concentric cavity of waist $w_0\approx13\,\mu\mathrm{m}$, a light atomic gas such as $^6$Li yields a relatively large transverse spatial extent. For the above parameters, we obtain a transverse size ratio $
\zeta \,=\, \sqrt2 \, x_0 / w_0$ in the range $
[{0.65},{0.98}]$, indicating that the atomic cloud can resolve the fine spatial structure of a laser-speckle pattern, so the light–matter overlaps fluctuate strongly across modes and the couplings $g_{ij,k}$ approach the desired random limit \cite{cavity_proposal}.

Further, realistic values of the light--matter coupling of the $m$-th cavity mode can be on the order of $\abs{\Omega_m}/2\pi \range{1}{5}\, \mathrm{MHz}$.
Choosing the laser driving the $\ket g\leftrightarrow\ket e$ transition in the range $
\abs{\Omega_{\mathrm{d}}}/2\pi\range{0.01}{1}\,\mathrm{GHz}$ and
$|\Delta_{\mathrm{da}}|/2\pi \range{5}{10}\,\mathrm{GHz}$,
the hierarchy in Eq.~\eqref{eq:YSYK_scales} is comfortably satisfied. The boson mass scale $\omega_0$ can be set by choosing an accessible cavity–laser detuning range $\abs{\Delta_{m}}/2\pi \approx \abs{\Delta_{\mathrm{cd}}}/2\pi\range{50}{200}\,\mathrm{MHz}$. Plugging these values into Eq.~\eqref{Eq:three-leg}, one finds that the Yukawa coupling scale can remain in the Megahertz regime, 
$\frac{g}{\sqrt{2 \omega_0}} \sim \frac{\abs{\Omega_{\mathrm{d}}}\,\abs{\Omega_m}}{\abs{\Delta_{\mathrm{da}}}} \sim 2 \pi \times 
1 \mathrm{MHz}$. If we proceed with the adiabatic elimination of the photons, the effective SYK$_4$ coupling can scale as $J \sim \frac{\abs{\Omega_{\mathrm{d}}}^2\,\abs{\Omega_m}^2}{\abs{\Delta_{\mathrm{da}}}^2\,\abs{\Delta_{\mathrm{cd}}}}
  \sim 2 \pi \times 
  20 \mathrm{kHz}$. 
  Thus, four‑fermion interactions emerge at the kilohertz scale.
\subsection{Dissipation}
Cavity-QED systems are inherently open quantum systems. 
To observe coherent dynamics, the achieved interaction energy scale must exceed the relevant loss rates. 
There are two dominant decoherence channels: spontaneous decay with rate $\Gamma$ from the excited atomic state and photon losses from the cavity with rate $\kappa$. The associated Lindblad dynamics are outlined in Appendix~\ref{App:dressed-channels}. 
As in the dispersive regime the excited fermionic states are only weakly populated, the decoherence rate corresponding to excited-state decay acting on the ground-state population is effectively suppressed to $\tilde\Gamma 
    = \Gamma\,\frac{\abs{\Omega_{\mathrm{d}}}^2}{\abs{\Delta_{\mathrm{da}}}^2}$.

When in addition the detuning from the cavity photon is large (limit of large boson mass), the cavity photons are only virtually excited. In this regime, the effective four-point fermion interactions and the effective decoherence rate deriving from photon loss acting on the ground-state fermions are  
\begin{equation}
J = \frac{|\Omega_{\mathrm{d}}|^2\,|\Omega_m|^2}
               {|\Delta_{\mathrm{da}}|^2\,\bigl|\Delta_{\mathrm{cd}}+i\,\kappa/2\bigr|},
\quad
    \tilde\kappa 
    = \kappa\,\frac{\abs{\Omega_{\mathrm{d}}}^2\,\abs{\Omega_m}^2}
                     {\abs{\Delta_{\mathrm{da}}}^2\bigl(\abs{\Delta_{\mathrm{cd}}}^2+\kappa^2/4\bigr)}.
\label{J_dissipative}
\end{equation}

The severity of the decoherence versus the achieved interaction strengths can be quantified by the dimensionless figures of merit $g^2/(\kappa\, \tilde\Gamma)$ (Yukawa-SYK model) and $J^2/(\tilde\kappa\, \tilde\Gamma)$ (SYK$_4$ limit). 
Assuming $|\Delta_{\mathrm{cd}}|\gg \kappa/2$, so $\bigl|\Delta_{\mathrm{cd}}+i\kappa/2\bigr|\approx|\Delta_{\mathrm{cd}}|$, these ratios become
\begin{equation}
    \frac{g^2}{\kappa\,\tilde\Gamma}
    =
    \frac{J^2}{\tilde\kappa\,\tilde\Gamma}
    = \frac{\abs{\Omega_m}^2}{\kappa\,\Gamma}
    \;\equiv\;\frac{\eta}{4}.
\end{equation}
They are set entirely by the single-atom cooperativity $\eta$. With state-of-the-art parameters ($\kappa/2\pi\!\simeq\!0.5\,{\rm MHz}$, $\Gamma/2\pi\!\simeq\!6\,{\rm MHz}$, $\abs{\Omega_m}/2\pi\!\simeq\!2.6\,{\rm MHz}$) one finds $\eta\sim10$, so that both $g^2/(\kappa\,\tilde\Gamma)$ and $J^2/(\tilde\kappa\,\tilde\Gamma)$ are well above unity.  In other words, both the Yukawa-mediated and SYK$_4$-type scrambling processes can outpace photon leakage and spontaneous emission. 
By tuning the laser frequency $\omega_{\rm d}$ entering in the detuning $\Delta_{\rm da}$
, the YSYK realization can probe chaotic dynamics over a broad parameter window, also in regimes outside the SYK$_4$ hierarchy of Eq.~\eqref{eq:SYK4_scales}.

\section{Conclusion}

\label{sec:conclusion}

In this work, we introduced a spinless version of the Yukawa--SYK model, where fermions interact via boson-mediated couplings, and showed that it can naturally interpolate between single-particle and many-body chaos. By tuning the ratio $\omega_0/g^{2/3}$, which controls the competition between the Yukawa interaction and the boson mass scale, we identified a crossover from SYK$_2$-like dynamics at small $\omega_0$ to SYK$_4$-like behavior at large $\omega_0$. The crossover itself is resolved through finite-size spectral and dynamical diagnostics, while its two asymptotic limits are anchored by perturbative analyses that explain the emergence of the effective quadratic and quartic descriptions. In addition, the rescaling framework introduced here provides clear relations for a quantitative comparison with the SYK$_2$ and SYK$_4$ limits appearing at strong, respectively, low coupling.

In the extreme strong-coupling regime $\omega_0/g^{2/3} \ll 1$, the model shows signatures of an integrable system, exhibiting Poissonian spectral statistics, a SFF with a superlinear ramp, and an OTOC showing a pre-scrambling plateau, with quantitative agreement with predictions from the SYK$_2$ model up to long times $t\sim \omega_0^{-5/2}$. At slightly lower couplings, the crossover from integrable behavior to that of an interacting system is sharply manifested in both the SFF as a jump of plateau time to the Heisenberg time, and in the OTOC with the absence of a finite-value saturation.

In the weak-coupling regime $\omega_0/g^{2/3} \gg 1$, the system progressively approaches the chaotic SYK$_4$-like regime. The dynamics are dominated by the lowest-energy band in the spectrum, corresponding to the zero-boson occupancy sector, while higher energy bands manifest themselves through superimposed oscillations,  whose magnitude consistently decreases as $\omega_0$ increases. This behavior provides a concrete signature of the adiabatic decoupling of the higher-boson sectors, which leads to an effective model described by a four-fermion interaction.

The richness of the Yukawa--SYK model is reflected in the emergence of multiple, distinct time scales. Beyond the characteristic timescales governing the SYK$_2$ and SYK$_4$ regimes, we identify a secondary scrambling process in the strong-coupling limit. Specifically, the OTOCs display a delayed decay at late times, indicating a second stage of scrambling occurring on a slower timescale. This behavior suggests that, when the system is nearly integrable, weakly broken symmetries constrain the dynamics to quasi-conserved sectors during early-time evolution, resulting in a pre-scrambling plateau in the OTOC. Only at much later times does the system explore the full Hilbert space, leading to complete information scrambling. Similar multi-stage scrambling dynamics have been observed in local random unitary circuits with charge conservation~\cite{Rakovszky2018} as well as in the mass-deformed SYK model~\cite{Dieplinger2023}, which has been studied as a model for the transition between ergodic and many-body localized phases, exhibiting intermediate extended non-ergodic regimes~\cite{Monteiro2021, Dieplinger2021}. Similarly, the YSYK model can serve as a promising framework for investigating how these crossover signatures evolve with system size, in dialogue with broader questions of localization and thermalization~\cite{Basko_2006,Kravtsov_2015,Abanin:2018yrt}. Complementary complexity-oriented diagnostics may also help sharpen that comparison in future work~\cite{Santra2025}. 
Furthermore, it would be worthwhile to investigate how our infinite-temperature results extend to finite temperatures.

One of the most appealing features of the YSYK model, as discussed in Sec.~\ref{sec:experiment}, is its potential for scalable implementation on quantum simulators based on ultracold atoms in an optical cavity, where the disordered interactions can be engineered using a laser with spatially varying intensity~\cite{cavity_proposal}.
Our numerical results provide a reference against which to benchmark such a quantum simulator.
An experimental realization of YSYK would have far-reaching implications across several areas of physics, from the controlled study of quantum chaos and localization-related crossover phenomena to the experimental exploration of quantum black hole dynamics through the holographic duality~\cite{Sachdev:2015efa,Cotler2016}.
Moreover, the fact that cavity QED systems are intrinsically open quantum systems compels us to extend our theoretical understanding of chaos~\cite{Denisov:2018nif,Sa:2020fpf} and holography beyond isolated settings~\cite{Jana:2020vyx,Pelliconi:2023ojb}, incorporating the role of dissipation and decoherence~\cite{Kulkarni:2021gtt,Hosseinabadi:2023mid}. As these topics are currently attracting considerable attention, our proposal not only offers a path toward realizing strongly interacting, disorder-dominated models in the laboratory, but also opens new avenues for studying the interplay between quantum chaos, holography, and open-quantum system dynamics.

\begin{acknowledgments}
We would like to thank Rahel Baumgartner, Rohit Prasad Bhatt, Jean-Philippe Brantut, Léa Dubois, Ekaterina Fedotova, Robin L\"{o}wenberg, Francesca Orsi, Pietro Pelliconi, Julian Sonner, and Yi-Neng Zhou for helpful discussions and comments. 

\paragraph*{Funding.}
This project has received funding by the European Union under Horizon Europe Programme - Grant Agreement 101080086 - NeQST and under NextGenerationEU via the ICSC – Centro Nazionale di Ricerca in HPC, Big Data and Quantum Computing. This project has been funded by the Caritro Foundation.
This work was supported by the Swiss State Secretariat for Education,
Research and Innovation (SERI) under contract number UeMO19-5.1.
This project has received funding from 
the European Union - Next Generation EU, Mission 4 Component 2 - CUP E53D23002240006 and
the Italian Ministry of University and Research (MUR) through the FARE grant for the project DAVNE (Grant R20PEX7Y3A), was supported by the Provincia Autonoma di Trento, and by Q@TN, the joint lab between the University of Trento, FBK-Fondazione Bruno Kessler, INFN-National Institute for Nuclear Physics and CNR-National Research Council. 

Views and opinions expressed are however those of the author(s) only and do not necessarily reflect those of the European Union or the European Commission. Neither the European Union nor the granting authority can be held responsible for them.
\end{acknowledgments}

\bibliographystyle{apsrev4-2}
\bibliography{references}

\begin{thebibliography}{136}%
\makeatletter
\providecommand \@ifxundefined [1]{%
 \@ifx{#1\undefined}
}%
\providecommand \@ifnum [1]{%
 \ifnum #1\expandafter \@firstoftwo
 \else \expandafter \@secondoftwo
 \fi
}%
\providecommand \@ifx [1]{%
 \ifx #1\expandafter \@firstoftwo
 \else \expandafter \@secondoftwo
 \fi
}%
\providecommand \natexlab [1]{#1}%
\providecommand \enquote  [1]{``#1''}%
\providecommand \bibnamefont  [1]{#1}%
\providecommand \bibfnamefont [1]{#1}%
\providecommand \citenamefont [1]{#1}%
\providecommand \href@noop [0]{\@secondoftwo}%
\providecommand \href [0]{\begingroup \@sanitize@url \@href}%
\providecommand \@href[1]{\@@startlink{#1}\@@href}%
\providecommand \@@href[1]{\endgroup#1\@@endlink}%
\providecommand \@sanitize@url [0]{\catcode `\\12\catcode `\$12\catcode
  `\&12\catcode `\#12\catcode `\^12\catcode `\_12\catcode `\%12\relax}%
\providecommand \@@startlink[1]{}%
\providecommand \@@endlink[0]{}%
\providecommand \url  [0]{\begingroup\@sanitize@url \@url }%
\providecommand \@url [1]{\endgroup\@href {#1}{\urlprefix }}%
\providecommand \urlprefix  [0]{URL }%
\providecommand \Eprint [0]{\href }%
\providecommand \doibase [0]{https://doi.org/}%
\providecommand \selectlanguage [0]{\@gobble}%
\providecommand \bibinfo  [0]{\@secondoftwo}%
\providecommand \bibfield  [0]{\@secondoftwo}%
\providecommand \translation [1]{[#1]}%
\providecommand \BibitemOpen [0]{}%
\providecommand \bibitemStop [0]{}%
\providecommand \bibitemNoStop [0]{.\EOS\space}%
\providecommand \EOS [0]{\spacefactor3000\relax}%
\providecommand \BibitemShut  [1]{\csname bibitem#1\endcsname}%
\let\auto@bib@innerbib\@empty
\bibitem [{\citenamefont {Berry}\ and\ \citenamefont
  {Tabor}(1977)}]{Berry1977}%
  \BibitemOpen
  \bibfield  {author} {\bibinfo {author} {\bibfnamefont {M.~V.}\ \bibnamefont
  {Berry}}\ and\ \bibinfo {author} {\bibfnamefont {M.}~\bibnamefont {Tabor}},\
  }\href {https://api.semanticscholar.org/CorpusID:123407512} {\bibfield
  {journal} {\bibinfo  {journal} {Proceedings of the Royal Society of London.
  A. Mathematical and Physical Sciences}\ }\textbf {\bibinfo {volume} {356}},\
  \bibinfo {pages} {375} (\bibinfo {year} {1977})}\BibitemShut {NoStop}%
\bibitem [{\citenamefont {Bohigas}\ \emph {et~al.}(1984)\citenamefont
  {Bohigas}, \citenamefont {Giannoni},\ and\ \citenamefont {Schmit}}]{BGS1984}%
  \BibitemOpen
  \bibfield  {author} {\bibinfo {author} {\bibfnamefont {O.}~\bibnamefont
  {Bohigas}}, \bibinfo {author} {\bibfnamefont {M.~J.}\ \bibnamefont
  {Giannoni}},\ and\ \bibinfo {author} {\bibfnamefont {C.}~\bibnamefont
  {Schmit}},\ }\href {https://doi.org/10.1103/PhysRevLett.52.1} {\bibfield
  {journal} {\bibinfo  {journal} {Phys. Rev. Lett.}\ }\textbf {\bibinfo
  {volume} {52}},\ \bibinfo {pages} {1} (\bibinfo {year} {1984})}\BibitemShut
  {NoStop}%
\bibitem [{\citenamefont {Haake}(1991)}]{haake_quantum_1991}%
  \BibitemOpen
  \bibfield  {author} {\bibinfo {author} {\bibfnamefont {F.}~\bibnamefont
  {Haake}},\ }in\ \href@noop {} {\emph {\bibinfo {booktitle} {Quantum coherence
  in mesoscopic systems}}}\ (\bibinfo  {publisher} {Springer},\ \bibinfo {year}
  {1991})\ pp.\ \bibinfo {pages} {583--595}\BibitemShut {NoStop}%
\bibitem [{\citenamefont {Mehta}(2004)}]{Mehta2004}%
  \BibitemOpen
  \bibfield  {author} {\bibinfo {author} {\bibfnamefont {M.~L.}\ \bibnamefont
  {Mehta}},\ }\href@noop {} {\emph {\bibinfo {title} {Random Matrices}}},\
  \bibinfo {edition} {3rd}\ ed.,\ Vol.\ \bibinfo {volume} {142}\ (\bibinfo
  {publisher} {Elsevier Academic Press},\ \bibinfo {year} {2004})\BibitemShut
  {NoStop}%
\bibitem [{\citenamefont {D'Alessio}\ \emph {et~al.}(2016)\citenamefont
  {D'Alessio}, \citenamefont {Kafri}, \citenamefont {Polkovnikov},\ and\
  \citenamefont {Rigol}}]{DAlessio2016}%
  \BibitemOpen
  \bibfield  {author} {\bibinfo {author} {\bibfnamefont {L.}~\bibnamefont
  {D'Alessio}}, \bibinfo {author} {\bibfnamefont {Y.}~\bibnamefont {Kafri}},
  \bibinfo {author} {\bibfnamefont {A.}~\bibnamefont {Polkovnikov}},\ and\
  \bibinfo {author} {\bibfnamefont {M.}~\bibnamefont {Rigol}},\ }\href
  {https://doi.org/10.1080/00018732.2016.1198134} {\bibfield  {journal}
  {\bibinfo  {journal} {Adv. Phys.}\ }\textbf {\bibinfo {volume} {65}},\
  \bibinfo {pages} {239} (\bibinfo {year} {2016})}\BibitemShut {NoStop}%
\bibitem [{\citenamefont {Sachdev}\ and\ \citenamefont {Ye}(1993)}]{SY93}%
  \BibitemOpen
  \bibfield  {author} {\bibinfo {author} {\bibfnamefont {S.}~\bibnamefont
  {Sachdev}}\ and\ \bibinfo {author} {\bibfnamefont {J.}~\bibnamefont {Ye}},\
  }\href {https://doi.org/10.1103/PhysRevLett.70.3339} {\bibfield  {journal}
  {\bibinfo  {journal} {Phys. Rev. Lett.}\ }\textbf {\bibinfo {volume} {70}},\
  \bibinfo {pages} {3339} (\bibinfo {year} {1993})}\BibitemShut {NoStop}%
\bibitem [{\citenamefont {Kitaev}(2015)}]{Kitaev2015}%
  \BibitemOpen
  \bibfield  {author} {\bibinfo {author} {\bibfnamefont {A.}~\bibnamefont
  {Kitaev}},\ }\href@noop {} {\bibinfo {title} {A simple model of quantum
  holography}},\ \bibinfo {howpublished} {Talks at the Kavli Institute for
  Theoretical Physics (KITP), April 7 and May 27, 2015} (\bibinfo {year}
  {2015}),\ \bibinfo {note} {unpublished}\BibitemShut {NoStop}%
\bibitem [{\citenamefont {Deutsch}(1991)}]{Deutsch1991}%
  \BibitemOpen
  \bibfield  {author} {\bibinfo {author} {\bibfnamefont {J.~M.}\ \bibnamefont
  {Deutsch}},\ }\href {https://doi.org/10.1103/PhysRevA.43.2046} {\bibfield
  {journal} {\bibinfo  {journal} {Phys. Rev. A}\ }\textbf {\bibinfo {volume}
  {43}},\ \bibinfo {pages} {2046} (\bibinfo {year} {1991})}\BibitemShut
  {NoStop}%
\bibitem [{\citenamefont {Srednicki}(1994)}]{Srednicki1994}%
  \BibitemOpen
  \bibfield  {author} {\bibinfo {author} {\bibfnamefont {M.}~\bibnamefont
  {Srednicki}},\ }\href {https://doi.org/10.1103/PhysRevE.50.888} {\bibfield
  {journal} {\bibinfo  {journal} {Phys. Rev. E}\ }\textbf {\bibinfo {volume}
  {50}},\ \bibinfo {pages} {888} (\bibinfo {year} {1994})}\BibitemShut
  {NoStop}%
\bibitem [{\citenamefont {Rigol}\ \emph {et~al.}(2008)\citenamefont {Rigol},
  \citenamefont {Dunjko},\ and\ \citenamefont {Olshanii}}]{Rigol2008}%
  \BibitemOpen
  \bibfield  {author} {\bibinfo {author} {\bibfnamefont {M.}~\bibnamefont
  {Rigol}}, \bibinfo {author} {\bibfnamefont {V.}~\bibnamefont {Dunjko}},\ and\
  \bibinfo {author} {\bibfnamefont {M.}~\bibnamefont {Olshanii}},\ }\href
  {https://doi.org/10.1038/nature06838} {\bibfield  {journal} {\bibinfo
  {journal} {Nature}\ }\textbf {\bibinfo {volume} {452}},\ \bibinfo {pages}
  {854} (\bibinfo {year} {2008})}\BibitemShut {NoStop}%
\bibitem [{\citenamefont {Sonner}\ and\ \citenamefont
  {Vielma}(2017)}]{Sonner:2017hxc}%
  \BibitemOpen
  \bibfield  {author} {\bibinfo {author} {\bibfnamefont {J.}~\bibnamefont
  {Sonner}}\ and\ \bibinfo {author} {\bibfnamefont {M.}~\bibnamefont
  {Vielma}},\ }\href {https://doi.org/10.1007/JHEP11(2017)149} {\bibfield
  {journal} {\bibinfo  {journal} {J. High Energy Phys.}\ }\textbf {\bibinfo
  {volume} {11}},\ \bibinfo {pages} {149}}\BibitemShut {NoStop}%
\bibitem [{\citenamefont {Larzul}\ and\ \citenamefont
  {Schir\'o}(2022)}]{Larzul_2022}%
  \BibitemOpen
  \bibfield  {author} {\bibinfo {author} {\bibfnamefont {A.}~\bibnamefont
  {Larzul}}\ and\ \bibinfo {author} {\bibfnamefont {M.}~\bibnamefont
  {Schir\'o}},\ }\href {https://doi.org/10.1103/PhysRevB.105.045105} {\bibfield
   {journal} {\bibinfo  {journal} {Phys. Rev. B}\ }\textbf {\bibinfo {volume}
  {105}},\ \bibinfo {pages} {045105} (\bibinfo {year} {2022})}\BibitemShut
  {NoStop}%
\bibitem [{\citenamefont {Bandyopadhyay}\ \emph {et~al.}(2023)\citenamefont
  {Bandyopadhyay}, \citenamefont {Uhrich}, \citenamefont {Paviglianiti},\ and\
  \citenamefont {Hauke}}]{Bandyopadhyay_2023}%
  \BibitemOpen
  \bibfield  {author} {\bibinfo {author} {\bibfnamefont {S.}~\bibnamefont
  {Bandyopadhyay}}, \bibinfo {author} {\bibfnamefont {P.}~\bibnamefont
  {Uhrich}}, \bibinfo {author} {\bibfnamefont {A.}~\bibnamefont
  {Paviglianiti}},\ and\ \bibinfo {author} {\bibfnamefont {P.}~\bibnamefont
  {Hauke}},\ }\href {https://doi.org/10.22331/q-2023-05-24-1022} {\bibfield
  {journal} {\bibinfo  {journal} {{Quantum}}\ }\textbf {\bibinfo {volume}
  {7}},\ \bibinfo {pages} {1022} (\bibinfo {year} {2023})}\BibitemShut
  {NoStop}%
\bibitem [{\citenamefont {Louw}\ and\ \citenamefont
  {Kehrein}(2022)}]{Louw_2022}%
  \BibitemOpen
  \bibfield  {author} {\bibinfo {author} {\bibfnamefont {J.~C.}\ \bibnamefont
  {Louw}}\ and\ \bibinfo {author} {\bibfnamefont {S.}~\bibnamefont {Kehrein}},\
  }\href {https://doi.org/10.1103/PhysRevB.105.075117} {\bibfield  {journal}
  {\bibinfo  {journal} {Phys. Rev. B}\ }\textbf {\bibinfo {volume} {105}},\
  \bibinfo {pages} {075117} (\bibinfo {year} {2022})}\BibitemShut {NoStop}%
\bibitem [{\citenamefont {Paviglianiti}\ \emph {et~al.}(2023)\citenamefont
  {Paviglianiti}, \citenamefont {Bandyopadhyay}, \citenamefont {Uhrich},\ and\
  \citenamefont {Hauke}}]{Paviglianiti_2023}%
  \BibitemOpen
  \bibfield  {author} {\bibinfo {author} {\bibfnamefont {A.}~\bibnamefont
  {Paviglianiti}}, \bibinfo {author} {\bibfnamefont {S.}~\bibnamefont
  {Bandyopadhyay}}, \bibinfo {author} {\bibfnamefont {P.}~\bibnamefont
  {Uhrich}},\ and\ \bibinfo {author} {\bibfnamefont {P.}~\bibnamefont
  {Hauke}},\ }\bibfield  {journal} {\bibinfo  {journal} {Journal of High Energy
  Physics}\ }\textbf {\bibinfo {volume} {2023}},\ \href
  {https://doi.org/10.1007/jhep03(2023)126} {10.1007/jhep03(2023)126} (\bibinfo
  {year} {2023})\BibitemShut {NoStop}%
\bibitem [{\citenamefont {Jaramillo}\ \emph {et~al.}(2025)\citenamefont
  {Jaramillo}, \citenamefont {Jha},\ and\ \citenamefont
  {Kehrein}}]{Jaramillo_2025}%
  \BibitemOpen
  \bibfield  {author} {\bibinfo {author} {\bibfnamefont {S.~S.}\ \bibnamefont
  {Jaramillo}}, \bibinfo {author} {\bibfnamefont {R.}~\bibnamefont {Jha}},\
  and\ \bibinfo {author} {\bibfnamefont {S.}~\bibnamefont {Kehrein}},\ }\href
  {https://doi.org/10.1103/PhysRevB.111.195153} {\bibfield  {journal} {\bibinfo
   {journal} {Phys. Rev. B}\ }\textbf {\bibinfo {volume} {111}},\ \bibinfo
  {pages} {195153} (\bibinfo {year} {2025})}\BibitemShut {NoStop}%
\bibitem [{\citenamefont {Perugu}\ \emph {et~al.}(2025)\citenamefont {Perugu},
  \citenamefont {Haldar},\ and\ \citenamefont {Banerjee}}]{Perugu_2025}%
  \BibitemOpen
  \bibfield  {author} {\bibinfo {author} {\bibfnamefont {R.}~\bibnamefont
  {Perugu}}, \bibinfo {author} {\bibfnamefont {A.}~\bibnamefont {Haldar}},\
  and\ \bibinfo {author} {\bibfnamefont {S.}~\bibnamefont {Banerjee}},\ }\href
  {https://doi.org/10.1103/9cbp-7tvv} {\bibfield  {journal} {\bibinfo
  {journal} {Phys. Rev. B}\ }\textbf {\bibinfo {volume} {112}},\ \bibinfo
  {pages} {184301} (\bibinfo {year} {2025})}\BibitemShut {NoStop}%
\bibitem [{\citenamefont {Maldacena}\ and\ \citenamefont
  {Stanford}(2016)}]{Maldacena:2016hyu}%
  \BibitemOpen
  \bibfield  {author} {\bibinfo {author} {\bibfnamefont {J.}~\bibnamefont
  {Maldacena}}\ and\ \bibinfo {author} {\bibfnamefont {D.}~\bibnamefont
  {Stanford}},\ }\href {https://doi.org/10.1103/PhysRevD.94.106002} {\bibfield
  {journal} {\bibinfo  {journal} {Phys. Rev. D}\ }\textbf {\bibinfo {volume}
  {94}},\ \bibinfo {pages} {106002} (\bibinfo {year} {2016})}\BibitemShut
  {NoStop}%
\bibitem [{\citenamefont {Polchinski}\ and\ \citenamefont
  {Rosenhaus}(2016)}]{Polchinski:2016xgd}%
  \BibitemOpen
  \bibfield  {author} {\bibinfo {author} {\bibfnamefont {J.}~\bibnamefont
  {Polchinski}}\ and\ \bibinfo {author} {\bibfnamefont {V.}~\bibnamefont
  {Rosenhaus}},\ }\href {https://doi.org/10.1007/JHEP04(2016)001} {\bibfield
  {journal} {\bibinfo  {journal} {J. High Energy Phys.}\ }\textbf {\bibinfo
  {volume} {04}},\ \bibinfo {pages} {001}}\BibitemShut {NoStop}%
\bibitem [{\citenamefont {Maldacena}\ \emph
  {et~al.}(2016{\natexlab{a}})\citenamefont {Maldacena}, \citenamefont
  {Shenker},\ and\ \citenamefont {Stanford}}]{Maldacena_2016}%
  \BibitemOpen
  \bibfield  {author} {\bibinfo {author} {\bibfnamefont {J.}~\bibnamefont
  {Maldacena}}, \bibinfo {author} {\bibfnamefont {S.~H.}\ \bibnamefont
  {Shenker}},\ and\ \bibinfo {author} {\bibfnamefont {D.}~\bibnamefont
  {Stanford}},\ }\href {https://doi.org/10.1007/JHEP08(2016)106} {\bibfield
  {journal} {\bibinfo  {journal} {J. High Energy Phys.}\ }\textbf {\bibinfo
  {volume} {2016}}\bibinfo  {number} { (8)},\ \bibinfo {pages}
  {106}}\BibitemShut {NoStop}%
\bibitem [{\citenamefont {Chowdhury}\ \emph {et~al.}(2022)\citenamefont
  {Chowdhury}, \citenamefont {Georges}, \citenamefont {Parcollet},\ and\
  \citenamefont {Sachdev}}]{Chowdhury2022}%
  \BibitemOpen
\bibfield  {number} {  }\bibfield  {author} {\bibinfo {author} {\bibfnamefont
  {D.}~\bibnamefont {Chowdhury}}, \bibinfo {author} {\bibfnamefont
  {A.}~\bibnamefont {Georges}}, \bibinfo {author} {\bibfnamefont
  {O.}~\bibnamefont {Parcollet}},\ and\ \bibinfo {author} {\bibfnamefont
  {S.}~\bibnamefont {Sachdev}},\ }\href
  {https://doi.org/10.1103/RevModPhys.94.035004} {\bibfield  {journal}
  {\bibinfo  {journal} {Rev. Mod. Phys.}\ }\textbf {\bibinfo {volume} {94}},\
  \bibinfo {pages} {035004} (\bibinfo {year} {2022})}\BibitemShut {NoStop}%
\bibitem [{\citenamefont {Cotler}\ \emph {et~al.}(2017)\citenamefont {Cotler},
  \citenamefont {Gur-Ari}, \citenamefont {Hanada}, \citenamefont {Polchinski},
  \citenamefont {Saad}, \citenamefont {Shenker}, \citenamefont {Stanford},
  \citenamefont {Streicher},\ and\ \citenamefont {Tezuka}}]{Cotler2016}%
  \BibitemOpen
  \bibfield  {author} {\bibinfo {author} {\bibfnamefont {J.~S.}\ \bibnamefont
  {Cotler}}, \bibinfo {author} {\bibfnamefont {G.}~\bibnamefont {Gur-Ari}},
  \bibinfo {author} {\bibfnamefont {M.}~\bibnamefont {Hanada}}, \bibinfo
  {author} {\bibfnamefont {J.}~\bibnamefont {Polchinski}}, \bibinfo {author}
  {\bibfnamefont {P.}~\bibnamefont {Saad}}, \bibinfo {author} {\bibfnamefont
  {S.~H.}\ \bibnamefont {Shenker}}, \bibinfo {author} {\bibfnamefont
  {D.}~\bibnamefont {Stanford}}, \bibinfo {author} {\bibfnamefont
  {A.}~\bibnamefont {Streicher}},\ and\ \bibinfo {author} {\bibfnamefont
  {M.}~\bibnamefont {Tezuka}},\ }\href
  {https://doi.org/10.48550/arXiv.1611.04650} {\bibfield  {journal} {\bibinfo
  {journal} {J. High Energy Phys.}\ }\textbf {\bibinfo {volume} {05}},\
  \bibinfo {pages} {118}},\ \bibinfo {note} {[Erratum: JHEP 09, 002
  (2018)]}\BibitemShut {NoStop}%
\bibitem [{\citenamefont {Maldacena}\ \emph
  {et~al.}(2016{\natexlab{b}})\citenamefont {Maldacena}, \citenamefont
  {Stanford},\ and\ \citenamefont {Yang}}]{Maldacena:2016upp}%
  \BibitemOpen
  \bibfield  {author} {\bibinfo {author} {\bibfnamefont {J.}~\bibnamefont
  {Maldacena}}, \bibinfo {author} {\bibfnamefont {D.}~\bibnamefont
  {Stanford}},\ and\ \bibinfo {author} {\bibfnamefont {Z.}~\bibnamefont
  {Yang}},\ }\href {https://doi.org/10.1093/ptep/ptw124} {\bibfield  {journal}
  {\bibinfo  {journal} {Prog. Theor. Exp. Phys.}\ }\textbf {\bibinfo {volume}
  {2016}},\ \bibinfo {pages} {12C104} (\bibinfo {year}
  {2016}{\natexlab{b}})}\BibitemShut {NoStop}%
\bibitem [{\citenamefont {Jackiw}(1985)}]{Jackiw:1984je}%
  \BibitemOpen
  \bibfield  {author} {\bibinfo {author} {\bibfnamefont {R.}~\bibnamefont
  {Jackiw}},\ }\href {https://doi.org/10.1016/0550-3213(85)90448-1} {\bibfield
  {journal} {\bibinfo  {journal} {Nucl. Phys. B}\ }\textbf {\bibinfo {volume}
  {252}},\ \bibinfo {pages} {343} (\bibinfo {year} {1985})}\BibitemShut
  {NoStop}%
\bibitem [{\citenamefont {Teitelboim}(1983)}]{Teitelboim:1983ux}%
  \BibitemOpen
  \bibfield  {author} {\bibinfo {author} {\bibfnamefont {C.}~\bibnamefont
  {Teitelboim}},\ }\href {https://doi.org/10.1016/0370-2693(83)90012-6}
  {\bibfield  {journal} {\bibinfo  {journal} {Phys. Lett. B}\ }\textbf
  {\bibinfo {volume} {126}},\ \bibinfo {pages} {41} (\bibinfo {year}
  {1983})}\BibitemShut {NoStop}%
\bibitem [{\citenamefont {Turner}\ \emph {et~al.}(2018)\citenamefont {Turner},
  \citenamefont {Michailidis}, \citenamefont {Abanin}, \citenamefont {Serbyn},\
  and\ \citenamefont {Papic}}]{Turner:2017fxc}%
  \BibitemOpen
  \bibfield  {author} {\bibinfo {author} {\bibfnamefont {C.~J.}\ \bibnamefont
  {Turner}}, \bibinfo {author} {\bibfnamefont {A.~A.}\ \bibnamefont
  {Michailidis}}, \bibinfo {author} {\bibfnamefont {D.~A.}\ \bibnamefont
  {Abanin}}, \bibinfo {author} {\bibfnamefont {M.}~\bibnamefont {Serbyn}},\
  and\ \bibinfo {author} {\bibfnamefont {Z.}~\bibnamefont {Papic}},\ }\href
  {https://doi.org/10.1038/s41567-018-0137-5} {\bibfield  {journal} {\bibinfo
  {journal} {Nature Phys.}\ }\textbf {\bibinfo {volume} {14}},\ \bibinfo
  {pages} {745} (\bibinfo {year} {2018})}\BibitemShut {NoStop}%
\bibitem [{\citenamefont {\ifmmode~\check{S}\else \v{S}\fi{}untajs}\ \emph
  {et~al.}(2020)\citenamefont {\ifmmode~\check{S}\else \v{S}\fi{}untajs},
  \citenamefont {Bon\ifmmode~\check{c}\else \v{c}\fi{}a}, \citenamefont
  {Prosen},\ and\ \citenamefont {Vidmar}}]{Suntajs2020}%
  \BibitemOpen
  \bibfield  {author} {\bibinfo {author} {\bibfnamefont {J.}~\bibnamefont
  {\ifmmode~\check{S}\else \v{S}\fi{}untajs}}, \bibinfo {author} {\bibfnamefont
  {J.}~\bibnamefont {Bon\ifmmode~\check{c}\else \v{c}\fi{}a}}, \bibinfo
  {author} {\bibfnamefont {T.~c.~v.}\ \bibnamefont {Prosen}},\ and\ \bibinfo
  {author} {\bibfnamefont {L.}~\bibnamefont {Vidmar}},\ }\href
  {https://doi.org/10.1103/PhysRevE.102.062144} {\bibfield  {journal} {\bibinfo
   {journal} {Phys. Rev. E}\ }\textbf {\bibinfo {volume} {102}},\ \bibinfo
  {pages} {062144} (\bibinfo {year} {2020})}\BibitemShut {NoStop}%
\bibitem [{\citenamefont {Sels}\ and\ \citenamefont
  {Polkovnikov}(2021)}]{Sels2021}%
  \BibitemOpen
  \bibfield  {author} {\bibinfo {author} {\bibfnamefont {D.}~\bibnamefont
  {Sels}}\ and\ \bibinfo {author} {\bibfnamefont {A.}~\bibnamefont
  {Polkovnikov}},\ }\href {https://doi.org/10.1103/PhysRevE.104.054105}
  {\bibfield  {journal} {\bibinfo  {journal} {Phys. Rev. E}\ }\textbf {\bibinfo
  {volume} {104}},\ \bibinfo {pages} {054105} (\bibinfo {year}
  {2021})}\BibitemShut {NoStop}%
\bibitem [{\citenamefont {Schiulaz}\ \emph {et~al.}(2019)\citenamefont
  {Schiulaz}, \citenamefont {Torres-Herrera},\ and\ \citenamefont
  {Santos}}]{Schiulaz2019}%
  \BibitemOpen
  \bibfield  {author} {\bibinfo {author} {\bibfnamefont {M.}~\bibnamefont
  {Schiulaz}}, \bibinfo {author} {\bibfnamefont {E.~J.}\ \bibnamefont
  {Torres-Herrera}},\ and\ \bibinfo {author} {\bibfnamefont {L.~F.}\
  \bibnamefont {Santos}},\ }\href {https://doi.org/10.1103/PhysRevB.99.174313}
  {\bibfield  {journal} {\bibinfo  {journal} {Phys. Rev. B}\ }\textbf {\bibinfo
  {volume} {99}},\ \bibinfo {pages} {174313} (\bibinfo {year}
  {2019})}\BibitemShut {NoStop}%
\bibitem [{\citenamefont {Basko}\ \emph {et~al.}(2006)\citenamefont {Basko},
  \citenamefont {Aleiner},\ and\ \citenamefont {Altshuler}}]{Basko_2006}%
  \BibitemOpen
  \bibfield  {author} {\bibinfo {author} {\bibfnamefont {D.}~\bibnamefont
  {Basko}}, \bibinfo {author} {\bibfnamefont {I.}~\bibnamefont {Aleiner}},\
  and\ \bibinfo {author} {\bibfnamefont {B.}~\bibnamefont {Altshuler}},\ }\href
  {https://doi.org/10.1016/j.aop.2005.11.014} {\bibfield  {journal} {\bibinfo
  {journal} {Annals of Physics}\ }\textbf {\bibinfo {volume} {321}},\ \bibinfo
  {pages} {1126–1205} (\bibinfo {year} {2006})}\BibitemShut {NoStop}%
\bibitem [{\citenamefont {Kravtsov}\ \emph {et~al.}(2015)\citenamefont
  {Kravtsov}, \citenamefont {Khaymovich}, \citenamefont {Cuevas},\ and\
  \citenamefont {Amini}}]{Kravtsov_2015}%
  \BibitemOpen
  \bibfield  {author} {\bibinfo {author} {\bibfnamefont {V.~E.}\ \bibnamefont
  {Kravtsov}}, \bibinfo {author} {\bibfnamefont {I.~M.}\ \bibnamefont
  {Khaymovich}}, \bibinfo {author} {\bibfnamefont {E.}~\bibnamefont {Cuevas}},\
  and\ \bibinfo {author} {\bibfnamefont {M.}~\bibnamefont {Amini}},\ }\href
  {https://doi.org/10.1088/1367-2630/17/12/122002} {\bibfield  {journal}
  {\bibinfo  {journal} {New Journal of Physics}\ }\textbf {\bibinfo {volume}
  {17}},\ \bibinfo {pages} {122002} (\bibinfo {year} {2015})}\BibitemShut
  {NoStop}%
\bibitem [{\citenamefont {Abanin}\ \emph
  {et~al.}(2019{\natexlab{a}})\citenamefont {Abanin}, \citenamefont {Altman},
  \citenamefont {Bloch},\ and\ \citenamefont {Serbyn}}]{Abanin:2018yrt}%
  \BibitemOpen
  \bibfield  {author} {\bibinfo {author} {\bibfnamefont {D.~A.}\ \bibnamefont
  {Abanin}}, \bibinfo {author} {\bibfnamefont {E.}~\bibnamefont {Altman}},
  \bibinfo {author} {\bibfnamefont {I.}~\bibnamefont {Bloch}},\ and\ \bibinfo
  {author} {\bibfnamefont {M.}~\bibnamefont {Serbyn}},\ }\href
  {https://doi.org/10.1103/revmodphys.91.021001} {\bibfield  {journal}
  {\bibinfo  {journal} {Rev. Mod. Phys.}\ }\textbf {\bibinfo {volume} {91}},\
  \bibinfo {pages} {021001} (\bibinfo {year} {2019}{\natexlab{a}})},\ \Eprint
  {https://arxiv.org/abs/1804.11065} {arXiv:1804.11065 [cond-mat.dis-nn]}
  \BibitemShut {NoStop}%
\bibitem [{\citenamefont {Esterlis}\ and\ \citenamefont
  {Schmalian}(2019)}]{Schmalian_YSYK}%
  \BibitemOpen
  \bibfield  {author} {\bibinfo {author} {\bibfnamefont {I.}~\bibnamefont
  {Esterlis}}\ and\ \bibinfo {author} {\bibfnamefont {J.}~\bibnamefont
  {Schmalian}},\ }\href {https://doi.org/10.1103/PhysRevB.100.115132}
  {\bibfield  {journal} {\bibinfo  {journal} {Phys. Rev. B}\ }\textbf {\bibinfo
  {volume} {100}},\ \bibinfo {pages} {115132} (\bibinfo {year}
  {2019})}\BibitemShut {NoStop}%
\bibitem [{\citenamefont {Wang}(2020)}]{Wan2020}%
  \BibitemOpen
  \bibfield  {author} {\bibinfo {author} {\bibfnamefont {Y.}~\bibnamefont
  {Wang}},\ }\href {https://doi.org/10.1103/PhysRevLett.124.017002} {\bibfield
  {journal} {\bibinfo  {journal} {Phys. Rev. Lett.}\ }\textbf {\bibinfo
  {volume} {124}},\ \bibinfo {pages} {017002} (\bibinfo {year}
  {2020})}\BibitemShut {NoStop}%
\bibitem [{\citenamefont {Wang}\ and\ \citenamefont
  {Chubukov}(2020)}]{wang_quantum_2020}%
  \BibitemOpen
  \bibfield  {author} {\bibinfo {author} {\bibfnamefont {Y.}~\bibnamefont
  {Wang}}\ and\ \bibinfo {author} {\bibfnamefont {A.~V.}\ \bibnamefont
  {Chubukov}},\ }\href {https://doi.org/10.1103/PhysRevResearch.2.033084}
  {\bibfield  {journal} {\bibinfo  {journal} {Phys. Rev. Res.}\ }\textbf
  {\bibinfo {volume} {2}},\ \bibinfo {pages} {033084} (\bibinfo {year}
  {2020})},\ \bibinfo {note} {publisher: American Physical Society}\BibitemShut
  {NoStop}%
\bibitem [{\citenamefont {Pan}\ \emph {et~al.}(2021)\citenamefont {Pan},
  \citenamefont {Wang}, \citenamefont {Davis}, \citenamefont {Wang},\ and\
  \citenamefont {Meng}}]{pan_yukawa-syk_2021}%
  \BibitemOpen
  \bibfield  {author} {\bibinfo {author} {\bibfnamefont {G.}~\bibnamefont
  {Pan}}, \bibinfo {author} {\bibfnamefont {W.}~\bibnamefont {Wang}}, \bibinfo
  {author} {\bibfnamefont {A.}~\bibnamefont {Davis}}, \bibinfo {author}
  {\bibfnamefont {Y.}~\bibnamefont {Wang}},\ and\ \bibinfo {author}
  {\bibfnamefont {Z.~Y.}\ \bibnamefont {Meng}},\ }\href
  {https://doi.org/10.1103/PhysRevResearch.3.013250} {\bibfield  {journal}
  {\bibinfo  {journal} {Phys. Rev. Res.}\ }\textbf {\bibinfo {volume} {3}},\
  \bibinfo {pages} {013250} (\bibinfo {year} {2021})}\BibitemShut {NoStop}%
\bibitem [{\citenamefont {Patel}\ \emph {et~al.}(2023)\citenamefont {Patel},
  \citenamefont {Guo}, \citenamefont {Esterlis},\ and\ \citenamefont
  {Sachdev}}]{patel_universal_2023}%
  \BibitemOpen
  \bibfield  {author} {\bibinfo {author} {\bibfnamefont {A.~A.}\ \bibnamefont
  {Patel}}, \bibinfo {author} {\bibfnamefont {H.}~\bibnamefont {Guo}}, \bibinfo
  {author} {\bibfnamefont {I.}~\bibnamefont {Esterlis}},\ and\ \bibinfo
  {author} {\bibfnamefont {S.}~\bibnamefont {Sachdev}},\ }\href
  {https://doi.org/10.1126/science.abq6011} {\bibfield  {journal} {\bibinfo
  {journal} {Science}\ }\textbf {\bibinfo {volume} {381}},\ \bibinfo {pages}
  {790} (\bibinfo {year} {2023})}\BibitemShut {NoStop}%
\bibitem [{\citenamefont {Li}\ \emph {et~al.}(2024)\citenamefont {Li},
  \citenamefont {Valentinis}, \citenamefont {Patel}, \citenamefont {Guo},
  \citenamefont {Schmalian}, \citenamefont {Sachdev},\ and\ \citenamefont
  {Esterlis}}]{Li:2024kxr}%
  \BibitemOpen
  \bibfield  {author} {\bibinfo {author} {\bibfnamefont {C.}~\bibnamefont
  {Li}}, \bibinfo {author} {\bibfnamefont {D.}~\bibnamefont {Valentinis}},
  \bibinfo {author} {\bibfnamefont {A.~A.}\ \bibnamefont {Patel}}, \bibinfo
  {author} {\bibfnamefont {H.}~\bibnamefont {Guo}}, \bibinfo {author}
  {\bibfnamefont {J.}~\bibnamefont {Schmalian}}, \bibinfo {author}
  {\bibfnamefont {S.}~\bibnamefont {Sachdev}},\ and\ \bibinfo {author}
  {\bibfnamefont {I.}~\bibnamefont {Esterlis}},\ }\href
  {https://doi.org/10.1103/PhysRevLett.133.186502} {\bibfield  {journal}
  {\bibinfo  {journal} {Phys. Rev. Lett.}\ }\textbf {\bibinfo {volume} {133}},\
  \bibinfo {pages} {186502} (\bibinfo {year} {2024})}\BibitemShut {NoStop}%
\bibitem [{\citenamefont {Marcus}\ and\ \citenamefont
  {Vandoren}(2019)}]{Marcus_2019}%
  \BibitemOpen
  \bibfield  {author} {\bibinfo {author} {\bibfnamefont {E.}~\bibnamefont
  {Marcus}}\ and\ \bibinfo {author} {\bibfnamefont {S.}~\bibnamefont
  {Vandoren}},\ }\bibfield  {journal} {\bibinfo  {journal} {Journal of High
  Energy Physics}\ }\textbf {\bibinfo {volume} {2019}},\ \href
  {https://doi.org/10.1007/JHEP01(2019)166} {10.1007/JHEP01(2019)166} (\bibinfo
  {year} {2019})\BibitemShut {NoStop}%
\bibitem [{\citenamefont {Davis}\ and\ \citenamefont
  {Wang}(2023)}]{davis_quantum_2023}%
  \BibitemOpen
  \bibfield  {author} {\bibinfo {author} {\bibfnamefont {A.}~\bibnamefont
  {Davis}}\ and\ \bibinfo {author} {\bibfnamefont {Y.}~\bibnamefont {Wang}},\
  }\href {https://doi.org/10.1103/PhysRevB.107.205122} {\bibfield  {journal}
  {\bibinfo  {journal} {Phys. Rev. B}\ }\textbf {\bibinfo {volume} {107}},\
  \bibinfo {pages} {205122} (\bibinfo {year} {2023})}\BibitemShut {NoStop}%
\bibitem [{\citenamefont {Lunkin}\ \emph {et~al.}(2020)\citenamefont {Lunkin},
  \citenamefont {Kitaev},\ and\ \citenamefont {Feigel'man}}]{Lunkin_2020}%
  \BibitemOpen
  \bibfield  {author} {\bibinfo {author} {\bibfnamefont {A.~V.}\ \bibnamefont
  {Lunkin}}, \bibinfo {author} {\bibfnamefont {A.~Y.}\ \bibnamefont {Kitaev}},\
  and\ \bibinfo {author} {\bibfnamefont {M.~V.}\ \bibnamefont {Feigel'man}},\
  }\href {https://doi.org/10.1103/PhysRevLett.125.196602} {\bibfield  {journal}
  {\bibinfo  {journal} {Phys. Rev. Lett.}\ }\textbf {\bibinfo {volume} {125}},\
  \bibinfo {pages} {196602} (\bibinfo {year} {2020})}\BibitemShut {NoStop}%
\bibitem [{\citenamefont {Hauck}\ \emph {et~al.}(2020)\citenamefont {Hauck},
  \citenamefont {Klug}, \citenamefont {Esterlis},\ and\ \citenamefont
  {Schmalian}}]{HAUCK_YSYK}%
  \BibitemOpen
  \bibfield  {author} {\bibinfo {author} {\bibfnamefont {D.}~\bibnamefont
  {Hauck}}, \bibinfo {author} {\bibfnamefont {M.~J.}\ \bibnamefont {Klug}},
  \bibinfo {author} {\bibfnamefont {I.}~\bibnamefont {Esterlis}},\ and\
  \bibinfo {author} {\bibfnamefont {J.}~\bibnamefont {Schmalian}},\ }\href
  {https://doi.org/10.1016/j.aop.2020.168120} {\bibfield  {journal} {\bibinfo
  {journal} {Annals of Physics}\ }\textbf {\bibinfo {volume} {417}},\ \bibinfo
  {pages} {168120} (\bibinfo {year} {2020})},\ \bibinfo {note} {eliashberg
  theory at 60: Strong-coupling superconductivity and beyond}\BibitemShut
  {NoStop}%
\bibitem [{\citenamefont {Inkof}\ \emph {et~al.}(2022)\citenamefont {Inkof},
  \citenamefont {Schalm},\ and\ \citenamefont
  {Schmalian}}]{inkof_quantum_2022}%
  \BibitemOpen
  \bibfield  {author} {\bibinfo {author} {\bibfnamefont {G.-A.}\ \bibnamefont
  {Inkof}}, \bibinfo {author} {\bibfnamefont {K.}~\bibnamefont {Schalm}},\ and\
  \bibinfo {author} {\bibfnamefont {J.}~\bibnamefont {Schmalian}},\ }\href
  {https://doi.org/10.1038/s41535-022-00460-8} {\bibfield  {journal} {\bibinfo
  {journal} {npj Quantum Mater.}\ }\textbf {\bibinfo {volume} {7}},\ \bibinfo
  {pages} {56} (\bibinfo {year} {2022})}\BibitemShut {NoStop}%
\bibitem [{\citenamefont {Schmalian}(2022)}]{Schmalian:2022web}%
  \BibitemOpen
  \bibfield  {author} {\bibinfo {author} {\bibfnamefont {J.}~\bibnamefont
  {Schmalian}}\ }\href {https://doi.org/10.48550/arXiv.2209.00474}
  {10.48550/arXiv.2209.00474} (\bibinfo {year} {2022}),\ \Eprint
  {https://arxiv.org/abs/2209.00474} {arXiv:2209.00474 [cond-mat.str-el]}
  \BibitemShut {NoStop}%
\bibitem [{\citenamefont {Esterlis}\ and\ \citenamefont
  {Schmalian}(2025)}]{esterlis2025quantum}%
  \BibitemOpen
  \bibfield  {author} {\bibinfo {author} {\bibfnamefont {I.}~\bibnamefont
  {Esterlis}}\ and\ \bibinfo {author} {\bibfnamefont {J.}~\bibnamefont
  {Schmalian}},\ }\href {https://doi.org/10.48550/arXiv.2506.11952} {\bibinfo
  {title} {Quantum critical eliashberg theory}} (\bibinfo {year} {2025}),\
  \Eprint {https://arxiv.org/abs/2506.11952} {arXiv:2506.11952
  [cond-mat.str-el]} \BibitemShut {NoStop}%
\bibitem [{\citenamefont {Hartnoll}\ \emph {et~al.}(2008)\citenamefont
  {Hartnoll}, \citenamefont {Herzog},\ and\ \citenamefont
  {Horowitz}}]{Hartnoll:2008kx}%
  \BibitemOpen
  \bibfield  {author} {\bibinfo {author} {\bibfnamefont {S.~A.}\ \bibnamefont
  {Hartnoll}}, \bibinfo {author} {\bibfnamefont {C.~P.}\ \bibnamefont
  {Herzog}},\ and\ \bibinfo {author} {\bibfnamefont {G.~T.}\ \bibnamefont
  {Horowitz}},\ }\href {https://doi.org/10.1088/1126-6708/2008/12/015}
  {\bibfield  {journal} {\bibinfo  {journal} {Journal of High Energy Physics}\
  }\textbf {\bibinfo {volume} {2008}},\ \bibinfo {pages} {015} (\bibinfo {year}
  {2008})}\BibitemShut {NoStop}%
\bibitem [{\citenamefont {Wigner}(1951)}]{Wigner_1951}%
  \BibitemOpen
  \bibfield  {author} {\bibinfo {author} {\bibfnamefont {E.~P.}\ \bibnamefont
  {Wigner}},\ }\href {https://doi.org/10.1017/S0305004100027237} {\bibfield
  {journal} {\bibinfo  {journal} {Mathematical Proceedings of the Cambridge
  Philosophical Society}\ }\textbf {\bibinfo {volume} {47}},\ \bibinfo {pages}
  {790–798} (\bibinfo {year} {1951})}\BibitemShut {NoStop}%
\bibitem [{\citenamefont {Brody}\ \emph {et~al.}(1981)\citenamefont {Brody},
  \citenamefont {Flores}, \citenamefont {French}, \citenamefont {Mello},
  \citenamefont {Pandey},\ and\ \citenamefont {Wong}}]{Brody81}%
  \BibitemOpen
  \bibfield  {author} {\bibinfo {author} {\bibfnamefont {T.~A.}\ \bibnamefont
  {Brody}}, \bibinfo {author} {\bibfnamefont {J.}~\bibnamefont {Flores}},
  \bibinfo {author} {\bibfnamefont {J.~B.}\ \bibnamefont {French}}, \bibinfo
  {author} {\bibfnamefont {P.~A.}\ \bibnamefont {Mello}}, \bibinfo {author}
  {\bibfnamefont {A.}~\bibnamefont {Pandey}},\ and\ \bibinfo {author}
  {\bibfnamefont {S.~S.~M.}\ \bibnamefont {Wong}},\ }\href
  {https://doi.org/10.1103/RevModPhys.53.385} {\bibfield  {journal} {\bibinfo
  {journal} {Rev. Mod. Phys.}\ }\textbf {\bibinfo {volume} {53}},\ \bibinfo
  {pages} {385} (\bibinfo {year} {1981})}\BibitemShut {NoStop}%
\bibitem [{\citenamefont {Oganesyan}\ and\ \citenamefont
  {Huse}(2007)}]{Oganesyan2007}%
  \BibitemOpen
  \bibfield  {author} {\bibinfo {author} {\bibfnamefont {V.}~\bibnamefont
  {Oganesyan}}\ and\ \bibinfo {author} {\bibfnamefont {D.~A.}\ \bibnamefont
  {Huse}},\ }\href {https://doi.org/10.1103/PhysRevB.75.155111} {\bibfield
  {journal} {\bibinfo  {journal} {Phys. Rev. B}\ }\textbf {\bibinfo {volume}
  {75}},\ \bibinfo {pages} {155111} (\bibinfo {year} {2007})}\BibitemShut
  {NoStop}%
\bibitem [{\citenamefont {Atas}\ \emph {et~al.}(2013)\citenamefont {Atas},
  \citenamefont {Bogomolny}, \citenamefont {Giraud},\ and\ \citenamefont
  {Roux}}]{Atas2013}%
  \BibitemOpen
  \bibfield  {author} {\bibinfo {author} {\bibfnamefont {Y.~Y.}\ \bibnamefont
  {Atas}}, \bibinfo {author} {\bibfnamefont {E.}~\bibnamefont {Bogomolny}},
  \bibinfo {author} {\bibfnamefont {O.}~\bibnamefont {Giraud}},\ and\ \bibinfo
  {author} {\bibfnamefont {G.}~\bibnamefont {Roux}},\ }\href
  {https://doi.org/10.1103/PhysRevLett.110.084101} {\bibfield  {journal}
  {\bibinfo  {journal} {Phys. Rev. Lett.}\ }\textbf {\bibinfo {volume} {110}},\
  \bibinfo {pages} {084101} (\bibinfo {year} {2013})}\BibitemShut {NoStop}%
\bibitem [{\citenamefont {Gharibyan}\ \emph {et~al.}(2018)\citenamefont
  {Gharibyan}, \citenamefont {Hanada}, \citenamefont {Shenker},\ and\
  \citenamefont {Tezuka}}]{Gharibyan2018}%
  \BibitemOpen
  \bibfield  {author} {\bibinfo {author} {\bibfnamefont {H.}~\bibnamefont
  {Gharibyan}}, \bibinfo {author} {\bibfnamefont {M.}~\bibnamefont {Hanada}},
  \bibinfo {author} {\bibfnamefont {S.~H.}\ \bibnamefont {Shenker}},\ and\
  \bibinfo {author} {\bibfnamefont {M.}~\bibnamefont {Tezuka}},\ }\bibfield
  {journal} {\bibinfo  {journal} {Journal of High Energy Physics}\ }\textbf
  {\bibinfo {volume} {2018}},\ \href {https://doi.org/10.1007/JHEP07(2018)124}
  {10.1007/JHEP07(2018)124} (\bibinfo {year} {2018})\BibitemShut {NoStop}%
\bibitem [{\citenamefont {Saad}\ \emph {et~al.}(2019)\citenamefont {Saad},
  \citenamefont {Shenker},\ and\ \citenamefont {Stanford}}]{Saad:2018bqo}%
  \BibitemOpen
  \bibfield  {author} {\bibinfo {author} {\bibfnamefont {P.}~\bibnamefont
  {Saad}}, \bibinfo {author} {\bibfnamefont {S.~H.}\ \bibnamefont {Shenker}},\
  and\ \bibinfo {author} {\bibfnamefont {D.}~\bibnamefont {Stanford}},\ }\href
  {https://doi.org/10.48550/arXiv.1806.06840} {\bibinfo {title} {A
  semiclassical ramp in syk and in gravity}} (\bibinfo {year} {2019}),\ \Eprint
  {https://arxiv.org/abs/1806.06840} {arXiv:1806.06840 [hep-th]} \BibitemShut
  {NoStop}%
\bibitem [{\citenamefont {Larkin}\ and\ \citenamefont
  {Ovchinnikov}(1969)}]{Larkin1969}%
  \BibitemOpen
  \bibfield  {author} {\bibinfo {author} {\bibfnamefont {A.}~\bibnamefont
  {Larkin}}\ and\ \bibinfo {author} {\bibfnamefont {Y.}~\bibnamefont
  {Ovchinnikov}},\ }\href@noop {} {\bibfield  {journal} {\bibinfo  {journal}
  {Soviet Physics JETP}\ }\textbf {\bibinfo {volume} {28}},\ \bibinfo {pages}
  {1200} (\bibinfo {year} {1969})}\BibitemShut {NoStop}%
\bibitem [{\citenamefont {Shenker}\ and\ \citenamefont
  {Stanford}(2014)}]{Shenker_2014}%
  \BibitemOpen
  \bibfield  {author} {\bibinfo {author} {\bibfnamefont {S.~H.}\ \bibnamefont
  {Shenker}}\ and\ \bibinfo {author} {\bibfnamefont {D.}~\bibnamefont
  {Stanford}},\ }\bibfield  {journal} {\bibinfo  {journal} {Journal of High
  Energy Physics}\ }\textbf {\bibinfo {volume} {2014}},\ \href
  {https://doi.org/10.1007/JHEP03(2014)067} {10.1007/JHEP03(2014)067} (\bibinfo
  {year} {2014})\BibitemShut {NoStop}%
\bibitem [{\citenamefont {Shenker}\ and\ \citenamefont
  {Stanford}(2015)}]{Shenker:2014cwa}%
  \BibitemOpen
  \bibfield  {author} {\bibinfo {author} {\bibfnamefont {S.~H.}\ \bibnamefont
  {Shenker}}\ and\ \bibinfo {author} {\bibfnamefont {D.}~\bibnamefont
  {Stanford}},\ }\href {https://doi.org/10.48550/arXiv.1412.6087} {\bibfield
  {journal} {\bibinfo  {journal} {Journal of High Energy Physics}\ }\textbf
  {\bibinfo {volume} {05}},\ \bibinfo {pages} {132} (\bibinfo {year}
  {2015})}\BibitemShut {NoStop}%
\bibitem [{\citenamefont {Bi}\ \emph {et~al.}(2017)\citenamefont {Bi},
  \citenamefont {Jian}, \citenamefont {You}, \citenamefont {Pawlak},\ and\
  \citenamefont {Xu}}]{Bi:2017yvx}%
  \BibitemOpen
  \bibfield  {author} {\bibinfo {author} {\bibfnamefont {Z.}~\bibnamefont
  {Bi}}, \bibinfo {author} {\bibfnamefont {C.-M.}\ \bibnamefont {Jian}},
  \bibinfo {author} {\bibfnamefont {Y.-Z.}\ \bibnamefont {You}}, \bibinfo
  {author} {\bibfnamefont {K.~A.}\ \bibnamefont {Pawlak}},\ and\ \bibinfo
  {author} {\bibfnamefont {C.}~\bibnamefont {Xu}},\ }\href
  {https://doi.org/10.1103/PhysRevB.95.205105} {\bibfield  {journal} {\bibinfo
  {journal} {Phys. Rev. B}\ }\textbf {\bibinfo {volume} {95}},\ \bibinfo
  {pages} {205105} (\bibinfo {year} {2017})}\BibitemShut {NoStop}%
\bibitem [{\citenamefont {Kim}\ \emph {et~al.}(2020)\citenamefont {Kim},
  \citenamefont {Cao},\ and\ \citenamefont {Altman}}]{Kim_2020}%
  \BibitemOpen
  \bibfield  {author} {\bibinfo {author} {\bibfnamefont {J.}~\bibnamefont
  {Kim}}, \bibinfo {author} {\bibfnamefont {X.}~\bibnamefont {Cao}},\ and\
  \bibinfo {author} {\bibfnamefont {E.}~\bibnamefont {Altman}},\ }\href
  {https://doi.org/10.1103/PhysRevB.101.125112} {\bibfield  {journal} {\bibinfo
   {journal} {Phys. Rev. B}\ }\textbf {\bibinfo {volume} {101}},\ \bibinfo
  {pages} {125112} (\bibinfo {year} {2020})}\BibitemShut {NoStop}%
\bibitem [{\citenamefont {Sachdev}(2015)}]{Sachdev:2015efa}%
  \BibitemOpen
  \bibfield  {author} {\bibinfo {author} {\bibfnamefont {S.}~\bibnamefont
  {Sachdev}},\ }\href {https://doi.org/10.1103/PhysRevX.5.041025} {\bibfield
  {journal} {\bibinfo  {journal} {Phys. Rev. X}\ }\textbf {\bibinfo {volume}
  {5}},\ \bibinfo {pages} {041025} (\bibinfo {year} {2015})}\BibitemShut
  {NoStop}%
\bibitem [{\citenamefont {Davison}\ \emph {et~al.}(2017)\citenamefont
  {Davison}, \citenamefont {Fu}, \citenamefont {Georges}, \citenamefont {Gu},
  \citenamefont {Jensen},\ and\ \citenamefont {Sachdev}}]{Davison:2016ngz}%
  \BibitemOpen
  \bibfield  {author} {\bibinfo {author} {\bibfnamefont {R.~A.}\ \bibnamefont
  {Davison}}, \bibinfo {author} {\bibfnamefont {W.}~\bibnamefont {Fu}},
  \bibinfo {author} {\bibfnamefont {A.}~\bibnamefont {Georges}}, \bibinfo
  {author} {\bibfnamefont {Y.}~\bibnamefont {Gu}}, \bibinfo {author}
  {\bibfnamefont {K.}~\bibnamefont {Jensen}},\ and\ \bibinfo {author}
  {\bibfnamefont {S.}~\bibnamefont {Sachdev}},\ }\href
  {https://doi.org/10.1103/PhysRevB.95.155131} {\bibfield  {journal} {\bibinfo
  {journal} {Phys. Rev. B}\ }\textbf {\bibinfo {volume} {95}},\ \bibinfo
  {pages} {155131} (\bibinfo {year} {2017})}\BibitemShut {NoStop}%
\bibitem [{\citenamefont {Gu}\ \emph {et~al.}(2020)\citenamefont {Gu},
  \citenamefont {Kitaev}, \citenamefont {Sachdev},\ and\ \citenamefont
  {Tarnopolsky}}]{Gu_2020}%
  \BibitemOpen
  \bibfield  {author} {\bibinfo {author} {\bibfnamefont {Y.}~\bibnamefont
  {Gu}}, \bibinfo {author} {\bibfnamefont {A.}~\bibnamefont {Kitaev}}, \bibinfo
  {author} {\bibfnamefont {S.}~\bibnamefont {Sachdev}},\ and\ \bibinfo {author}
  {\bibfnamefont {G.}~\bibnamefont {Tarnopolsky}},\ }\bibfield  {journal}
  {\bibinfo  {journal} {Journal of High Energy Physics}\ }\textbf {\bibinfo
  {volume} {2020}},\ \href {https://doi.org/10.1007/jhep02(2020)157}
  {10.1007/jhep02(2020)157} (\bibinfo {year} {2020})\BibitemShut {NoStop}%
\bibitem [{\citenamefont {Liao}\ \emph {et~al.}(2020)\citenamefont {Liao},
  \citenamefont {Vikram},\ and\ \citenamefont {Galitski}}]{Liao2020}%
  \BibitemOpen
  \bibfield  {author} {\bibinfo {author} {\bibfnamefont {Y.}~\bibnamefont
  {Liao}}, \bibinfo {author} {\bibfnamefont {A.}~\bibnamefont {Vikram}},\ and\
  \bibinfo {author} {\bibfnamefont {V.}~\bibnamefont {Galitski}},\ }\href
  {https://doi.org/10.1103/PhysRevLett.125.250601} {\bibfield  {journal}
  {\bibinfo  {journal} {Phys. Rev. Lett.}\ }\textbf {\bibinfo {volume} {125}},\
  \bibinfo {pages} {250601} (\bibinfo {year} {2020})}\BibitemShut {NoStop}%
\bibitem [{\citenamefont {Winer}\ \emph {et~al.}(2020)\citenamefont {Winer},
  \citenamefont {Jian},\ and\ \citenamefont {Swingle}}]{Winer2020}%
  \BibitemOpen
  \bibfield  {author} {\bibinfo {author} {\bibfnamefont {M.}~\bibnamefont
  {Winer}}, \bibinfo {author} {\bibfnamefont {S.-K.}\ \bibnamefont {Jian}},\
  and\ \bibinfo {author} {\bibfnamefont {B.}~\bibnamefont {Swingle}},\ }\href
  {https://doi.org/10.1103/PhysRevLett.125.250602} {\bibfield  {journal}
  {\bibinfo  {journal} {Phys. Rev. Lett.}\ }\textbf {\bibinfo {volume} {125}},\
  \bibinfo {pages} {250602} (\bibinfo {year} {2020})}\BibitemShut {NoStop}%
\bibitem [{\citenamefont {Legramandi}\ \emph {et~al.}(2024)\citenamefont
  {Legramandi}, \citenamefont {Bandyopadhyay},\ and\ \citenamefont
  {Hauke}}]{Legramandi2024}%
  \BibitemOpen
  \bibfield  {author} {\bibinfo {author} {\bibfnamefont {A.}~\bibnamefont
  {Legramandi}}, \bibinfo {author} {\bibfnamefont {S.}~\bibnamefont
  {Bandyopadhyay}},\ and\ \bibinfo {author} {\bibfnamefont {P.}~\bibnamefont
  {Hauke}}\ }\href {https://doi.org/10.48550/arXiv.2412.14280}
  {10.48550/arXiv.2412.14280} (\bibinfo {year} {2024}),\ \Eprint
  {https://arxiv.org/abs/2412.14280} {arXiv:2412.14280 [cond-mat.stat-mech]}
  \BibitemShut {NoStop}%
\bibitem [{\citenamefont {Garc\'{\i}a-Garc\'{\i}a}\ \emph
  {et~al.}(2018)\citenamefont {Garc\'{\i}a-Garc\'{\i}a}, \citenamefont
  {Loureiro}, \citenamefont {Romero-Berm\'udez},\ and\ \citenamefont
  {Tezuka}}]{Garcia-Garcia2018}%
  \BibitemOpen
  \bibfield  {author} {\bibinfo {author} {\bibfnamefont {A.~M.}\ \bibnamefont
  {Garc\'{\i}a-Garc\'{\i}a}}, \bibinfo {author} {\bibfnamefont
  {B.}~\bibnamefont {Loureiro}}, \bibinfo {author} {\bibfnamefont
  {A.}~\bibnamefont {Romero-Berm\'udez}},\ and\ \bibinfo {author}
  {\bibfnamefont {M.}~\bibnamefont {Tezuka}},\ }\href
  {https://doi.org/10.1103/PhysRevLett.120.241603} {\bibfield  {journal}
  {\bibinfo  {journal} {Phys. Rev. Lett.}\ }\textbf {\bibinfo {volume} {120}},\
  \bibinfo {pages} {241603} (\bibinfo {year} {2018})}\BibitemShut {NoStop}%
\bibitem [{\citenamefont {Huang}\ \emph {et~al.}(2019)\citenamefont {Huang},
  \citenamefont {Brand\~ao},\ and\ \citenamefont {Zhang}}]{Finite_size_OTOC}%
  \BibitemOpen
  \bibfield  {author} {\bibinfo {author} {\bibfnamefont {Y.}~\bibnamefont
  {Huang}}, \bibinfo {author} {\bibfnamefont {F.~G. S.~L.}\ \bibnamefont
  {Brand\~ao}},\ and\ \bibinfo {author} {\bibfnamefont {Y.-L.}\ \bibnamefont
  {Zhang}},\ }\href {https://doi.org/10.1103/PhysRevLett.123.010601} {\bibfield
   {journal} {\bibinfo  {journal} {Phys. Rev. Lett.}\ }\textbf {\bibinfo
  {volume} {123}},\ \bibinfo {pages} {010601} (\bibinfo {year}
  {2019})}\BibitemShut {NoStop}%
\bibitem [{\citenamefont {Kukuljan}\ \emph {et~al.}(2017)\citenamefont
  {Kukuljan}, \citenamefont {Grozdanov},\ and\ \citenamefont
  {Prosen}}]{weak_quantum_chaos}%
  \BibitemOpen
  \bibfield  {author} {\bibinfo {author} {\bibfnamefont {I.}~\bibnamefont
  {Kukuljan}}, \bibinfo {author} {\bibfnamefont {S.~c.~v.}\ \bibnamefont
  {Grozdanov}},\ and\ \bibinfo {author} {\bibfnamefont {T.~c.~v.}\ \bibnamefont
  {Prosen}},\ }\href {https://doi.org/10.1103/PhysRevB.96.060301} {\bibfield
  {journal} {\bibinfo  {journal} {Phys. Rev. B}\ }\textbf {\bibinfo {volume}
  {96}},\ \bibinfo {pages} {060301} (\bibinfo {year} {2017})}\BibitemShut
  {NoStop}%
\bibitem [{\citenamefont {He}\ and\ \citenamefont {Lu}(2017)}]{He_MBL_OTOC}%
  \BibitemOpen
  \bibfield  {author} {\bibinfo {author} {\bibfnamefont {R.-Q.}\ \bibnamefont
  {He}}\ and\ \bibinfo {author} {\bibfnamefont {Z.-Y.}\ \bibnamefont {Lu}},\
  }\href {https://doi.org/10.1103/PhysRevB.95.054201} {\bibfield  {journal}
  {\bibinfo  {journal} {Phys. Rev. B}\ }\textbf {\bibinfo {volume} {95}},\
  \bibinfo {pages} {054201} (\bibinfo {year} {2017})}\BibitemShut {NoStop}%
\bibitem [{\citenamefont {Swingle}\ and\ \citenamefont
  {Chowdhury}(2017)}]{Swingle_slow_scrambling}%
  \BibitemOpen
  \bibfield  {author} {\bibinfo {author} {\bibfnamefont {B.}~\bibnamefont
  {Swingle}}\ and\ \bibinfo {author} {\bibfnamefont {D.}~\bibnamefont
  {Chowdhury}},\ }\href {https://doi.org/10.1103/PhysRevB.95.060201} {\bibfield
   {journal} {\bibinfo  {journal} {Phys. Rev. B}\ }\textbf {\bibinfo {volume}
  {95}},\ \bibinfo {pages} {060201} (\bibinfo {year} {2017})}\BibitemShut
  {NoStop}%
\bibitem [{\citenamefont {Lin}\ and\ \citenamefont
  {Motrunich}(2018)}]{OTOC_quantum_Ising_chain}%
  \BibitemOpen
  \bibfield  {author} {\bibinfo {author} {\bibfnamefont {C.-J.}\ \bibnamefont
  {Lin}}\ and\ \bibinfo {author} {\bibfnamefont {O.~I.}\ \bibnamefont
  {Motrunich}},\ }\href {https://doi.org/10.1103/PhysRevB.97.144304} {\bibfield
   {journal} {\bibinfo  {journal} {Phys. Rev. B}\ }\textbf {\bibinfo {volume}
  {97}},\ \bibinfo {pages} {144304} (\bibinfo {year} {2018})}\BibitemShut
  {NoStop}%
\bibitem [{\citenamefont {Rakovszky}\ \emph {et~al.}(2018)\citenamefont
  {Rakovszky}, \citenamefont {Pollmann},\ and\ \citenamefont {von
  Keyserlingk}}]{Rakovszky2018}%
  \BibitemOpen
  \bibfield  {author} {\bibinfo {author} {\bibfnamefont {T.}~\bibnamefont
  {Rakovszky}}, \bibinfo {author} {\bibfnamefont {F.}~\bibnamefont
  {Pollmann}},\ and\ \bibinfo {author} {\bibfnamefont {C.~W.}\ \bibnamefont
  {von Keyserlingk}},\ }\href {https://doi.org/10.1103/PhysRevX.8.031058}
  {\bibfield  {journal} {\bibinfo  {journal} {Phys. Rev. X}\ }\textbf {\bibinfo
  {volume} {8}},\ \bibinfo {pages} {031058} (\bibinfo {year}
  {2018})}\BibitemShut {NoStop}%
\bibitem [{\citenamefont {Nandy}\ \emph {et~al.}(2022)\citenamefont {Nandy},
  \citenamefont {\ifmmode \check{C}\else \v{C}\fi{}ade\ifmmode~\check{z}\else
  \v{z}\fi{}}, \citenamefont {Dietz}, \citenamefont {Andreanov},\ and\
  \citenamefont {Rosa}}]{Nandy22}%
  \BibitemOpen
  \bibfield  {author} {\bibinfo {author} {\bibfnamefont {D.~K.}\ \bibnamefont
  {Nandy}}, \bibinfo {author} {\bibfnamefont {T.}~\bibnamefont {\ifmmode
  \check{C}\else \v{C}\fi{}ade\ifmmode~\check{z}\else \v{z}\fi{}}}, \bibinfo
  {author} {\bibfnamefont {B.}~\bibnamefont {Dietz}}, \bibinfo {author}
  {\bibfnamefont {A.}~\bibnamefont {Andreanov}},\ and\ \bibinfo {author}
  {\bibfnamefont {D.}~\bibnamefont {Rosa}},\ }\href
  {https://doi.org/10.1103/PhysRevB.106.245147} {\bibfield  {journal} {\bibinfo
   {journal} {Phys. Rev. B}\ }\textbf {\bibinfo {volume} {106}},\ \bibinfo
  {pages} {245147} (\bibinfo {year} {2022})}\BibitemShut {NoStop}%
\bibitem [{\citenamefont {Dieplinger}\ and\ \citenamefont
  {Bera}(2023)}]{Dieplinger2023}%
  \BibitemOpen
  \bibfield  {author} {\bibinfo {author} {\bibfnamefont {J.}~\bibnamefont
  {Dieplinger}}\ and\ \bibinfo {author} {\bibfnamefont {S.}~\bibnamefont
  {Bera}},\ }\href {https://doi.org/10.1103/PhysRevB.107.224207} {\bibfield
  {journal} {\bibinfo  {journal} {Phys. Rev. B}\ }\textbf {\bibinfo {volume}
  {107}},\ \bibinfo {pages} {224207} (\bibinfo {year} {2023})}\BibitemShut
  {NoStop}%
\bibitem [{\citenamefont {Baumgartner}\ \emph
  {et~al.}(2024{\natexlab{a}})\citenamefont {Baumgartner}, \citenamefont
  {Delacr{\'e}taz}, \citenamefont {Nayak},\ and\ \citenamefont
  {Sonner}}]{Baumgartner:2024orz}%
  \BibitemOpen
  \bibfield  {author} {\bibinfo {author} {\bibfnamefont {R.~L.}\ \bibnamefont
  {Baumgartner}}, \bibinfo {author} {\bibfnamefont {L.~V.}\ \bibnamefont
  {Delacr{\'e}taz}}, \bibinfo {author} {\bibfnamefont {P.}~\bibnamefont
  {Nayak}},\ and\ \bibinfo {author} {\bibfnamefont {J.}~\bibnamefont {Sonner}}\
  }\href {https://doi.org/10.48550/arXiv.2405.19260}
  {10.48550/arXiv.2405.19260} (\bibinfo {year} {2024}{\natexlab{a}}),\ \Eprint
  {https://arxiv.org/abs/2405.19260} {arXiv:2405.19260 [cond-mat.stat-mech]}
  \BibitemShut {NoStop}%
\bibitem [{\citenamefont {Mivehvar}\ \emph {et~al.}(2021)\citenamefont
  {Mivehvar}, \citenamefont {Piazza}, \citenamefont {Donner},\ and\
  \citenamefont {Ritsch}}]{Mivehvar02012021}%
  \BibitemOpen
  \bibfield  {author} {\bibinfo {author} {\bibfnamefont {F.}~\bibnamefont
  {Mivehvar}}, \bibinfo {author} {\bibfnamefont {F.}~\bibnamefont {Piazza}},
  \bibinfo {author} {\bibfnamefont {T.}~\bibnamefont {Donner}},\ and\ \bibinfo
  {author} {\bibfnamefont {H.}~\bibnamefont {Ritsch}},\ }\href
  {https://doi.org/10.1080/00018732.2021.1969727} {\bibfield  {journal}
  {\bibinfo  {journal} {Advances in Physics}\ }\textbf {\bibinfo {volume}
  {70}},\ \bibinfo {pages} {1} (\bibinfo {year} {2021})}\BibitemShut {NoStop}%
\bibitem [{\citenamefont {Uhrich}\ \emph {et~al.}(2023)\citenamefont {Uhrich},
  \citenamefont {Bandyopadhyay}, \citenamefont {Sauerwein}, \citenamefont
  {Sonner}, \citenamefont {Brantut},\ and\ \citenamefont
  {Hauke}}]{cavity_proposal}%
  \BibitemOpen
  \bibfield  {author} {\bibinfo {author} {\bibfnamefont {P.}~\bibnamefont
  {Uhrich}}, \bibinfo {author} {\bibfnamefont {S.}~\bibnamefont
  {Bandyopadhyay}}, \bibinfo {author} {\bibfnamefont {N.}~\bibnamefont
  {Sauerwein}}, \bibinfo {author} {\bibfnamefont {J.}~\bibnamefont {Sonner}},
  \bibinfo {author} {\bibfnamefont {J.-P.}\ \bibnamefont {Brantut}},\ and\
  \bibinfo {author} {\bibfnamefont {P.}~\bibnamefont {Hauke}},\ }\href
  {https://doi.org/10.48550/arXiv.2303.11343} {\bibinfo {title} {A cavity
  quantum electrodynamics implementation of the sachdev--ye--kitaev model}}
  (\bibinfo {year} {2023}),\ \Eprint {https://arxiv.org/abs/2303.11343}
  {arXiv:2303.11343 [quant-ph]} \BibitemShut {NoStop}%
\bibitem [{\citenamefont {Baumgartner}\ \emph
  {et~al.}(2024{\natexlab{b}})\citenamefont {Baumgartner}, \citenamefont
  {Pelliconi}, \citenamefont {Bandyopadhyay}, \citenamefont {Orsi},
  \citenamefont {Sauerwein}, \citenamefont {Hauke}, \citenamefont {Brantut},\
  and\ \citenamefont {Sonner}}]{Baumgartner:2024ysk}%
  \BibitemOpen
  \bibfield  {author} {\bibinfo {author} {\bibfnamefont {R.}~\bibnamefont
  {Baumgartner}}, \bibinfo {author} {\bibfnamefont {P.}~\bibnamefont
  {Pelliconi}}, \bibinfo {author} {\bibfnamefont {S.}~\bibnamefont
  {Bandyopadhyay}}, \bibinfo {author} {\bibfnamefont {F.}~\bibnamefont {Orsi}},
  \bibinfo {author} {\bibfnamefont {N.}~\bibnamefont {Sauerwein}}, \bibinfo
  {author} {\bibfnamefont {P.}~\bibnamefont {Hauke}}, \bibinfo {author}
  {\bibfnamefont {J.-P.}\ \bibnamefont {Brantut}},\ and\ \bibinfo {author}
  {\bibfnamefont {J.}~\bibnamefont {Sonner}},\ }\href
  {https://doi.org/10.48550/arXiv.2411.17802} {\bibinfo {title} {Quantum
  simulation of the sachdev-ye-kitaev model using time-dependent disorder in
  optical cavities}} (\bibinfo {year} {2024}{\natexlab{b}}),\ \Eprint
  {https://arxiv.org/abs/2411.17802} {arXiv:2411.17802 [quant-ph]} \BibitemShut
  {NoStop}%
\bibitem [{\citenamefont {Baghdad}\ \emph {et~al.}(2023)\citenamefont
  {Baghdad}, \citenamefont {Bourdel}, \citenamefont {Schwartz}, \citenamefont
  {Ferri}, \citenamefont {Reichel},\ and\ \citenamefont {Long}}]{Baghdad_2023}%
  \BibitemOpen
  \bibfield  {author} {\bibinfo {author} {\bibfnamefont {M.}~\bibnamefont
  {Baghdad}}, \bibinfo {author} {\bibfnamefont {P.-A.}\ \bibnamefont
  {Bourdel}}, \bibinfo {author} {\bibfnamefont {S.}~\bibnamefont {Schwartz}},
  \bibinfo {author} {\bibfnamefont {F.}~\bibnamefont {Ferri}}, \bibinfo
  {author} {\bibfnamefont {J.}~\bibnamefont {Reichel}},\ and\ \bibinfo {author}
  {\bibfnamefont {R.}~\bibnamefont {Long}},\ }\href
  {https://doi.org/10.1038/s41567-023-02035-1} {\bibfield  {journal} {\bibinfo
  {journal} {Nature Physics}\ }\textbf {\bibinfo {volume} {19}},\ \bibinfo
  {pages} {1104–1109} (\bibinfo {year} {2023})}\BibitemShut {NoStop}%
\bibitem [{\citenamefont {Sauerwein}\ \emph {et~al.}(2023)\citenamefont
  {Sauerwein}, \citenamefont {Orsi}, \citenamefont {Uhrich}, \citenamefont
  {Bandyopadhyay}, \citenamefont {Mattiotti}, \citenamefont {Cantat-Moltrecht},
  \citenamefont {Pupillo}, \citenamefont {Hauke},\ and\ \citenamefont
  {Brantut}}]{sauerwein_engineering_2023}%
  \BibitemOpen
  \bibfield  {author} {\bibinfo {author} {\bibfnamefont {N.}~\bibnamefont
  {Sauerwein}}, \bibinfo {author} {\bibfnamefont {F.}~\bibnamefont {Orsi}},
  \bibinfo {author} {\bibfnamefont {P.}~\bibnamefont {Uhrich}}, \bibinfo
  {author} {\bibfnamefont {S.}~\bibnamefont {Bandyopadhyay}}, \bibinfo {author}
  {\bibfnamefont {F.}~\bibnamefont {Mattiotti}}, \bibinfo {author}
  {\bibfnamefont {T.}~\bibnamefont {Cantat-Moltrecht}}, \bibinfo {author}
  {\bibfnamefont {G.}~\bibnamefont {Pupillo}}, \bibinfo {author} {\bibfnamefont
  {P.}~\bibnamefont {Hauke}},\ and\ \bibinfo {author} {\bibfnamefont {J.-P.}\
  \bibnamefont {Brantut}},\ }\href {https://doi.org/10.1038/s41567-023-02033-3}
  {\bibfield  {journal} {\bibinfo  {journal} {Nature Physics}\ }\textbf
  {\bibinfo {volume} {19}},\ \bibinfo {pages} {1128} (\bibinfo {year}
  {2023})}\BibitemShut {NoStop}%
\bibitem [{\citenamefont {Orsi}\ \emph {et~al.}(2024)\citenamefont {Orsi},
  \citenamefont {Sauerwein}, \citenamefont {Bhatt}, \citenamefont {Faltinath},
  \citenamefont {Fedotova}, \citenamefont {Reiter}, \citenamefont
  {Cantat-Moltrecht},\ and\ \citenamefont {Brantut}}]{Forsi_cavity_control24}%
  \BibitemOpen
  \bibfield  {author} {\bibinfo {author} {\bibfnamefont {F.}~\bibnamefont
  {Orsi}}, \bibinfo {author} {\bibfnamefont {N.}~\bibnamefont {Sauerwein}},
  \bibinfo {author} {\bibfnamefont {R.~P.}\ \bibnamefont {Bhatt}}, \bibinfo
  {author} {\bibfnamefont {J.}~\bibnamefont {Faltinath}}, \bibinfo {author}
  {\bibfnamefont {E.}~\bibnamefont {Fedotova}}, \bibinfo {author}
  {\bibfnamefont {N.}~\bibnamefont {Reiter}}, \bibinfo {author} {\bibfnamefont
  {T.}~\bibnamefont {Cantat-Moltrecht}},\ and\ \bibinfo {author} {\bibfnamefont
  {J.-P.}\ \bibnamefont {Brantut}},\ }\href
  {https://doi.org/10.1103/PRXQuantum.5.040333} {\bibfield  {journal} {\bibinfo
   {journal} {PRX Quantum}\ }\textbf {\bibinfo {volume} {5}},\ \bibinfo {pages}
  {040333} (\bibinfo {year} {2024})}\BibitemShut {NoStop}%
\bibitem [{\citenamefont {Garc\'{\i}a-Garc\'{\i}a}\ and\ \citenamefont
  {Verbaarschot}(2016)}]{Garcia-Garcia2016}%
  \BibitemOpen
  \bibfield  {author} {\bibinfo {author} {\bibfnamefont {A.~M.}\ \bibnamefont
  {Garc\'{\i}a-Garc\'{\i}a}}\ and\ \bibinfo {author} {\bibfnamefont {J.~J.~M.}\
  \bibnamefont {Verbaarschot}},\ }\href
  {https://doi.org/10.1103/PhysRevD.94.126010} {\bibfield  {journal} {\bibinfo
  {journal} {Phys. Rev. D}\ }\textbf {\bibinfo {volume} {94}},\ \bibinfo
  {pages} {126010} (\bibinfo {year} {2016})}\BibitemShut {NoStop}%
\bibitem [{\citenamefont {Garc\'{\i}a-Garc\'{\i}a}\ and\ \citenamefont
  {Verbaarschot}(2017)}]{Garcia-Garcia2017}%
  \BibitemOpen
  \bibfield  {author} {\bibinfo {author} {\bibfnamefont {A.~M.}\ \bibnamefont
  {Garc\'{\i}a-Garc\'{\i}a}}\ and\ \bibinfo {author} {\bibfnamefont {J.~J.~M.}\
  \bibnamefont {Verbaarschot}},\ }\href
  {https://doi.org/10.1103/PhysRevD.96.066012} {\bibfield  {journal} {\bibinfo
  {journal} {Phys. Rev. D}\ }\textbf {\bibinfo {volume} {96}},\ \bibinfo
  {pages} {066012} (\bibinfo {year} {2017})}\BibitemShut {NoStop}%
\bibitem [{\citenamefont {Bhattacharya}\ \emph {et~al.}(2017)\citenamefont
  {Bhattacharya}, \citenamefont {Chakrabarti}, \citenamefont {Jatkar},\ and\
  \citenamefont {Kundu}}]{Bhattacharya:2017vaz}%
  \BibitemOpen
  \bibfield  {author} {\bibinfo {author} {\bibfnamefont {R.}~\bibnamefont
  {Bhattacharya}}, \bibinfo {author} {\bibfnamefont {S.}~\bibnamefont
  {Chakrabarti}}, \bibinfo {author} {\bibfnamefont {D.~P.}\ \bibnamefont
  {Jatkar}},\ and\ \bibinfo {author} {\bibfnamefont {A.}~\bibnamefont
  {Kundu}},\ }\bibfield  {journal} {\bibinfo  {journal} {Journal of High Energy
  Physics}\ }\textbf {\bibinfo {volume} {2017}},\ \href
  {https://doi.org/10.1007/JHEP11(2017)180} {10.1007/JHEP11(2017)180} (\bibinfo
  {year} {2017})\BibitemShut {NoStop}%
\bibitem [{\citenamefont {Rigol}\ \emph {et~al.}(2006)\citenamefont {Rigol},
  \citenamefont {Muramatsu},\ and\ \citenamefont {Olshanii}}]{Rigol2006}%
  \BibitemOpen
  \bibfield  {author} {\bibinfo {author} {\bibfnamefont {M.}~\bibnamefont
  {Rigol}}, \bibinfo {author} {\bibfnamefont {A.}~\bibnamefont {Muramatsu}},\
  and\ \bibinfo {author} {\bibfnamefont {M.}~\bibnamefont {Olshanii}},\ }\href
  {https://doi.org/10.1103/PhysRevA.74.053616} {\bibfield  {journal} {\bibinfo
  {journal} {Phys. Rev. A}\ }\textbf {\bibinfo {volume} {74}},\ \bibinfo
  {pages} {053616} (\bibinfo {year} {2006})}\BibitemShut {NoStop}%
\bibitem [{\citenamefont {Calabrese}\ and\ \citenamefont
  {Cardy}(2005)}]{Calabrese_2005}%
  \BibitemOpen
  \bibfield  {author} {\bibinfo {author} {\bibfnamefont {P.}~\bibnamefont
  {Calabrese}}\ and\ \bibinfo {author} {\bibfnamefont {J.}~\bibnamefont
  {Cardy}},\ }\href {https://doi.org/10.1088/1742-5468/2005/04/p04010}
  {\bibfield  {journal} {\bibinfo  {journal} {Journal of Statistical Mechanics:
  Theory and Experiment}\ }\textbf {\bibinfo {volume} {2005}},\ \bibinfo
  {pages} {P04010} (\bibinfo {year} {2005})}\BibitemShut {NoStop}%
\bibitem [{\citenamefont {Kim}\ and\ \citenamefont {Huse}(2013)}]{Kim_2013}%
  \BibitemOpen
  \bibfield  {author} {\bibinfo {author} {\bibfnamefont {H.}~\bibnamefont
  {Kim}}\ and\ \bibinfo {author} {\bibfnamefont {D.~A.}\ \bibnamefont {Huse}},\
  }\href {https://doi.org/10.1103/PhysRevLett.111.127205} {\bibfield  {journal}
  {\bibinfo  {journal} {Phys. Rev. Lett.}\ }\textbf {\bibinfo {volume} {111}},\
  \bibinfo {pages} {127205} (\bibinfo {year} {2013})}\BibitemShut {NoStop}%
\bibitem [{\citenamefont {Nandkishore}\ and\ \citenamefont
  {Huse}(2015)}]{Nandkishore_2015}%
  \BibitemOpen
  \bibfield  {author} {\bibinfo {author} {\bibfnamefont {R.}~\bibnamefont
  {Nandkishore}}\ and\ \bibinfo {author} {\bibfnamefont {D.~A.}\ \bibnamefont
  {Huse}},\ }\href {https://doi.org/10.1146/annurev-conmatphys-031214-014726}
  {\bibfield  {journal} {\bibinfo  {journal} {Annual Review of Condensed Matter
  Physics}\ }\textbf {\bibinfo {volume} {6}},\ \bibinfo {pages} {15–38}
  (\bibinfo {year} {2015})}\BibitemShut {NoStop}%
\bibitem [{\citenamefont {Abanin}\ \emph
  {et~al.}(2019{\natexlab{b}})\citenamefont {Abanin}, \citenamefont {Altman},
  \citenamefont {Bloch},\ and\ \citenamefont {Serbyn}}]{Dima_2019}%
  \BibitemOpen
  \bibfield  {author} {\bibinfo {author} {\bibfnamefont {D.~A.}\ \bibnamefont
  {Abanin}}, \bibinfo {author} {\bibfnamefont {E.}~\bibnamefont {Altman}},
  \bibinfo {author} {\bibfnamefont {I.}~\bibnamefont {Bloch}},\ and\ \bibinfo
  {author} {\bibfnamefont {M.}~\bibnamefont {Serbyn}},\ }\href
  {https://doi.org/10.1103/RevModPhys.91.021001} {\bibfield  {journal}
  {\bibinfo  {journal} {Rev. Mod. Phys.}\ }\textbf {\bibinfo {volume} {91}},\
  \bibinfo {pages} {021001} (\bibinfo {year} {2019}{\natexlab{b}})}\BibitemShut
  {NoStop}%
\bibitem [{\citenamefont {Lukin}\ \emph {et~al.}(2019)\citenamefont {Lukin},
  \citenamefont {Rispoli}, \citenamefont {Schittko}, \citenamefont {Tai},
  \citenamefont {Kaufman}, \citenamefont {Choi}, \citenamefont {Khemani},
  \citenamefont {Léonard},\ and\ \citenamefont {Greiner}}]{Lukin_2019}%
  \BibitemOpen
  \bibfield  {author} {\bibinfo {author} {\bibfnamefont {A.}~\bibnamefont
  {Lukin}}, \bibinfo {author} {\bibfnamefont {M.}~\bibnamefont {Rispoli}},
  \bibinfo {author} {\bibfnamefont {R.}~\bibnamefont {Schittko}}, \bibinfo
  {author} {\bibfnamefont {M.~E.}\ \bibnamefont {Tai}}, \bibinfo {author}
  {\bibfnamefont {A.~M.}\ \bibnamefont {Kaufman}}, \bibinfo {author}
  {\bibfnamefont {S.}~\bibnamefont {Choi}}, \bibinfo {author} {\bibfnamefont
  {V.}~\bibnamefont {Khemani}}, \bibinfo {author} {\bibfnamefont
  {J.}~\bibnamefont {Léonard}},\ and\ \bibinfo {author} {\bibfnamefont
  {M.}~\bibnamefont {Greiner}},\ }\href
  {https://doi.org/10.1126/science.aau0818} {\bibfield  {journal} {\bibinfo
  {journal} {Science}\ }\textbf {\bibinfo {volume} {364}},\ \bibinfo {pages}
  {256–260} (\bibinfo {year} {2019})}\BibitemShut {NoStop}%
\bibitem [{\citenamefont {Thouless}(1974)}]{Thouless1974}%
  \BibitemOpen
  \bibfield  {author} {\bibinfo {author} {\bibfnamefont {D.}~\bibnamefont
  {Thouless}},\ }\href
  {https://doi.org/https://doi.org/10.1016/0370-1573(74)90029-5} {\bibfield
  {journal} {\bibinfo  {journal} {Physics Reports}\ }\textbf {\bibinfo {volume}
  {13}},\ \bibinfo {pages} {93} (\bibinfo {year} {1974})}\BibitemShut {NoStop}%
\bibitem [{\citenamefont {Altshuler}\ and\ \citenamefont
  {Shklovskii}(1986)}]{Altshuler1986}%
  \BibitemOpen
  \bibfield  {author} {\bibinfo {author} {\bibfnamefont {B.~L.}\ \bibnamefont
  {Altshuler}}\ and\ \bibinfo {author} {\bibfnamefont {B.~I.}\ \bibnamefont
  {Shklovskii}},\ }\href@noop {} {\bibfield  {journal} {\bibinfo  {journal}
  {Sov. Phys. JETP}\ }\textbf {\bibinfo {volume} {64}},\ \bibinfo {pages} {127}
  (\bibinfo {year} {1986})},\ \bibinfo {note} {originally published as Zh.
  Eksp. Teor. Fiz. {\bf 91}, 220–\textmd{tar} (1986)}\BibitemShut {NoStop}%
\bibitem [{\citenamefont {Zotos}\ \emph {et~al.}(1997)\citenamefont {Zotos},
  \citenamefont {Naef},\ and\ \citenamefont {Prelovsek}}]{Zotos_1997}%
  \BibitemOpen
  \bibfield  {author} {\bibinfo {author} {\bibfnamefont {X.}~\bibnamefont
  {Zotos}}, \bibinfo {author} {\bibfnamefont {F.}~\bibnamefont {Naef}},\ and\
  \bibinfo {author} {\bibfnamefont {P.}~\bibnamefont {Prelovsek}},\ }\href
  {https://doi.org/10.1103/PhysRevB.55.11029} {\bibfield  {journal} {\bibinfo
  {journal} {Phys. Rev. B}\ }\textbf {\bibinfo {volume} {55}},\ \bibinfo
  {pages} {11029} (\bibinfo {year} {1997})}\BibitemShut {NoStop}%
\bibitem [{\citenamefont {Heidrich-Meisner}\ \emph {et~al.}(2003)\citenamefont
  {Heidrich-Meisner}, \citenamefont {Honecker}, \citenamefont {Cabra},\ and\
  \citenamefont {Brenig}}]{H-Meisner_2003}%
  \BibitemOpen
  \bibfield  {author} {\bibinfo {author} {\bibfnamefont {F.}~\bibnamefont
  {Heidrich-Meisner}}, \bibinfo {author} {\bibfnamefont {A.}~\bibnamefont
  {Honecker}}, \bibinfo {author} {\bibfnamefont {D.~C.}\ \bibnamefont
  {Cabra}},\ and\ \bibinfo {author} {\bibfnamefont {W.}~\bibnamefont
  {Brenig}},\ }\href {https://doi.org/10.1103/PhysRevB.68.134436} {\bibfield
  {journal} {\bibinfo  {journal} {Phys. Rev. B}\ }\textbf {\bibinfo {volume}
  {68}},\ \bibinfo {pages} {134436} (\bibinfo {year} {2003})}\BibitemShut
  {NoStop}%
\bibitem [{\citenamefont {Bertini}\ \emph {et~al.}(2021)\citenamefont
  {Bertini}, \citenamefont {Heidrich-Meisner}, \citenamefont {Karrasch},
  \citenamefont {Prosen}, \citenamefont {Steinigeweg},\ and\ \citenamefont
  {\ifmmode \check{Z}\else \v{Z}\fi{}nidari\ifmmode~\check{c}\else
  \v{c}\fi{}}}]{Bertini_2021}%
  \BibitemOpen
  \bibfield  {author} {\bibinfo {author} {\bibfnamefont {B.}~\bibnamefont
  {Bertini}}, \bibinfo {author} {\bibfnamefont {F.}~\bibnamefont
  {Heidrich-Meisner}}, \bibinfo {author} {\bibfnamefont {C.}~\bibnamefont
  {Karrasch}}, \bibinfo {author} {\bibfnamefont {T.}~\bibnamefont {Prosen}},
  \bibinfo {author} {\bibfnamefont {R.}~\bibnamefont {Steinigeweg}},\ and\
  \bibinfo {author} {\bibfnamefont {M.}~\bibnamefont {\ifmmode \check{Z}\else
  \v{Z}\fi{}nidari\ifmmode~\check{c}\else \v{c}\fi{}}},\ }\href
  {https://doi.org/10.1103/RevModPhys.93.025003} {\bibfield  {journal}
  {\bibinfo  {journal} {Rev. Mod. Phys.}\ }\textbf {\bibinfo {volume} {93}},\
  \bibinfo {pages} {025003} (\bibinfo {year} {2021})}\BibitemShut {NoStop}%
\bibitem [{\citenamefont {Wigner}(1958)}]{Wigner1958}%
  \BibitemOpen
  \bibfield  {author} {\bibinfo {author} {\bibfnamefont {E.~P.}\ \bibnamefont
  {Wigner}},\ }\href {https://doi.org/10.2307/1970008} {\bibfield  {journal}
  {\bibinfo  {journal} {Annals of Mathematics}\ }\textbf {\bibinfo {volume}
  {67}},\ \bibinfo {pages} {325} (\bibinfo {year} {1958})}\BibitemShut
  {NoStop}%
\bibitem [{\citenamefont {Dyson}(1962)}]{Dyson1962}%
  \BibitemOpen
  \bibfield  {author} {\bibinfo {author} {\bibfnamefont {F.~J.}\ \bibnamefont
  {Dyson}},\ }\href {https://doi.org/10.1063/1.1703773} {\bibfield  {journal}
  {\bibinfo  {journal} {Journal of Mathematical Physics}\ }\textbf {\bibinfo
  {volume} {3}},\ \bibinfo {pages} {140} (\bibinfo {year} {1962})}\BibitemShut
  {NoStop}%
\bibitem [{\citenamefont {Luitz}\ \emph {et~al.}(2015)\citenamefont {Luitz},
  \citenamefont {Laflorencie},\ and\ \citenamefont {Alet}}]{Luitz2015}%
  \BibitemOpen
  \bibfield  {author} {\bibinfo {author} {\bibfnamefont {D.~J.}\ \bibnamefont
  {Luitz}}, \bibinfo {author} {\bibfnamefont {N.}~\bibnamefont {Laflorencie}},\
  and\ \bibinfo {author} {\bibfnamefont {F.}~\bibnamefont {Alet}},\ }\href
  {https://doi.org/10.1103/PhysRevB.91.081103} {\bibfield  {journal} {\bibinfo
  {journal} {Phys. Rev. B}\ }\textbf {\bibinfo {volume} {91}},\ \bibinfo
  {pages} {081103} (\bibinfo {year} {2015})}\BibitemShut {NoStop}%
\bibitem [{\citenamefont {T\'avora}\ \emph {et~al.}(2016)\citenamefont
  {T\'avora}, \citenamefont {Torres-Herrera},\ and\ \citenamefont
  {Santos}}]{Tavora2016}%
  \BibitemOpen
  \bibfield  {author} {\bibinfo {author} {\bibfnamefont {M.}~\bibnamefont
  {T\'avora}}, \bibinfo {author} {\bibfnamefont {E.~J.}\ \bibnamefont
  {Torres-Herrera}},\ and\ \bibinfo {author} {\bibfnamefont {L.~F.}\
  \bibnamefont {Santos}},\ }\href {https://doi.org/10.1103/PhysRevA.94.041603}
  {\bibfield  {journal} {\bibinfo  {journal} {Phys. Rev. A}\ }\textbf {\bibinfo
  {volume} {94}},\ \bibinfo {pages} {041603} (\bibinfo {year}
  {2016})}\BibitemShut {NoStop}%
\bibitem [{\citenamefont {Torres-Herrera}\ and\ \citenamefont
  {Santos}(2017)}]{Herrera2017}%
  \BibitemOpen
  \bibfield  {author} {\bibinfo {author} {\bibfnamefont {E.~J.}\ \bibnamefont
  {Torres-Herrera}}\ and\ \bibinfo {author} {\bibfnamefont {L.~F.}\
  \bibnamefont {Santos}},\ }\href {https://doi.org/10.1098/rsta.2016.0434}
  {\bibfield  {journal} {\bibinfo  {journal} {Philosophical Transactions of the
  Royal Society A: Mathematical, Physical and Engineering Sciences}\ }\textbf
  {\bibinfo {volume} {375}},\ \bibinfo {pages} {20160434} (\bibinfo {year}
  {2017})}\BibitemShut {NoStop}%
\bibitem [{\citenamefont {Torres-Herrera}\ \emph {et~al.}(2018)\citenamefont
  {Torres-Herrera}, \citenamefont {Garc\'{\i}a-Garc\'{\i}a},\ and\
  \citenamefont {Santos}}]{Herrera2018}%
  \BibitemOpen
  \bibfield  {author} {\bibinfo {author} {\bibfnamefont {E.~J.}\ \bibnamefont
  {Torres-Herrera}}, \bibinfo {author} {\bibfnamefont {A.~M.}\ \bibnamefont
  {Garc\'{\i}a-Garc\'{\i}a}},\ and\ \bibinfo {author} {\bibfnamefont {L.~F.}\
  \bibnamefont {Santos}},\ }\href {https://doi.org/10.1103/PhysRevB.97.060303}
  {\bibfield  {journal} {\bibinfo  {journal} {Phys. Rev. B}\ }\textbf {\bibinfo
  {volume} {97}},\ \bibinfo {pages} {060303} (\bibinfo {year}
  {2018})}\BibitemShut {NoStop}%
\bibitem [{\citenamefont {Santos}\ \emph {et~al.}(2020)\citenamefont {Santos},
  \citenamefont {P\'erez-Bernal},\ and\ \citenamefont
  {Torres-Herrera}}]{Santos2020}%
  \BibitemOpen
  \bibfield  {author} {\bibinfo {author} {\bibfnamefont {L.~F.}\ \bibnamefont
  {Santos}}, \bibinfo {author} {\bibfnamefont {F.}~\bibnamefont
  {P\'erez-Bernal}},\ and\ \bibinfo {author} {\bibfnamefont {E.~J.}\
  \bibnamefont {Torres-Herrera}},\ }\href
  {https://doi.org/10.1103/PhysRevResearch.2.043034} {\bibfield  {journal}
  {\bibinfo  {journal} {Phys. Rev. Res.}\ }\textbf {\bibinfo {volume} {2}},\
  \bibinfo {pages} {043034} (\bibinfo {year} {2020})}\BibitemShut {NoStop}%
\bibitem [{\citenamefont {Altshuler}\ \emph {et~al.}(1988)\citenamefont
  {Altshuler}, \citenamefont {Zharekeshev}, \citenamefont {Kotochigova},\ and\
  \citenamefont {Shklovskii}}]{Altshuler1988}%
  \BibitemOpen
  \bibfield  {author} {\bibinfo {author} {\bibfnamefont {B.~L.}\ \bibnamefont
  {Altshuler}}, \bibinfo {author} {\bibfnamefont {I.~K.}\ \bibnamefont
  {Zharekeshev}}, \bibinfo {author} {\bibfnamefont {S.~A.}\ \bibnamefont
  {Kotochigova}},\ and\ \bibinfo {author} {\bibfnamefont {B.~I.}\ \bibnamefont
  {Shklovskii}},\ }\href@noop {} {\bibfield  {journal} {\bibinfo  {journal}
  {Sov. Phys. JETP}\ }\textbf {\bibinfo {volume} {67}},\ \bibinfo {pages} {625}
  (\bibinfo {year} {1988})},\ \bibinfo {note} {originally published as Zh.
  Eksp. Teor. Fiz. {\bf 94}, 343 (1988)}\BibitemShut {NoStop}%
\bibitem [{\citenamefont {Evers}\ and\ \citenamefont
  {Mirlin}(2008)}]{Evers_2008}%
  \BibitemOpen
  \bibfield  {author} {\bibinfo {author} {\bibfnamefont {F.}~\bibnamefont
  {Evers}}\ and\ \bibinfo {author} {\bibfnamefont {A.~D.}\ \bibnamefont
  {Mirlin}},\ }\href {https://doi.org/10.1103/RevModPhys.80.1355} {\bibfield
  {journal} {\bibinfo  {journal} {Rev. Mod. Phys.}\ }\textbf {\bibinfo {volume}
  {80}},\ \bibinfo {pages} {1355} (\bibinfo {year} {2008})}\BibitemShut
  {NoStop}%
\bibitem [{\citenamefont {Liu}(2018)}]{Liu_2018}%
  \BibitemOpen
  \bibfield  {author} {\bibinfo {author} {\bibfnamefont {J.}~\bibnamefont
  {Liu}},\ }\href {https://doi.org/10.1103/PhysRevD.98.086026} {\bibfield
  {journal} {\bibinfo  {journal} {Phys. Rev. D}\ }\textbf {\bibinfo {volume}
  {98}},\ \bibinfo {pages} {086026} (\bibinfo {year} {2018})}\BibitemShut
  {NoStop}%
\bibitem [{\citenamefont {Chan}\ \emph {et~al.}(2018)\citenamefont {Chan},
  \citenamefont {De~Luca},\ and\ \citenamefont {Chalker}}]{Chan2018}%
  \BibitemOpen
  \bibfield  {author} {\bibinfo {author} {\bibfnamefont {A.}~\bibnamefont
  {Chan}}, \bibinfo {author} {\bibfnamefont {A.}~\bibnamefont {De~Luca}},\ and\
  \bibinfo {author} {\bibfnamefont {J.~T.}\ \bibnamefont {Chalker}},\ }\href
  {https://doi.org/10.1103/PhysRevLett.121.060601} {\bibfield  {journal}
  {\bibinfo  {journal} {Phys. Rev. Lett.}\ }\textbf {\bibinfo {volume} {121}},\
  \bibinfo {pages} {060601} (\bibinfo {year} {2018})}\BibitemShut {NoStop}%
\bibitem [{\citenamefont {Prakash}\ \emph {et~al.}(2021)\citenamefont
  {Prakash}, \citenamefont {Pixley},\ and\ \citenamefont
  {Kulkarni}}]{Prakash2021}%
  \BibitemOpen
  \bibfield  {author} {\bibinfo {author} {\bibfnamefont {A.}~\bibnamefont
  {Prakash}}, \bibinfo {author} {\bibfnamefont {J.~H.}\ \bibnamefont
  {Pixley}},\ and\ \bibinfo {author} {\bibfnamefont {M.}~\bibnamefont
  {Kulkarni}},\ }\href {https://doi.org/10.1103/PhysRevResearch.3.L012019}
  {\bibfield  {journal} {\bibinfo  {journal} {Phys. Rev. Res.}\ }\textbf
  {\bibinfo {volume} {3}},\ \bibinfo {pages} {L012019} (\bibinfo {year}
  {2021})}\BibitemShut {NoStop}%
\bibitem [{\citenamefont {Gopalakrishnan}\ \emph {et~al.}(2016)\citenamefont
  {Gopalakrishnan}, \citenamefont {Agarwal}, \citenamefont {Demler},
  \citenamefont {Huse},\ and\ \citenamefont {Knap}}]{Gopalakrishnan2016}%
  \BibitemOpen
  \bibfield  {author} {\bibinfo {author} {\bibfnamefont {S.}~\bibnamefont
  {Gopalakrishnan}}, \bibinfo {author} {\bibfnamefont {K.}~\bibnamefont
  {Agarwal}}, \bibinfo {author} {\bibfnamefont {E.~A.}\ \bibnamefont {Demler}},
  \bibinfo {author} {\bibfnamefont {D.~A.}\ \bibnamefont {Huse}},\ and\
  \bibinfo {author} {\bibfnamefont {M.}~\bibnamefont {Knap}},\ }\href
  {https://doi.org/10.1103/PhysRevB.93.134206} {\bibfield  {journal} {\bibinfo
  {journal} {Phys. Rev. B}\ }\textbf {\bibinfo {volume} {93}},\ \bibinfo
  {pages} {134206} (\bibinfo {year} {2016})}\BibitemShut {NoStop}%
\bibitem [{\citenamefont {Serbyn}\ and\ \citenamefont
  {Moore}(2016)}]{Serbyn2016}%
  \BibitemOpen
  \bibfield  {author} {\bibinfo {author} {\bibfnamefont {M.}~\bibnamefont
  {Serbyn}}\ and\ \bibinfo {author} {\bibfnamefont {J.~E.}\ \bibnamefont
  {Moore}},\ }\href {https://doi.org/10.1103/PhysRevB.93.041424} {\bibfield
  {journal} {\bibinfo  {journal} {Phys. Rev. B}\ }\textbf {\bibinfo {volume}
  {93}},\ \bibinfo {pages} {041424} (\bibinfo {year} {2016})}\BibitemShut
  {NoStop}%
\bibitem [{\citenamefont {Serbyn}\ \emph {et~al.}(2017)\citenamefont {Serbyn},
  \citenamefont {Papi\ifmmode~\acute{c}\else \'{c}\fi{}},\ and\ \citenamefont
  {Abanin}}]{Serbyn2017}%
  \BibitemOpen
  \bibfield  {author} {\bibinfo {author} {\bibfnamefont {M.}~\bibnamefont
  {Serbyn}}, \bibinfo {author} {\bibfnamefont {Z.}~\bibnamefont
  {Papi\ifmmode~\acute{c}\else \'{c}\fi{}}},\ and\ \bibinfo {author}
  {\bibfnamefont {D.~A.}\ \bibnamefont {Abanin}},\ }\href
  {https://doi.org/10.1103/PhysRevB.96.104201} {\bibfield  {journal} {\bibinfo
  {journal} {Phys. Rev. B}\ }\textbf {\bibinfo {volume} {96}},\ \bibinfo
  {pages} {104201} (\bibinfo {year} {2017})}\BibitemShut {NoStop}%
\bibitem [{\citenamefont {Prosen}(2002)}]{Prosen2002}%
  \BibitemOpen
  \bibfield  {author} {\bibinfo {author} {\bibfnamefont {T.~c.~v.}\
  \bibnamefont {Prosen}},\ }\href {https://doi.org/10.1103/PhysRevE.65.036208}
  {\bibfield  {journal} {\bibinfo  {journal} {Phys. Rev. E}\ }\textbf {\bibinfo
  {volume} {65}},\ \bibinfo {pages} {036208} (\bibinfo {year}
  {2002})}\BibitemShut {NoStop}%
\bibitem [{\citenamefont {Altland}\ and\ \citenamefont
  {Bagrets}(2018)}]{Altland2018}%
  \BibitemOpen
  \bibfield  {author} {\bibinfo {author} {\bibfnamefont {A.}~\bibnamefont
  {Altland}}\ and\ \bibinfo {author} {\bibfnamefont {D.}~\bibnamefont
  {Bagrets}},\ }\href
  {https://doi.org/https://doi.org/10.1016/j.nuclphysb.2018.02.015} {\bibfield
  {journal} {\bibinfo  {journal} {Nuclear Physics B}\ }\textbf {\bibinfo
  {volume} {930}},\ \bibinfo {pages} {45} (\bibinfo {year} {2018})}\BibitemShut
  {NoStop}%
\bibitem [{\citenamefont {Roberts}\ \emph {et~al.}(2018)\citenamefont
  {Roberts}, \citenamefont {Stanford},\ and\ \citenamefont
  {Streicher}}]{Roberts:2018mnp}%
  \BibitemOpen
  \bibfield  {author} {\bibinfo {author} {\bibfnamefont {D.~A.}\ \bibnamefont
  {Roberts}}, \bibinfo {author} {\bibfnamefont {D.}~\bibnamefont {Stanford}},\
  and\ \bibinfo {author} {\bibfnamefont {A.}~\bibnamefont {Streicher}},\
  }\bibfield  {journal} {\bibinfo  {journal} {Journal of High Energy Physics}\
  }\textbf {\bibinfo {volume} {2018}},\ \href
  {https://doi.org/10.1007/JHEP06(2018)122} {10.1007/JHEP06(2018)122} (\bibinfo
  {year} {2018})\BibitemShut {NoStop}%
\bibitem [{\citenamefont {Cotler}\ \emph {et~al.}(2018)\citenamefont {Cotler},
  \citenamefont {Ding},\ and\ \citenamefont
  {Penington}}]{Cotler_butterfly_effect}%
  \BibitemOpen
  \bibfield  {author} {\bibinfo {author} {\bibfnamefont {J.~S.}\ \bibnamefont
  {Cotler}}, \bibinfo {author} {\bibfnamefont {D.}~\bibnamefont {Ding}},\ and\
  \bibinfo {author} {\bibfnamefont {G.~R.}\ \bibnamefont {Penington}},\ }\href
  {https://doi.org/https://doi.org/10.1016/j.aop.2018.07.020} {\bibfield
  {journal} {\bibinfo  {journal} {Annals of Physics}\ }\textbf {\bibinfo
  {volume} {396}},\ \bibinfo {pages} {318} (\bibinfo {year}
  {2018})}\BibitemShut {NoStop}%
\bibitem [{\citenamefont {Varikuti}\ \emph {et~al.}(2024)\citenamefont
  {Varikuti}, \citenamefont {Sahu}, \citenamefont {Lakshminarayan},\ and\
  \citenamefont {Madhok}}]{Varikuti_2024}%
  \BibitemOpen
  \bibfield  {author} {\bibinfo {author} {\bibfnamefont {N.~D.}\ \bibnamefont
  {Varikuti}}, \bibinfo {author} {\bibfnamefont {A.}~\bibnamefont {Sahu}},
  \bibinfo {author} {\bibfnamefont {A.}~\bibnamefont {Lakshminarayan}},\ and\
  \bibinfo {author} {\bibfnamefont {V.}~\bibnamefont {Madhok}},\ }\href
  {https://doi.org/10.1103/PhysRevE.109.014209} {\bibfield  {journal} {\bibinfo
   {journal} {Phys. Rev. E}\ }\textbf {\bibinfo {volume} {109}},\ \bibinfo
  {pages} {014209} (\bibinfo {year} {2024})}\BibitemShut {NoStop}%
\bibitem [{\citenamefont {Sekino}\ and\ \citenamefont
  {Susskind}(2008)}]{Fast_scramblers}%
  \BibitemOpen
  \bibfield  {author} {\bibinfo {author} {\bibfnamefont {Y.}~\bibnamefont
  {Sekino}}\ and\ \bibinfo {author} {\bibfnamefont {L.}~\bibnamefont
  {Susskind}},\ }\href {https://doi.org/10.1088/1126-6708/2008/10/065}
  {\bibfield  {journal} {\bibinfo  {journal} {Journal of High Energy Physics}\
  }\textbf {\bibinfo {volume} {2008}},\ \bibinfo {pages} {065} (\bibinfo {year}
  {2008})}\BibitemShut {NoStop}%
\bibitem [{\citenamefont {Dowling}\ \emph {et~al.}(2023)\citenamefont
  {Dowling}, \citenamefont {Kos},\ and\ \citenamefont {Modi}}]{Dowling_2023}%
  \BibitemOpen
  \bibfield  {author} {\bibinfo {author} {\bibfnamefont {N.}~\bibnamefont
  {Dowling}}, \bibinfo {author} {\bibfnamefont {P.}~\bibnamefont {Kos}},\ and\
  \bibinfo {author} {\bibfnamefont {K.}~\bibnamefont {Modi}},\ }\href
  {https://doi.org/10.1103/PhysRevLett.131.180403} {\bibfield  {journal}
  {\bibinfo  {journal} {Phys. Rev. Lett.}\ }\textbf {\bibinfo {volume} {131}},\
  \bibinfo {pages} {180403} (\bibinfo {year} {2023})}\BibitemShut {NoStop}%
\bibitem [{\citenamefont {Xu}\ \emph {et~al.}(2020)\citenamefont {Xu},
  \citenamefont {Scaffidi},\ and\ \citenamefont {Cao}}]{Scrambling_and_chaos}%
  \BibitemOpen
  \bibfield  {author} {\bibinfo {author} {\bibfnamefont {T.}~\bibnamefont
  {Xu}}, \bibinfo {author} {\bibfnamefont {T.}~\bibnamefont {Scaffidi}},\ and\
  \bibinfo {author} {\bibfnamefont {X.}~\bibnamefont {Cao}},\ }\href
  {https://doi.org/10.1103/PhysRevLett.124.140602} {\bibfield  {journal}
  {\bibinfo  {journal} {Phys. Rev. Lett.}\ }\textbf {\bibinfo {volume} {124}},\
  \bibinfo {pages} {140602} (\bibinfo {year} {2020})}\BibitemShut {NoStop}%
\bibitem [{\citenamefont {García-Mata}\ \emph {et~al.}(2023)\citenamefont
  {García-Mata}, \citenamefont {Jalabert},\ and\ \citenamefont
  {Wisniacki}}]{OTOC_and_quantumchaos}%
  \BibitemOpen
  \bibfield  {author} {\bibinfo {author} {\bibfnamefont {I.}~\bibnamefont
  {García-Mata}}, \bibinfo {author} {\bibfnamefont {R.}~\bibnamefont
  {Jalabert}},\ and\ \bibinfo {author} {\bibfnamefont {D.}~\bibnamefont
  {Wisniacki}},\ }\href {https://doi.org/10.4249/scholarpedia.55237} {\bibfield
   {journal} {\bibinfo  {journal} {Scholarpedia}\ }\textbf {\bibinfo {volume}
  {18}},\ \bibinfo {pages} {55237} (\bibinfo {year} {2023})},\ \bibinfo {note}
  {revision \#204529}\BibitemShut {NoStop}%
\bibitem [{\citenamefont {Hummel}\ \emph {et~al.}(2019)\citenamefont {Hummel},
  \citenamefont {Geiger}, \citenamefont {Urbina},\ and\ \citenamefont
  {Richter}}]{Hummel_2019}%
  \BibitemOpen
  \bibfield  {author} {\bibinfo {author} {\bibfnamefont {Q.}~\bibnamefont
  {Hummel}}, \bibinfo {author} {\bibfnamefont {B.}~\bibnamefont {Geiger}},
  \bibinfo {author} {\bibfnamefont {J.~D.}\ \bibnamefont {Urbina}},\ and\
  \bibinfo {author} {\bibfnamefont {K.}~\bibnamefont {Richter}},\ }\href
  {https://doi.org/10.1103/PhysRevLett.123.160401} {\bibfield  {journal}
  {\bibinfo  {journal} {Phys. Rev. Lett.}\ }\textbf {\bibinfo {volume} {123}},\
  \bibinfo {pages} {160401} (\bibinfo {year} {2019})}\BibitemShut {NoStop}%
\bibitem [{\citenamefont {Kidd}\ \emph {et~al.}(2021)\citenamefont {Kidd},
  \citenamefont {Safavi-Naini},\ and\ \citenamefont {Corney}}]{Kidd_2021}%
  \BibitemOpen
  \bibfield  {author} {\bibinfo {author} {\bibfnamefont {R.~A.}\ \bibnamefont
  {Kidd}}, \bibinfo {author} {\bibfnamefont {A.}~\bibnamefont {Safavi-Naini}},\
  and\ \bibinfo {author} {\bibfnamefont {J.~F.}\ \bibnamefont {Corney}},\
  }\href {https://doi.org/10.1103/PhysRevA.103.033304} {\bibfield  {journal}
  {\bibinfo  {journal} {Phys. Rev. A}\ }\textbf {\bibinfo {volume} {103}},\
  \bibinfo {pages} {033304} (\bibinfo {year} {2021})}\BibitemShut {NoStop}%
\bibitem [{\citenamefont {Garc\'{\i}a-Garc\'{\i}a}\ \emph
  {et~al.}(2025)\citenamefont {Garc\'{\i}a-Garc\'{\i}a}, \citenamefont {Liu},
  \citenamefont {S\'a}, \citenamefont {Verbaarschot},\ and\ \citenamefont
  {Zheng}}]{garcíagarcía2025}%
  \BibitemOpen
  \bibfield  {author} {\bibinfo {author} {\bibfnamefont {A.~M.}\ \bibnamefont
  {Garc\'{\i}a-Garc\'{\i}a}}, \bibinfo {author} {\bibfnamefont
  {C.}~\bibnamefont {Liu}}, \bibinfo {author} {\bibfnamefont {L.}~\bibnamefont
  {S\'a}}, \bibinfo {author} {\bibfnamefont {J.~J.~M.}\ \bibnamefont
  {Verbaarschot}},\ and\ \bibinfo {author} {\bibfnamefont {J.-p.}\ \bibnamefont
  {Zheng}},\ }\href {https://doi.org/10.1103/j589-dszc} {\bibfield  {journal}
  {\bibinfo  {journal} {Phys. Rev. E}\ }\textbf {\bibinfo {volume} {112}},\
  \bibinfo {pages} {054203} (\bibinfo {year} {2025})}\BibitemShut {NoStop}%
\bibitem [{\citenamefont {D\'ora}\ and\ \citenamefont
  {Moessner}(2017)}]{Dora_2017}%
  \BibitemOpen
  \bibfield  {author} {\bibinfo {author} {\bibfnamefont {B.}~\bibnamefont
  {D\'ora}}\ and\ \bibinfo {author} {\bibfnamefont {R.}~\bibnamefont
  {Moessner}},\ }\href {https://doi.org/10.1103/PhysRevLett.119.026802}
  {\bibfield  {journal} {\bibinfo  {journal} {Phys. Rev. Lett.}\ }\textbf
  {\bibinfo {volume} {119}},\ \bibinfo {pages} {026802} (\bibinfo {year}
  {2017})}\BibitemShut {NoStop}%
\bibitem [{\citenamefont {Nosaka}\ \emph {et~al.}(2018)\citenamefont {Nosaka},
  \citenamefont {Rosa},\ and\ \citenamefont {Yoon}}]{Nosaka:2018iat}%
  \BibitemOpen
  \bibfield  {author} {\bibinfo {author} {\bibfnamefont {T.}~\bibnamefont
  {Nosaka}}, \bibinfo {author} {\bibfnamefont {D.}~\bibnamefont {Rosa}},\ and\
  \bibinfo {author} {\bibfnamefont {J.}~\bibnamefont {Yoon}},\ }\bibfield
  {journal} {\bibinfo  {journal} {Journal of High Energy Physics}\ }\textbf
  {\bibinfo {volume} {2018}},\ \href {https://doi.org/10.1007/JHEP09(2018)041}
  {10.1007/JHEP09(2018)041} (\bibinfo {year} {2018})\BibitemShut {NoStop}%
\bibitem [{\citenamefont {Goldman}\ and\ \citenamefont
  {Dalibard}(2014)}]{Goldman:2014xja}%
  \BibitemOpen
  \bibfield  {author} {\bibinfo {author} {\bibfnamefont {N.}~\bibnamefont
  {Goldman}}\ and\ \bibinfo {author} {\bibfnamefont {J.}~\bibnamefont
  {Dalibard}},\ }\href {https://doi.org/10.1103/PhysRevX.4.031027} {\bibfield
  {journal} {\bibinfo  {journal} {Phys. Rev. X}\ }\textbf {\bibinfo {volume}
  {4}},\ \bibinfo {pages} {031027} (\bibinfo {year} {2014})},\ \bibinfo {note}
  {[Erratum: Phys.Rev.X 5, 029902 (2015)]}\BibitemShut {NoStop}%
\bibitem [{\citenamefont {Eckardt}\ and\ \citenamefont
  {Anisimovas}(2015)}]{Eckardt:2015mtt}%
  \BibitemOpen
  \bibfield  {author} {\bibinfo {author} {\bibfnamefont {A.}~\bibnamefont
  {Eckardt}}\ and\ \bibinfo {author} {\bibfnamefont {E.}~\bibnamefont
  {Anisimovas}},\ }\href {https://doi.org/10.1088/1367-2630/17/9/093039}
  {\bibfield  {journal} {\bibinfo  {journal} {New J. Phys.}\ }\textbf {\bibinfo
  {volume} {17}},\ \bibinfo {pages} {093039} (\bibinfo {year}
  {2015})}\BibitemShut {NoStop}%
\bibitem [{\citenamefont {Johansen}\ \emph {et~al.}(2022)\citenamefont
  {Johansen}, \citenamefont {Lang}, \citenamefont {Morales}, \citenamefont
  {Baumgärtner}, \citenamefont {Donner},\ and\ \citenamefont
  {Piazza}}]{Johansen2022}%
  \BibitemOpen
  \bibfield  {author} {\bibinfo {author} {\bibfnamefont {C.~H.}\ \bibnamefont
  {Johansen}}, \bibinfo {author} {\bibfnamefont {J.}~\bibnamefont {Lang}},
  \bibinfo {author} {\bibfnamefont {A.}~\bibnamefont {Morales}}, \bibinfo
  {author} {\bibfnamefont {A.}~\bibnamefont {Baumgärtner}}, \bibinfo {author}
  {\bibfnamefont {T.}~\bibnamefont {Donner}},\ and\ \bibinfo {author}
  {\bibfnamefont {F.}~\bibnamefont {Piazza}},\ }\href
  {https://doi.org/10.21468/SciPostPhys.12.3.094} {\bibfield  {journal}
  {\bibinfo  {journal} {SciPost Phys.}\ }\textbf {\bibinfo {volume} {12}},\
  \bibinfo {pages} {094} (\bibinfo {year} {2022})}\BibitemShut {NoStop}%
\bibitem [{\citenamefont {Monteiro}\ \emph {et~al.}(2021)\citenamefont
  {Monteiro}, \citenamefont {Micklitz}, \citenamefont {Tezuka},\ and\
  \citenamefont {Altland}}]{Monteiro2021}%
  \BibitemOpen
  \bibfield  {author} {\bibinfo {author} {\bibfnamefont {F.}~\bibnamefont
  {Monteiro}}, \bibinfo {author} {\bibfnamefont {T.}~\bibnamefont {Micklitz}},
  \bibinfo {author} {\bibfnamefont {M.}~\bibnamefont {Tezuka}},\ and\ \bibinfo
  {author} {\bibfnamefont {A.}~\bibnamefont {Altland}},\ }\href
  {https://doi.org/10.1103/PhysRevResearch.3.013023} {\bibfield  {journal}
  {\bibinfo  {journal} {Phys. Rev. Res.}\ }\textbf {\bibinfo {volume} {3}},\
  \bibinfo {pages} {013023} (\bibinfo {year} {2021})}\BibitemShut {NoStop}%
\bibitem [{\citenamefont {Dieplinger}\ \emph {et~al.}(2021)\citenamefont
  {Dieplinger}, \citenamefont {Bera},\ and\ \citenamefont
  {Evers}}]{Dieplinger2021}%
  \BibitemOpen
  \bibfield  {author} {\bibinfo {author} {\bibfnamefont {J.}~\bibnamefont
  {Dieplinger}}, \bibinfo {author} {\bibfnamefont {S.}~\bibnamefont {Bera}},\
  and\ \bibinfo {author} {\bibfnamefont {F.}~\bibnamefont {Evers}},\ }\href
  {https://doi.org/10.1016/j.aop.2021.168503} {\bibfield  {journal} {\bibinfo
  {journal} {Annals of Physics}\ }\textbf {\bibinfo {volume} {435}},\ \bibinfo
  {pages} {168503} (\bibinfo {year} {2021})},\ \bibinfo {note} {special Issue
  on Localisation 2020}\BibitemShut {NoStop}%
\bibitem [{\citenamefont {Santra}\ \emph {et~al.}(2025)\citenamefont {Santra},
  \citenamefont {Windey}, \citenamefont {Bandyopadhyay}, \citenamefont
  {Legramandi},\ and\ \citenamefont {Hauke}}]{Santra2025}%
  \BibitemOpen
  \bibfield  {author} {\bibinfo {author} {\bibfnamefont {G.~C.}\ \bibnamefont
  {Santra}}, \bibinfo {author} {\bibfnamefont {A.}~\bibnamefont {Windey}},
  \bibinfo {author} {\bibfnamefont {S.}~\bibnamefont {Bandyopadhyay}}, \bibinfo
  {author} {\bibfnamefont {A.}~\bibnamefont {Legramandi}},\ and\ \bibinfo
  {author} {\bibfnamefont {P.}~\bibnamefont {Hauke}},\ }\href
  {https://doi.org/10.48550/arXiv.2505.09707} {\bibinfo {title} {Complexity
  transitions in chaotic quantum systems: Nonstabilizerness, entanglement, and
  fractal dimension in syk and random matrix models}} (\bibinfo {year}
  {2025}),\ \Eprint {https://arxiv.org/abs/2505.09707} {arXiv:2505.09707
  [quant-ph]} \BibitemShut {NoStop}%
\bibitem [{\citenamefont {Denisov}\ \emph {et~al.}(2019)\citenamefont
  {Denisov}, \citenamefont {Laptyeva}, \citenamefont {Tarnowski}, \citenamefont
  {Chru\ifmmode \acute{s}\else \'{s}\fi{}ci\ifmmode~\acute{n}\else
  \'{n}\fi{}ski},\ and\ \citenamefont {\ifmmode~\dot{Z}\else
  \.{Z}\fi{}yczkowski}}]{Denisov:2018nif}%
  \BibitemOpen
  \bibfield  {author} {\bibinfo {author} {\bibfnamefont {S.}~\bibnamefont
  {Denisov}}, \bibinfo {author} {\bibfnamefont {T.}~\bibnamefont {Laptyeva}},
  \bibinfo {author} {\bibfnamefont {W.}~\bibnamefont {Tarnowski}}, \bibinfo
  {author} {\bibfnamefont {D.}~\bibnamefont {Chru\ifmmode \acute{s}\else
  \'{s}\fi{}ci\ifmmode~\acute{n}\else \'{n}\fi{}ski}},\ and\ \bibinfo {author}
  {\bibfnamefont {K.}~\bibnamefont {\ifmmode~\dot{Z}\else
  \.{Z}\fi{}yczkowski}},\ }\href
  {https://doi.org/10.1103/PhysRevLett.123.140403} {\bibfield  {journal}
  {\bibinfo  {journal} {Phys. Rev. Lett.}\ }\textbf {\bibinfo {volume} {123}},\
  \bibinfo {pages} {140403} (\bibinfo {year} {2019})}\BibitemShut {NoStop}%
\bibitem [{\citenamefont {S\'a}\ \emph {et~al.}(2020)\citenamefont {S\'a},
  \citenamefont {Ribeiro},\ and\ \citenamefont {Prosen}}]{Sa:2020fpf}%
  \BibitemOpen
  \bibfield  {author} {\bibinfo {author} {\bibfnamefont {L.}~\bibnamefont
  {S\'a}}, \bibinfo {author} {\bibfnamefont {P.}~\bibnamefont {Ribeiro}},\ and\
  \bibinfo {author} {\bibfnamefont {T.~c.~v.}\ \bibnamefont {Prosen}},\ }\href
  {https://doi.org/10.1103/PhysRevX.10.021019} {\bibfield  {journal} {\bibinfo
  {journal} {Phys. Rev. X}\ }\textbf {\bibinfo {volume} {10}},\ \bibinfo
  {pages} {021019} (\bibinfo {year} {2020})}\BibitemShut {NoStop}%
\bibitem [{\citenamefont {Jana}\ \emph {et~al.}(2020)\citenamefont {Jana},
  \citenamefont {Loganayagam},\ and\ \citenamefont {Rangamani}}]{Jana:2020vyx}%
  \BibitemOpen
  \bibfield  {author} {\bibinfo {author} {\bibfnamefont {C.}~\bibnamefont
  {Jana}}, \bibinfo {author} {\bibfnamefont {R.}~\bibnamefont {Loganayagam}},\
  and\ \bibinfo {author} {\bibfnamefont {M.}~\bibnamefont {Rangamani}},\
  }\bibfield  {journal} {\bibinfo  {journal} {Journal of High Energy Physics}\
  }\textbf {\bibinfo {volume} {2020}},\ \href
  {https://doi.org/10.1007/JHEP07(2020)242} {10.1007/JHEP07(2020)242} (\bibinfo
  {year} {2020})\BibitemShut {NoStop}%
\bibitem [{\citenamefont {Pelliconi}\ and\ \citenamefont
  {Sonner}(2024)}]{Pelliconi:2023ojb}%
  \BibitemOpen
  \bibfield  {author} {\bibinfo {author} {\bibfnamefont {P.}~\bibnamefont
  {Pelliconi}}\ and\ \bibinfo {author} {\bibfnamefont {J.}~\bibnamefont
  {Sonner}},\ }\href {https://doi.org/10.1007/JHEP06(2024)185} {\bibfield
  {journal} {\bibinfo  {journal} {Journal of High Energy Physics}\ }\textbf
  {\bibinfo {volume} {06}},\ \bibinfo {pages} {185} (\bibinfo {year}
  {2024})}\BibitemShut {NoStop}%
\bibitem [{\citenamefont {Kulkarni}\ \emph {et~al.}(2022)\citenamefont
  {Kulkarni}, \citenamefont {Numasawa},\ and\ \citenamefont
  {Ryu}}]{Kulkarni:2021gtt}%
  \BibitemOpen
  \bibfield  {author} {\bibinfo {author} {\bibfnamefont {A.}~\bibnamefont
  {Kulkarni}}, \bibinfo {author} {\bibfnamefont {T.}~\bibnamefont {Numasawa}},\
  and\ \bibinfo {author} {\bibfnamefont {S.}~\bibnamefont {Ryu}},\ }\href
  {https://doi.org/10.1103/PhysRevB.106.075138} {\bibfield  {journal} {\bibinfo
   {journal} {Phys. Rev. B}\ }\textbf {\bibinfo {volume} {106}},\ \bibinfo
  {pages} {075138} (\bibinfo {year} {2022})}\BibitemShut {NoStop}%
\bibitem [{\citenamefont {Hosseinabadi}\ \emph {et~al.}(2023)\citenamefont
  {Hosseinabadi}, \citenamefont {Kelly}, \citenamefont {Schmalian},\ and\
  \citenamefont {Marino}}]{Hosseinabadi:2023mid}%
  \BibitemOpen
  \bibfield  {author} {\bibinfo {author} {\bibfnamefont {H.}~\bibnamefont
  {Hosseinabadi}}, \bibinfo {author} {\bibfnamefont {S.~P.}\ \bibnamefont
  {Kelly}}, \bibinfo {author} {\bibfnamefont {J.}~\bibnamefont {Schmalian}},\
  and\ \bibinfo {author} {\bibfnamefont {J.}~\bibnamefont {Marino}},\ }\href
  {https://doi.org/10.1103/PhysRevB.108.104319} {\bibfield  {journal} {\bibinfo
   {journal} {Phys. Rev. B}\ }\textbf {\bibinfo {volume} {108}},\ \bibinfo
  {pages} {104319} (\bibinfo {year} {2023})}\BibitemShut {NoStop}%
\bibitem [{\citenamefont {Fu}\ \emph {et~al.}(2017)\citenamefont {Fu},
  \citenamefont {Gaiotto}, \citenamefont {Maldacena},\ and\ \citenamefont
  {Sachdev}}]{Fu_2017}%
  \BibitemOpen
  \bibfield  {author} {\bibinfo {author} {\bibfnamefont {W.}~\bibnamefont
  {Fu}}, \bibinfo {author} {\bibfnamefont {D.}~\bibnamefont {Gaiotto}},
  \bibinfo {author} {\bibfnamefont {J.}~\bibnamefont {Maldacena}},\ and\
  \bibinfo {author} {\bibfnamefont {S.}~\bibnamefont {Sachdev}},\ }\href
  {https://doi.org/10.1103/PhysRevD.95.026009} {\bibfield  {journal} {\bibinfo
  {journal} {Phys. Rev. D}\ }\textbf {\bibinfo {volume} {95}},\ \bibinfo
  {pages} {026009} (\bibinfo {year} {2017})}\BibitemShut {NoStop}%
\bibitem [{\citenamefont {Breuer}\ and\ \citenamefont
  {Petruccione}(2007)}]{OpenSystems_Petruccione_book}%
  \BibitemOpen
  \bibfield  {author} {\bibinfo {author} {\bibfnamefont {H.-P.}\ \bibnamefont
  {Breuer}}\ and\ \bibinfo {author} {\bibfnamefont {F.}~\bibnamefont
  {Petruccione}}\ }(\bibinfo  {publisher} {Oxford University Press},\ \bibinfo
  {year} {2007})\BibitemShut {NoStop}%
\end{thebibliography}%

\clearpage
\onecolumngrid
\appendix

\section{Numerical implementation and methods}\label{Appendix: numerical_implementation}

In this Appendix, we summarize the numerical conventions underpinning our simulations and provide more in depth details on the numerical implementation. 

\subsection{Hamiltonian construction and exact diagonalization}

We work with $N$ fermionic modes at half-filling and $M$ bosonic modes truncated to $n_k\in$ $\{0,\dots,N_b\}$. The Hilbert space factorizes as $\mathcal H=\mathcal H_{\mathrm f}\otimes\mathcal H_{\mathrm b}$ with total dimension $D=\binom{N}{N/2}(N_b+1)^M$. The fermionic basis is generated lexicographically as bitstrings with Hamming weight $N/2$, improving computational efficiency in time and memory. 

The Hamiltonian in Eq.~\eqref{eq:hamiltonian_explicit}
is assembled directly as a sparse matrix. For each bosonic basis element, the diagonal term $\omega_0\big(\sum_k n_k+M/2\big)$ is added uniformly to all fermionic rows. The interaction term in Eq.~\eqref{eq:hamiltonian_explicit} couples neighboring bosonic states: for each boson mode $k$ with occupancy $n_k$, the Hamiltonian is non-zero only for states with occupancy $n_k \pm 1$.
Each of these non-zero matrix elements carries a fermionic block $\sum_{ij} g_{ij,k} c_i^\dagger c_j$ which is computed on the fly by evaluating all possible $c_i^\dagger c_j$ hopping terms acting on the fermionic bitstring. Invalid hops (empty initial or filled final fermionic state) are discarded.

For diagonalization, dense Hermitian routines are used when $D\lesssim 2\times10^3$, converting to a dense array to obtain the full spectrum. For larger $D$, the Hamiltonian is kept sparse and a Lanczos solver is used to find the eigenvalues of the lowest-energy cluster in the DOS. The requested number of eigenvalues is chosen adaptively to balance cost and resolution.

\subsection{Spectral form factor, ramp and plateau time (SYK$_2$-like regime)}

The results in Fig.~\ref{fig:syk2_sff_full} have been obtained by calculating the disorder-averaged SFF $K(t)$ at $\beta=0$ for different values of $\omega_0$ and rescaling the time according to Eq.~\eqref{eq:SYK2_time_rescaling}.

For each $\omega_0$, the plateau onset time is calculated by marking the earliest $t$, $\alpha t>\sqrt{N}$, at which $K(t)$ has a constant value $K_{\rm pl}$ within a certain tolerance $\delta_{\rm pl}$, 
\begin{equation}
\frac{|K(t)-K_{\rm pl}|}{K_{\rm pl}}<\delta_{\rm pl} \, ,
\end{equation}
over a contiguous time window of time  $ \Delta t_{\rm pl}$. The plateau time $t_{\rm pl}$ is then defined as the left edge of this window. For calculating the plateau times in Fig.~\ref{fig:syk2_sff_full} (circles), we set $\delta_{\rm pl}=0.1$ and $\alpha \Delta t_{\rm pl}=1.9$, which are large compared to the grid spacing, yet small compared with $t_\mathrm{H}$. This choice suppresses spurious detections caused by finite-size fluctuations and yields stable figures under resampling of the time axis.

The ramp onset time $t_{\rm r}$ is determined as follows. We first define a reference ramp corresponding to $\omega_0 = 0.5$, i.e., in the regime where the ramp is fully developed. In the interval $\alpha t \in [2N,D]$, $K(t)$ is well described by a power law
\begin{equation}
K_{\rm ref}(t)=A_{\rm ref}\,t^{\, \beta_{\rm ref}} \, ,
\end{equation}
with $\beta_{\rm ref} \sim 1$. Here, we take into account that small deviations from perfect linearity in the ramp can arise due to finite-size effects and incomplete RMT-like spectral rigidity.

For each $\omega_0$, we then scan $\alpha t\in[2N,D]$ and declare the ramp onset $t_{\rm r}$ as the first time where $K(t)$ matches $K_{\rm ref}(t)$ within a tight relative tolerance,
\begin{equation}
\frac{|K(t)-K_{\rm ref}(t)|}{K_{\rm ref}(t)}<\delta_{\rm r},\qquad \delta_{\rm r}=5\times10^{-3} \, .
\end{equation}
This definition is insensitive to the absolute normalization of $K$ and to the detailed short-time slope. 
For each $\omega_0$, the ramp times $t_{\rm r}$ are marked in Fig.~\ref{fig:syk2_sff_full} using squares.

\subsection{OTOC functions}
\label{Appendix: OTOC_numerics}

Here, we provide additional  details on the methods used to compute the OTOCs presented in the main text. Based on the system sizes explored in this work, we found it computationally most efficient to evaluate the OTOCs of the full spectrum by directly exponentiating the YSYK Hamiltonian to get the time evolution $U(t) = e^{i H t}$. The OTOC is then obtained through straightforward matrix multiplication 
\begin{equation}
    F(t)  = \Re\left[\mathrm{Tr}\,(\rho_\beta U(t) A^\dag U(t)^\dag B^\dag U(t) A U(t)^\dag B)\right], 
\end{equation}
where $\rho_0 = \mathrm{1}/D$, and $A = 2 n_1 - 1$ and $B = 2 n_2 - 1$ are sparse matrices. 

In contrast, 
when restricting to the lowest-energy peak of the DOS, the OTOC is computed using the Lehmann representation and exact diagonalization. More precisely, we compute the commutator squared $C(t)$, 
which can be written as
\begin{equation}
\begin{split}
        C(t) = \bigg[ \sum_{n,m} e^{-\beta E_n / 2} \bigg( e^{i(E_k - E_m)t}\langle n | B | k\rangle \langle k | A | m \rangle - e^{i(E_n-E_k)t} \langle n | A | k \rangle \langle k | B | m \rangle \bigg) \bigg]^2,
\end{split}
\end{equation}
and use Eq.~\eqref{Eq:F_1_minus_2C} to compute the OTOC from it. 
This approach gives $F(t)$ for the low-energy peaks in Fig.~\ref{Fig: OTOC_cSYK4_regime_combined} of Sec.~\ref{sec:weak}.

\begin{figure}[t!]
  \centering
  \includegraphics[width=0.6\linewidth]{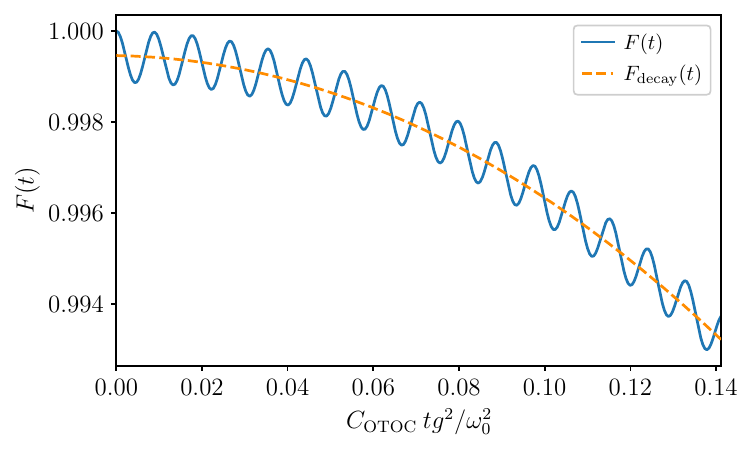}
  \caption{Early-time disorder-averaged YSYK OTOC for $N=8, M=4$, $N_b=1$, and $\omega_0=10$, averaged over $1000$ samples, alongside the filtered decay $F_{\rm decay}(t)$. The chosen numerical filter gives good results to isolate the oscillations from the decaying part of $F(t)$.}
  \label{Fig: OTOC_w10_oscillations_filter}
\end{figure}

Lastly, for filtering out the oscillations arising in the weak-coupling regime of the YSYK model, we use a numerical low-pass filter (specifically, a Savitzky--Golay filter) on the full OTOC that attenuates high-frequency components and passes low-frequency components.
The result of this procedure is given in Fig.~\ref{Fig: OTOC_w10_oscillations_filter} for $\omega_0=10$, which allows us to nicely differentiate the decaying part of the OTOC $F_{\rm decay}(t)$ from the full oscillating part. Substracting this decay from the full $F(t)$ for different values of $\omega_0$ gives the collapse in Fig.~\ref{Fig: OTOC_cSYK4_regime_combined}{\bf b} upon rescaling the axes.    

\section{Comparison with complex SYK$_q$ coupling constant} \label{Appendix: comparison coupling constants}

The YSYK model exhibits clear signatures of both single-body and many-body chaos. Yet, a precise comparison with the complex SYK$_q$ model for $q =2,4$ in their respective limiting regimes requires rescaling the coupling constants to match the characteristic energy scales. To this end, we adopt the operational procedure described in Sec.~\ref{sec:time_comparison}
and calculate the time rescaling $\alpha_\text{SFF}$ and $\alpha_\text{OTOC}$ necessary for comparing YSYK with complex SYK$_q$ in the small and large $\omega_0$ limit, respectively.

\subsection{Small-$\omega_0$ regime}
\label{Appendix: comparison coupling constants_small omega}

In this regime, the benchmark model is complex SYK with $q=2$, as defined in Sec.~\ref{sec:CSYK}. We begin by considering the calculation of $\alpha_\text{SFF}$. In order to calculate $\sigma_\mathrm{SYK_2}$, we first need to calculate the first (averaged) moments of the Hamiltonian. With some straightforward combinatorics, one can show that
\begin{align}
    \frac{\overline{\mathrm{Tr}\, H_{\text{SYK}_2}^2} }{\mathrm{Tr}\, \mathds{1}} = \frac{2J^2}{N} \binom{N/2+1}{2} \, , \qquad
     \frac{\overline{(\mathrm{Tr}\, H_{\text{SYK}_2})^2} }{(\mathrm{Tr}\, \mathds{1})^2} &= J^2 \, \,\frac{\binom{N-1}{N/2}^2}{\binom{N}{N/2}^2} \, ,
\end{align}
where all calculations have been performed in the half-filling sector.
From these results, we have
\begin{equation}
    \sigma_{\text{SYK}_2}^2 = \frac{J^2}{4} (N+1) \, . \label{eq:SYK2_norm}
\end{equation}

This value has to be put into relation to the one of the YSYK Hamiltonian $H$.  
For small $\omega_0$, the boson mass term becomes negligible and we can write
\begin{equation}
    H(\omega_0 \to 0) \sim \frac{1}{\sqrt{2\omega_0 MN}} \sum_{i,j=1}^N \sum_{k=1}^M g_{ij,k}\, c_i^\dagger c_j\, \left( a_k + a_k^\dagger \right) \, . 
\label{eq:H_omega_0=0}
\end{equation}
In this limit, $\mathrm{Tr}\, H(\omega_0 \to 0)=0$ and we just have to calculate $\mathrm{Tr}\, H(\omega_0 \to 0)^2$. Using the fact that $\overline{g_{ij,k} g_{i'j',k'}} = \delta_{kk'}\delta_{ij'}\delta_{j i'}$, we get
\begin{equation}
        \frac{\overline{\mathrm{Tr}\, H(\omega_0 \to 0)^2} }{\mathrm{Tr}\, \mathds{1}} = \frac{g^2}{\omega_0 N} \binom{N/2+1}{2} \frac{1}{M} \sum_{k} \frac{\mathrm{Tr}_B (a_k+a^\dag_k)^2}{\mathrm{Tr}_B \mathds{1}},
\end{equation}
where we separated the fermionic from the bosonic trace. While the fermionic trace has been directly computed leading to the same result as for the SYK$_2$ case, the bosonic trace $\mathrm{Tr}_B$ must be taken with care. In presence of a finite cutoff $N_b$ of the boson modes, we have
\begin{equation}
    \frac{1}{M} \sum_{k} \frac{\mathrm{Tr}_B (a_k+a^\dag_k)^2}{\mathrm{Tr}_B \mathds{1}} = \frac{(N_b+1)^M N_b}{(N_b+1)^M},
    \label{eq:boson_trace}
\end{equation}
which leads to
\begin{equation}
    \sigma(\omega_0 \to 0)^2=\frac{\overline{\mathrm{Tr}\, H(\omega_0 \to 0)^2} }{\mathrm{Tr}\, \mathds{1}} = \frac{g^2 N_b}{\omega_0 N} \binom{N/2+1}{2} \, . \label{eq:YSYK_norm_small_omega0}
\end{equation}

From Eqs.~\eqref{eq:SYK2_norm} and~\eqref{eq:YSYK_norm_small_omega0}, we obtain 
\begin{equation}
   \alpha_\text{SFF} = \frac{\sigma_{\rm SYK_2}}{\sigma(\omega_0 \to 0)} =  g \sqrt{\frac{N_b}{2 \omega_0} \frac{N+2}{N+1}} \approx g \sqrt{\frac{N_b}{2 \omega_0}} \, ,
\label{Eq: SYK2_time_rescaling}
\end{equation}
where we set $J=1$, which is equivalent to expressing the SYK dynamics in terms of the dimensionless time $Jt$. 

We now proceed with the calculation of the time rescaling of the OTOC. The operators used in the main text are of the form $A= 2 c^\dag_1 c_1 -1$ and $B= 2 c^\dag_2 c_2 -1$, which are unitary and Hermitian. For SYK$_2$, we have
\begin{align}
    \frac{\mathrm{Tr}\, \left( \big[[H_{\text{SYK}_2},A],B\big]^2 \right)}{\mathrm{Tr}\, \mathds{1}} = \frac{J^2}{N} W \, ,\qquad
    W = \sum_{i,j} \frac{\mathrm{Tr}\, \left( \big[[c_i^\dagger c_i,A],B\big] \big[[c_j^\dagger c_i,A],B\big] \right)}{\mathrm{Tr}\, \mathds{1}} \, .
\end{align}
An analogous calculation for YSYK leads to
\begin{equation}
    \frac{\mathrm{Tr}\, \left( \big[[H(\omega_0 \to 0),A],B\big]^2 \right)}{\mathrm{Tr}\, \mathds{1}} = \frac{g^2}{2 \omega_0 N} W N_b \, ,
\end{equation}
where we have again used Eq.~\eqref{eq:boson_trace}.
We can now combine these two results to compute $\alpha_\text{OTOC}$ as defined in Eq.~\eqref{eq:C_OTOC_def}, yielding
\begin{equation}
   \alpha_\text{OTOC} = g \sqrt{\frac{N_b}{2 \omega_0}} \, .
\label{Eq: SYK2_time_rescaling}
\end{equation}

\subsection{Large-$\omega_0$ regime}
\label{Appendix:large_omega_rescaling}

In the large-$\omega_0$ limit, the relevant benchmark model is complex SYK$_4$. Again, using the definition in Sec.~\ref{sec:CSYK}, we can determine the SFF time-rescaling factor $\alpha_\text{SFF}$ by first calculating the averaged Hamiltonian moments
\begin{equation}
   \frac{\overline{\mathrm{Tr}\, H_{\text{SYK}_4}^2} }{\mathrm{Tr}\, \mathds{1}} = \frac{12 J^2}{N^3} \binom{N/2+2}{4} \, , \qquad
    \frac{\overline{(\mathrm{Tr}\, H_{\text{SYK}_4})^2} }{(\mathrm{Tr}\, \mathds{1})^2} = \frac{2 J^2}{N^3} \, \, \binom{N}{2} \, \,\frac{\binom{N-2}{N/2}^2}{\binom{N}{N/2}^2} \, . 
    \label{eq:SYK4_trH2_SYK4_tr2H}
\end{equation}

In the YSYK model, as discussed in the main text, the large-$\omega_0$ dynamics are dominated by the lowest energy sector, which is captured by the effective Hamiltonian in Eq.~\eqref{eq:H_eff}, obtained via adiabatic elimination of the boson field. We begin by calculating its second moment, 
\begin{equation}
        \frac{\overline{\mathrm{Tr}\, H_{\text{eff.}}^2} }{\mathrm{Tr}\, \mathds{1}} = \frac{1}{4 \omega_0^4 M^2 N^2} \sum_{ii'jj'kk'll'} \sum_{mm'} \overline{g_{ij,m}g_{kl,m}g_{i'j',m'}g_{k'l',m'}} \, \frac{\mathrm{Tr}\,{c_i^\dagger c_j c_{k}^\dagger c_{l} c_{i'}^\dagger c_{j'} c_{k'}^\dagger c_{l'}}}{\mathrm{Tr}\, \mathds{1}} \, .
\end{equation}
Considering all possible Wick contractions, we find
\begin{equation}
        \overline{g_{ij,m}g_{kl,m}g_{i'j',m'}g_{k'l',m'}} = g^4\Big[ \delta_{mm'} \left(\delta_{i j'} \delta_{j i'} \delta_{k l'} \delta_{l k'} + \delta_{i l'}\delta_{j k'}\delta_{k j'}\delta_{l i'}\right)  + \delta_{il} \delta_{jk} \delta_{i'l'} \delta_{j'k'}\Big] \, .
\end{equation}
Notice that the last term above does not contain any delta function over the bosonic indices of $g$. Consequently, the summation over $m, m'$ yields a term that scales quadratically with $M$, unlike the previous terms, which are only linear. Since $M \propto N$, this contribution becomes leading and super-extensive, arising from the fact that, unlike in the standard SYK model, certain fermionic couplings have a nonzero mean. A careful evaluation of the trace gives
\begin{equation}
    \frac{\overline{\mathrm{Tr}\, H_{\text{eff.}}^2} }{\mathrm{Tr}\, \mathds{1}} = \frac{g^4}{4 \omega_0^4 M N^2}\left( 4 (M+2) \binom{N/2+1}{2}^2 - (N+1) \left( \frac{N}{2} \right)^2 \right) \, . 
\label{eq:SYK4_trHeff2}
\end{equation}
With a similar calculation, we can derive the averaged square of the first moment
\begin{equation}
\frac{\overline{(\mathrm{Tr}\, H_{\text{eff.}})^2} }{(\mathrm{Tr}\, \mathds{1})^2} = \frac{g^4}{4 \omega_0^4 M N^2}\left( 4 M \binom{N/2+1}{2}^2 + \frac{N^2 (N (5 N-8)+4)}{4 (N-1)^2}\right). \label{eq:SYK4_tr2Heff}
\end{equation}
Notice that the first term exactly cancels the super-extensive contribution of Eq.~\eqref{eq:SYK4_trHeff2} in the expression for $\sigma_{H_{\text{eff.}}}^2$, leading to a variance that scales extensively, as one would expect. 
We found that it is important to keep subleading contributions in the above expressions in order to obtain a quantitative agreement for the relatively small system sizes accessible in exact numerics. 

Combining Eqs.~\eqref{eq:SYK4_trH2_SYK4_tr2H},~\eqref{eq:SYK4_trHeff2}, and~\eqref{eq:SYK4_tr2Heff}, one can obtain an analytic expression for $\alpha_\text{SFF}$,
\begin{equation}
    \alpha_\text{SFF} =C_\text{SFF} \frac{g^2}{\omega_0^2} \, ,
\label{Eq: SYK4_time_rescaling_SFF}
\end{equation}
where $C_\text{SFF}$ contains all the $N$ and $M$ dependence. The following table presents the numerical values of $C_\text{SFF}$ for $N=8$ fermions used in Fig.~\ref{fig:syk4_limit}:
\begin{table}[t]
\centering
\small
\setlength{\tabcolsep}{4pt}
\begin{tabular}{lcccccccccc}
\hline\hline
$M$ & 1 & 2 & 3 & 4 & 5 & 6 & 7 & 8 & 9 & 10 \\
$C_{\mathrm{SFF}}$ & 2.537 & 1.749 & 1.465 & 1.269 & 1.134 & 1.036 & 0.959 & 0.897 & 0.846 & 0.802 \\
\hline\hline
\end{tabular}
\caption{Numerical values of $C_{\mathrm{SFF}}$ for $N=8$ fermions used in Fig.~\ref{fig:syk4_limit}.}
\label{Eq:SYK4_C_SFF}
\end{table}

We now move to the calculation for $\alpha_\text{OTOC}$. By inspecting the Hamiltonian, it becomes clear that we again have
\begin{equation}
    \alpha_\text{OTOC} =C_\text{OTOC} \frac{g^2}{\omega_0^2} \, ,
\label{Eq: SYK4_time_rescaling_OTOC}
\end{equation}
where $C_\text{OTOC}$ contains all the $N$ and $M$ dependence. 
In this case, however, rather than computing the commutator analytically, which would result in an overly cumbersome expression, we evaluate it numerically. To meaningfully compare the YSYK model with the complex SYK, we must isolate the physics described by the adiabatically eliminated Hamiltonian. This corresponds to taking large values of $\omega_0$ and restricting the Hilbert space to the lower-energy band, as shown in Fig.~\ref{Fig: OTOC_cSYK4_regime_combined}{\bf a}. 
By doing so, we neglect the fast oscillations that become increasingly suppressed as $\omega_0$ grows, but still influence the early-time dynamics. After projecting the operators $A= 2 c^\dag_1 c_1 -1$ and $B= 2 c^\dag_2 c_2 -1$ onto the lowest-energy band, we compute $C_\text{OTOC}$ numerically for fixed values of $N$ and $M$. For $N=8$ and $M=4$ we obtain, averaging over $1000$ samples,
\begin{equation}
    C_\text{OTOC} \approx 1.4118 \, .
\label{Eq: C_OTOC_factor}
\end{equation}
The above value, which has been computed for $\omega_0 = 10^4$, becomes $\omega_0$-independent at large $\omega_0$.

\section{Perturbative analysis}

In this Appendix, we elaborate on the technical details of the perturbative techniques used in Secs.~\ref{sec:strong} and~\ref{sec:weak}, demonstrating, on the one hand, the derivation of the timescale associated with the late-time scrambling process in the SYK$_2$-like regime, and on the other hand, the emergence of the SYK$_4$-like behavior.

\subsection{Small-$\omega_0$ regime: Magnus expansion}
\label{Appendix: Magnus_expansion_small_omega0}

We calculate the corrections to the leading part of the Hamiltonian in the small-$\omega_0$ regime, given in Eq.~\eqref{eq:H_omega_0=0}, in order to identify the emergence of new regimes at late times. To this end, we treat the rest of the YSYK Hamiltonian given in Eq.~\eqref{eq:hamiltonian_explicit} as a perturbation for small $\omega_0$. For simplicity, we restrict the calculation to hardcore bosons $N_b=1$. We start by rewriting the bosons in terms of spin-$1/2$ Pauli operators and perform a basis rotation on them, such that the Hamiltonian can be written as
\begin{equation}
    H = \frac{1}{\sqrt{2 \omega_0 M N}} \sum_{i,j,k} g_{ij,k} c^\dag_i c_j \sigma^z_k + \omega_0 \sum_k \frac{\sigma^x_k}{2},
\label{Eq:H_MagnusExpansion}
\end{equation}
where we ignored constant terms that do not impact the subsequent discussion.
The second term of Eq.~\eqref{Eq:H_MagnusExpansion}, 
\begin{equation}
    H_1 \equiv \frac{\omega_0}{2} \sum_k \sigma^x_k, 
\end{equation}
can be seen as a perturbation on top of the first one, 
\begin{equation}
    H_0 \equiv \sum_k \left( \sum_{i,j} \frac{g_{ij,k}}{\sqrt{2 \omega_0 M N}} c^\dag_i c_j \right) \sigma^z_k \equiv \sum_k \Omega_k \sigma^z_k.
\end{equation}
It is then useful to go into the interaction picture with respect to $H_0$, within which the perturbation acquires a time dependence, $H_{I,1}(t) = e^{i H_0 t} H_1 e^{-i H_0 t}$, with  
\begin{equation}
    H_{I,1}(t) = \frac{\omega_0}{2} \sum_k \left( e^{2 i \Omega_k t} \sigma^+_k + e^{-2 i \Omega_k t} \sigma^-_k \right),
\label{Eq:H_MagnusExpansion}
\end{equation}
where $\sigma^\pm = \frac{1}{2}(\sigma^x \pm i \sigma^y)$. 
Strictly speaking, the expression above should be treated with caution, as $\Omega_k$ is an operator. Nevertheless, since it is in general full rank, its inverse and all its functions are well defined, allowing us to easily track the scaling with $\omega_0$.  

In the interaction picture, the time evolution operator is split into a part given by $U_0(t)=$ $e^{-i H_0 t}$ and one generated by the perturbation, $U_I(t) = \mathcal{T} e^{-i \int_0^t H_{I,1}(t') dt'}$. Using Eq.~\eqref{Eq:H_MagnusExpansion}, we compute the first terms of $\log U_I$ using a Magnus expansion. The first term gives 
\begin{align}
    \int_0^t \! H_{I,1}(t_1) \dd{t}_1 \! &= \! - \sum_k \frac{i \omega_0}{4 \Omega_k} \left[ ( e^{2 i \Omega_k t}-1) \sigma^+_k + (1 - e^{-2 i \Omega_k t}) \sigma^-_k \right] \nonumber \\
    &= \frac{\omega_0}{4} \! \sum_k \! \frac{1}{\Omega_k} \left[\sin(2 \Omega_k t)\sigma^x_k \! + \! \cos(2 \Omega_k t) \sigma^y_k \! - \! \sigma^y_k \right]. \label{eq:first_term_magnus}
\end{align}
For short times, an expansion of Eq.~\eqref{eq:first_term_magnus} for $\Omega_k t \ll 1$, or equivalently, $t \ll \sqrt{\omega_0}$, suggests that the first-order correction goes as $\sim \omega_0$, which is subleading with respect to the time evolution generated by $H_0 \sim \omega_0^{-1/2}$. 
Similarly, for larger times $\Omega_k t \gtrsim 1$,
Eq.~\eqref{eq:first_term_magnus} quickly oscillates in time with a bounded amplitude that is suppressed as $\omega_0^{3/2}$.

More relevant contributions come from higher-order terms. The second term in the Magnus expansion is given by
\begin{equation}
    \frac{1}{2!} \!\int_0^t \!\dd{t}_1 \! \!\int_0^{t_1} \!\dd{t}_2\! \comm{H_{I,1}(t_1)}{H_{I,1}(t_2)} \!=\! \frac{i \omega_0^2}{8}\! \sum_k \!\frac{\sigma^z_k}{\Omega_k} \!\left(t \!- \!\frac{\sin(2 \Omega_k t)}{2 \Omega_k} \right).
    \label{eq:second_term_magnus}
\end{equation}
The oscillating second term on the right-hand side can again be neglected, since it is of bounded amplitude and suppressed at least as $\omega_0^{5/2}$ for any $t$. The situation is different for the first term, which becomes of order one for $t \sim \omega_0^{-5/2}$. For such late times, the time evolution generated by $H_0$ saturates to a constant value, as can be seen from the SYK$_2$ plateau in Fig.~\ref{Fig: OTOC_cSYK2_regime_latetimes}. For $t \sim \omega_0^{-5/2}$, the first term in Eq.~\eqref{eq:second_term_magnus} induces new dynamics, generating a new relevant time scale as can be seen from the collapse in Fig.~\ref{Fig: OTOC_cSYK2_regime_latetimes}.

To check that this is indeed the dominating behavior, we expand $\log U_I$ up to the third term: 
\begin{align}
    &\frac{1}{3!} \int_0^t \dd{t}_1 \int_0^{t_1} \dd{t}_2 \int_0^{t_3} \dd{t}_3 \bigg\{ \big[ [H_{I,1}(t_1), H_{I,1}(t_2)], H_{I,1}(t_3) \big] + \big[ [H_{I,1}(t_1), H_{I,1}(t_2)], H_{I,1}(t_3) \big] \bigg\}  \nonumber \\
    &= -\sum_k \frac{\omega_0^3}{192 \Omega_k^3} \left( e^{i \Omega_k t} \sigma^+_k  + e^{-i \Omega_k t} \sigma^-_k \right)  \bigg( 12 \Omega_k t \cos(\Omega_k t) - 9 \sin(\Omega_k t) - \sin(3 \Omega_k t) \bigg) \, .
\end{align}
Once again, the oscillating terms are bounded and have a strongly suppressed amplitude. Moreover, the remaining correction scales at most linearly in time, while the coefficient is suppressed in $\omega_0$ compared to the one calculated in Eq.~\eqref{eq:second_term_magnus}. This confirms that the leading-order correction scales as $\sim t \omega_0^{5/2}$, explaining the collapse in the inset of Fig.~\ref{Fig: OTOC_cSYK2_regime_latetimes}.

\subsection{Large-$\omega_0$ regime: Schrieffer--Wolff transformation}
\label{Appendix: SWT}

Our numerical results show that SYK$_4$-like chaos emerges in the weak-coupling limit, $\omega_0/g^{2/3} $ $\gg 1$. This behavior can be explained through a perturbative calculation using a Schrieffer--Wolff (SW) transformation. To this end, we decompose the Hamiltonian in Eq.~\eqref{eq:hamiltonian_explicit} into an unperturbed part $H_0$ and a small interaction $V$ (where now the roles are inverted as compared to the preceding section): 
\begin{align}
    H_0 &= \sum_{k=1}^M \omega_0 \left( a_k^\dagger a_k + \tfrac{1}{2} \right), \\
    V &= \frac{1}{\sqrt{2\omega_0 MN}} \sum_{i,j,k} g_{ij,k}\, c_i^\dagger c_j\, \left( a_k + a_k^\dagger \right).
\end{align}
The interaction $V$ couples states with different boson numbers. The SW transformation is a unitary rotation  $U=e^S$ of the Hamiltonian designed to eliminate this off-diagonal coupling to first order. This is implemented by imposing that the generator $S$ satisfies the condition $[S, H_0] = -V$, which yields
\begin{equation}
    S = \frac{1}{\omega_0 \sqrt{2\omega_0 MN}} \sum_{i,j,k} g_{ij,k}\, (c_i^\dagger c_j) (a_k^\dagger - a_k).
\end{equation}
The transformed Hamiltonian, expanded to second order in $\omega_0$, is $\tilde{H} = H_0 + \frac{1}{2}[S,V]$. The effective dynamics for the low-energy states, which live in the boson vacuum, are found by projecting $\tilde{H}$ onto this subspace with the projector $P_0 = \ket{\{0_k\}}\bra{\{0_k\}}$. 

The term $P_0[S,V]P_0$ generates an effective fermion--fermion interaction. A direct calculation yields the four-fermion interaction
\begin{equation}
    H_{\mathrm{eff}} = -\frac{1}{2\omega_0^2 MN} \sum_k \sum_{i,j,i',j'} g_{ij,k} g_{i'j',k} c_i^\dagger c_j c_{i'}^\dagger c_{j'}.
    \label{eq:H_eff}
\end{equation}
We can absorb the sum over bosonic modes into an effective coupling,
\begin{equation}
\label{eq:J_from_g}
    J_{ij,i'j'} \;=\; \frac{1}{M}\sum_{k=1}^M g_{ij,k}\,g_{i'j',k},
\end{equation}
so that the low-energy Hamiltonian  \eqref{eq:H_eff} becomes formally similar to the one of SYK$_4$, 
\begin{equation}
    H_{\mathrm{eff}}
    = -\frac{1}{2\,\omega_0^2\,N}
      \sum_{i,j,i',j'=1}^N
      J_{ij,i'j'}\;
      c_i^\dagger c_j\,c_{i'}^\dagger c_{j'}.
    \label{eq:H_eff_compact}
\end{equation}
As our numerics shows, these coupling indeed give rise to an emergent SYK$_4$ regime in the weak-coupling limit.

\section{Finite-size effects}

In this Appendix, we examine the impact of finite-size effects in our numerical results, focusing on both spectral and dynamical diagnostics.
We investigate how changes in the system size and the fermion–boson ratio $M/N$ affect the mean gap ratio and the out-of-time-ordered correlators (OTOCs), confirming the robustness of the features discussed in the main text. Moreover, we introduce a low-rank SYK model illustrating that the effect of normal ordering in a four-fermion Hamiltonian---specifically, the addition of a quadratic term summed over the bosonic degrees of freedom---becomes negligible for large enough boson modes when considering the spectral form factor (SFF).

\subsection{Sensitivity of average mean gap ratio to $M/N$} 
\label{Appendix:gap_ratio_finite_size}
\begin{figure*}[ht!]
    \centering
    \includegraphics[width=1.0\linewidth]{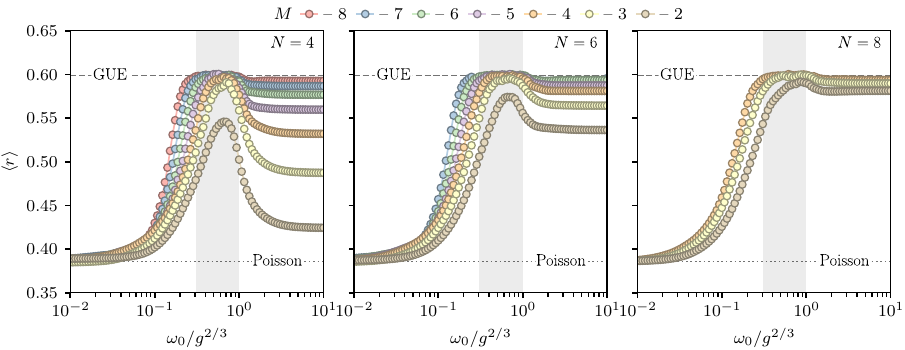}
    \caption{
\textbf{Disorder-averaged mean gap ratio $\langle r\rangle$ versus $\omega_0/g^{2/3}$.}
Data for $N_f\!=\!4,6,8$ (panels) and $M\!=\!2\!-\!8$ (colors). 
As $\omega_0/g^{2/3}$ increases, $\langle r\rangle$ crosses from Poisson statistics (dashed) to a GUE-like plateau (dotted); the gray band indicates the chaotic window as in Fig.~\ref{fig:avg_gap_ratio}. 
The slight decline at large $\omega_0/g^{2/3}$ stems from spectral clustering (separated boson-occupation bands), which introduces many small $r_n$ at band boundaries, while intra-band statistics remain GUE-like.
}
    \label{fig:mean_gap_ratio_panels}
\end{figure*}

Figure~\ref{fig:mean_gap_ratio_panels} shows the evolution of the mean gap ratio $\langle r\rangle$ with $\omega_0/g^{2/3}$ by varying $N$ (panels) and $M$ (colors).
For all $M>2$, we observe a crossover from Poisson statistics at small $\omega_0/g^{2/3}$ to a broad GUE-like plateau around $\langle r\rangle\simeq\langle r\rangle_{GUE}$, which is highlighted by a gray band.
At larger $\omega_0/g^{2/3}$, the curves display a weak non-monotonic decline. This behavior does not signal a loss of level repulsion; rather the clustered spectra characterized by well-separated boson-occupation bands generate many instances with small $r_n$ across band boundaries (due to a large denominator), while the statistics within a band remain GUE-like.

A finite-size scaling analysis, which would require $N\to \infty$ while keeping $M/N$ fixed is not feasible because the Hilbert space grows exponentially in both $N$ and $M$, quickly reaching the limits of exact diagonalization.
For this reason, in the main text we compare different ratios $M/N$ while keeping the total Hilbert-space dimension $D$ approximately fixed. For these data, we find similar crossovers, indicating that the chaotic window is insensitive to the fermion–boson balance. Moreover, at fixed $M/N$, increasing $D$ leaves the position and width of the plateau stable, with only a modest sharpening of the crossover.
These results indicate the robustness of the chaotic window against the ratio $M/N$ and Hilbert-space size.

\begin{figure}[h]
  \centering
  \includegraphics[width=0.5\linewidth]{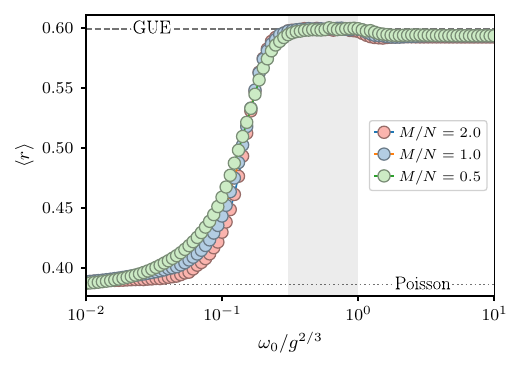} \caption{Averaged gap ratio $\langle r \rangle$ versus the interaction variable $\omega_0/g^{2/3}$ for boson–fermion mode ratios $M/N\in\{2.0,1.0,0.5\}$ (obtained for $(M,N)=(8,4),(6,6),(4,8)$, with bosonic cutoff $N_b=1$). Data points show averages over 500 disorder realizations for each $(M,N)$. 
  At small $\omega_0/g^{2/3}$, $\langle r \rangle$ is Poissonian (dotted line). It reaches the GUE value (dashed) in a broad crossover region $\omega_0/g^{2/3}\sim 0.3-1$ (grey band). 
  At larger $\omega_0/g^{2/3}$, it drops slightly due to DOS clustering. 
  } \label{fig:avg_gap_ratio}
\end{figure}

To complement the broader scan of Fig.~\ref{fig:mean_gap_ratio_panels}, it is useful to consider a reduced comparison closer to the one underlying the discussion in Sec.~\ref{sec:phase_diagram}. In Fig.~\ref{fig:avg_gap_ratio}, we therefore show $\langle r\rangle$ for three representative values of $M/N$, chosen so that the total Hilbert-space dimension remains approximately fixed. This compact view makes clear that the qualitative picture extracted from Fig.~\ref{fig:master_diagram_M4_N8} is stable against changing the balance between fermionic and bosonic degrees of freedom: $\langle r\rangle$ rises from the Poisson value at small $\omega_0/g^{2/3}$ to a broad plateau near the GUE value, and then decreases mildly once the DOS becomes strongly clustered.

\subsection{OTOCs at different system sizes}
\label{Appendix:finite_size_effects_OTOCs}

\begin{figure*}[ht!]
    \centering
    \includegraphics[width=1.0\linewidth]{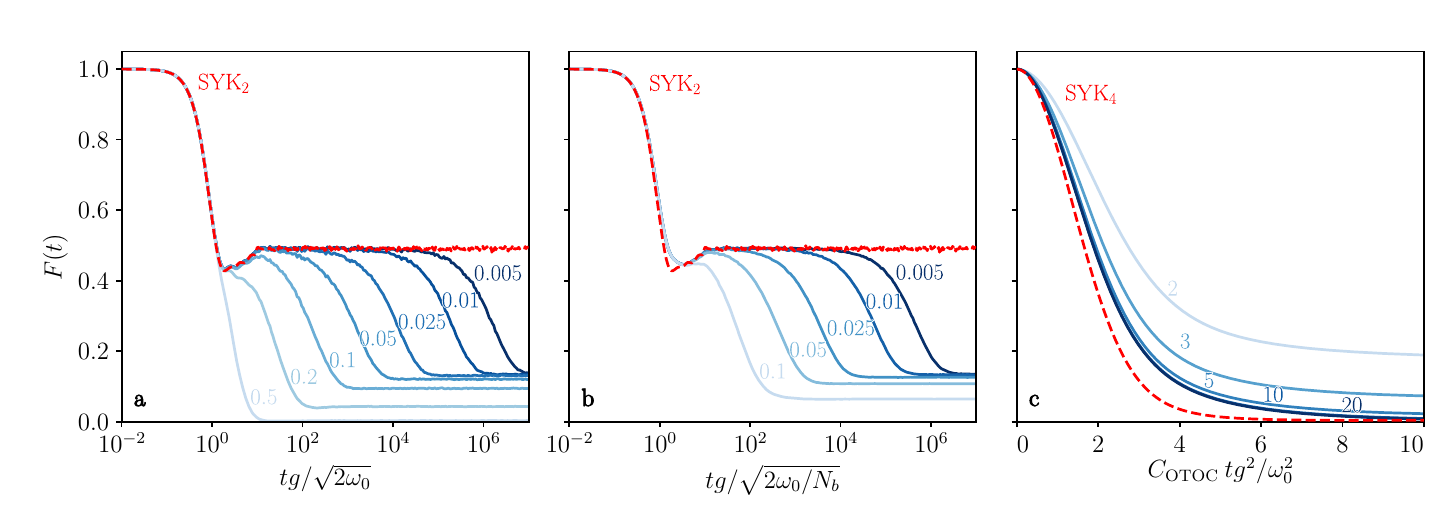}
    \caption{{\bf OTOCs for small-$\omega_0$ and large-$\omega_0$ regime.} 
({\bf a}) YSYK OTOCs for $N=6$, $M=3$, $N_b=1$, and several small values of $\omega_{0}$, averaged over 2000 samples. As in the main text results in Fig.~\ref{Fig: OTOC_cSYK2_regime_latetimes} ($N=8$, $M=4$, $N_b=1$), the YSYK OTOCs closely follow that of SYK$_2$ (here averaged over 10\,000 samples), up to an $\omega_{0}$-dependent time scale at which they undergo a second scrambling process. 
({\bf b}) Same as panel ({\bf a}) but for $N_b=3$ (YSYK OTOCs averaged over 1000 samples, SYK$_2$ over 10\,000). The qualitative behavior remains unaltered under increased boson-occupation cutoff. 
({\bf c}) YSYK OTOCs for $N=8$, $M=3$, $N_b=1$, and several large values of $\omega_{0}$ averaged over 1000 samples. As in main text Fig.~\ref{Fig: OTOC_cSYK4_regime_combined}{\bf c}, the results for increasing $\omega_0$ move toward the SYK$_4$ result. However, the smaller number of boson modes $M$ makes the agreement worse, indicating that increasing $M$ improves the behavior. 
To obtain good quantitative agreement, results in panels ({\bf a}) and ({\bf b}) are plotted versus $tg/\sqrt{2\omega_{0}/N_b}$, and those in panel ({\bf c}) versus $C_{\rm OTOC}\, t g^{2}/\omega_{0}^{2}$.}

    \label{fig:OTOC_finite_size_panel}
\end{figure*}

Starting with the small-$\omega_0$ regime, Fig.~\ref{fig:OTOC_finite_size_panel}a shows the OTOCs for different values of $\omega_0$ and a smaller system size of $N=6, M=3, N_b=1$ compared to the results in Fig.~\ref{Fig: OTOC_cSYK2_regime_latetimes} of Sec.~\ref{sec:strong} of the main text. Notably, the two-stage scrambling behavior of $F(t)$, with the intermediate plateau and eventual $\omega_0$-dependent deviation from it, is still present. However, the intermediate plateau value for $N=6$ is smaller than the one for $N=8$. This is most likely due to finite-size effects, as for complex SYK$_2$ we expect the OTOC to eventually (though slower than for SYK$_4$~\cite{Finite_size_OTOC}) saturate to zero in the limit of $N\to\infty$~\footnote{A similar behavior appears in the thermal $2$-point correlation functions of the complex SYK$_2$ model. There, in comparison with the saddle-point solutions, the finite-$N$ results initially deviate further from the large-$N$ solution for increasing, but small $N$. However, once $N$ becomes sufficiently large, the expected convergence toward the saddle-point solution is recovered. One may speculate that a similar effect might be at work at the plateau value in the OTOC due to the small number of fermion modes $N$.}. Moreover, comparing the curves with same boson mass $\omega_0$ for $N=6$ and $N=8$, one can notice that the saturation value for $N=8$ tends to be smaller. This result indicates that the observed non-zero saturation value may also be a finite-size effect and eventually the OTOC becomes fully scrambling at very late times.

Figure~\ref{fig:OTOC_finite_size_panel}b plots the OTOCs for the same system size as panel (a), but with an increased cut-off on the boson-occupation of $N_b=3$. Note that in order to match the results with the OTOC of complex SYK$_2$, we have to take into account the cut-off by rescaling the time according to Eq.~\eqref{Eq: SYK2_time_rescaling}. The qualitative behavior of the OTOCs remains unchanged. Together with the finite-size effects discussed above, this suggests that the main findings regarding the OTOCs discussed in Sec.~\ref{sec:strong} are robust and do not depend sensitively on the specific system sizes considered.

Moving to the large-$\omega_0$ regime, Fig.~\ref{fig:OTOC_finite_size_panel}{\bf c} plots the disorder-averaged YSYK OTOCs for $N=8, M=3$, and $N_b=1$ for the lowest energy peaks in the respective DOS. In this case, to match the OTOCs with those of the complex SYK$_4$, time is rescaled by the factor $C_{\rm OTOC} = 1.6277\ldots$, which we computed numerically for $\omega_0 = 10^4$ following the prescription in Appendix \ref{Appendix:large_omega_rescaling}. For large values of $\omega_0$, the OTOCs appear to converge; however, when compared with Fig.~\ref{Fig: OTOC_cSYK4_regime_combined}{\bf c} the large-$\omega_0$ YSYK OTOCs exhibit a poorer agreement with the OTOC of complex SYK$_4$. This comparison indicates that increasing the number of boson modes $M$ improves the agreement with respect to the complex SYK$_4$ target model. In contrast, since we restrict our analysis in the large-$\omega_0$ limit to the zero-boson occupancy sector, in this regime we do not expect the boson cut-off $N_b$ to have any significant effect on the OTOCs.

\subsection{Normal-ordered effective Hamiltonian}
\label{Appendix: toy_model}

\begin{figure}[h]
    \centering
    \includegraphics[width=0.6\linewidth]{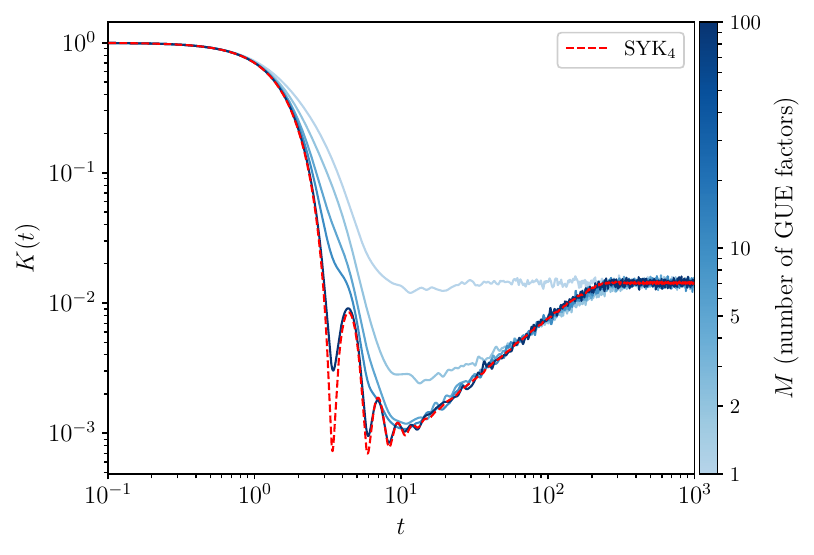}
    \caption{Spectral form factor $K(t)$ for the normal-ordered effective model. Solid curves from lighter to darker shades are for increasing $M$, i.e., increasing rank of the antisymmetric couplings $J_{ij;kl}$ defined in Eq.~\eqref{eq:J_NO_eff_model}. As $M$ increases, the curves closely approach the SYK$_4$ benchmark (dashed). Data for $N=8$ at half filling, averaged over $500$ realizations, and normalized by $D^2$ with $D=\binom{N}{N/2}$.}
    \label{fig:sff-gueprod-Mscan}
\end{figure}

In the weak fermion-boson coupling regime, the effective Hamiltonian in Eq.~\eqref{eq:Heff_SYK4_like} formally differs from the SYK$_4$ Hamiltonian in Eq.~\eqref{eq:SYKq_H}: the fermions are not normal ordered and the coupling $J$ is defined in terms of the YSYK coupling $g$ according to Eq.~\eqref{eq:J_from_g}. 
As discussed in Sec.~\ref{sec:strong}, normal ordering generates a quadratic term that is subleading in the large-$N$ limit. In this section, we therefore focus on examining how the form of the effective coupling influences the system's physics, illustrated by analyzing the SFF. 

The normal ordered effective Hamiltonian is defined as
\begin{equation}
H \;=\; \sum_{i<j \atop k<l} J_{ij;kl}\; c_i^\dagger c_j^\dagger c_k c_l\, ,
\end{equation}
with
\begin{equation}
\label{eq:J_NO_eff_model}
J_{ij;kl}
\;=\;
\frac{1}{M}\sum_{s=1}^{M}\,\frac{1}{2}\Big(g_{ik,s}\,g_{jl,s}-g_{il,s}\,g_{jk,s}\Big)\, ,
\quad (i<j,\;k<l)\, ,
\end{equation}
where the symmetries of the SYK$_4$ coupling $J_{ij;kl}$ are explicitly satisfied. Notice that this model differs from the SUSY SYK model with $\mathcal{N}=2$ supercharges by the definition of the coupling $J_{ij;kl}$~\cite{Fu_2017}.
Each $g_{ij,k}$ is an independent random variable, with zero mean and a variance that is set to $\overline{g_{ij,k}^2} = 2\sqrt{M/N^3}$ such that the variance of $J_{ij;kl}$ matches Eq.~\eqref{eq:SYK_q_variance}.

As illustrated in Fig.~\ref{fig:sff-gueprod-Mscan}, as $M$ increases, the spectral form factor of the above effective model converges to that of the SYK$_4$ model. This is analogous to what has been observed in the Majorana case \cite{Bi:2017yvx,Kim_2020}. For small $M$, the SFF ramp is suppressed, whereas for larger $M$, a clear dip–ramp–plateau structure emerges that aligns with the SYK$_4$ benchmark. These results suggest that the inexact correspondence in the SFF between the YSYK and the SYK$_4$ models shown in Fig.~\ref{fig:syk4_limit} is due to small system sizes and that it can be resolved by taking $M$ sufficiently large, as expected for a low-rank SYK model~\cite{Bi:2017yvx, Kim_2020}.

\section{Details on the experimental proposal}

In this Appendix, we provide a detailed derivation of the effective light–matter Hamiltonian underlying the proposed experimental realization of the YSYK model.
We show how adiabatic elimination of the excited state and cavity modes yields an effective fermion–boson Yukawa interaction, and how a two-tone drive can be used to cancel residual disordered Stark shifts.
We also discuss the decay channels and the scaling of the effective interaction strength.

\subsection{Derivation of the effective light-matter Hamiltonian}
\label{app:light_matter_Hamiltonian}

Expanding the squared amplitude in Eq.~\eqref{Eq:H_mb_AE_fermion}, 
we obtain
\begin{align}
\label{Eq:H_AE_fermion_collected}
H_{\mathrm{mb}} &= H_{\mathrm{kt}} +\sum_m\Delta_m\,a_m^\dagger a_m +\int d^2r \frac{|\Omega_{\mathrm{d}} g_{\mathrm{d}}(\mathbf{r})|^2}{\Delta_{\mathrm{da}}(\mathbf{r})} \psi_g^\dagger \psi_g \nonumber  \\
&+\frac{1}{2}\sum_m\int d^2r\frac{\Omega_{\mathrm{d}}^*\Omega_m\,g_{\mathrm{d}}^*(\mathbf{r}) g_m(\mathbf{r})}{\Delta_{\mathrm{da}}(\mathbf{r})} a_m^\dagger \, \psi_g^\dagger\psi_g + \mathrm{h.c.} \nonumber  \\
&+ \frac{1}{4}\sum_{m,m'}\int d^2r\frac{\Omega_m^*\Omega_{m'} g_m^*(\mathbf{r}) g_{m'}(\mathbf{r})}{\Delta_{\mathrm{da}}(\mathbf{r})}a_m^\dagger a_{m'} \psi_g^\dagger \psi_g \, . 
\end{align}
These three new terms correspond, respectively, to (i) a spatially dependent AC–Stark shift of the ground‐state atoms, (ii) a linear coupling between the cavity photons and the atomic density, and (iii) a hopping between photon modes that depends on the atomic ground-state density. 

Expanding the ground-state field operator in terms of eigenmodes of the trap $\psi_g(\mathbf{r})=$ $\sum_{\ell}\phi_{\ell}(\mathbf{r})\,c_{\ell},$, the many-body light-matter Hamiltonian in Eq.~\eqref{Eq:H_AE_fermion_collected} reads
\begin{align}
\label{eq:Hmicroscopic}
H_{\mathrm{mb}}&\simeq{}\sum_{\ell}\epsilon_{\ell}\,c_{\ell}^\dagger c_{\ell}
+\sum_m\Delta_m\,a_m^\dagger a_m
+\sum_{\ell,\ell'}\mathcal{M}^{(0)}_{\ell\ell'}\,c_{\ell}^\dagger c_{\ell'}\nonumber \\
&\;+\frac12\sum_{m,\ell,\ell'}(\mathcal{M}^{(1)}_{m,\ell\ell'}\,a_m^\dagger+h.c)\,c_{\ell}^\dagger c_{\ell'}\nonumber \\
&\;+\frac14\sum_{m,m',\ell,\ell'}\mathcal{M}^{(2)}_{mm',\ell\ell'}\,a_m^\dagger a_{m'}\,c_{\ell}^\dagger c_{\ell'} \, ,
\end{align}
where we have defined the overlap integrals
\begin{align}
\mathcal{M}^{(0)}_{\ell\ell'} &= \int d^2r\,\frac{|\Omega_{\mathrm{d}}\,g_{\mathrm{d}}(\mathbf{r})|^2}{\Delta_{\mathrm{da}}(\mathbf{r})}\,
\phi_{\ell}^*(\mathbf{r})\,\phi_{\ell'}(\mathbf{r})\, ,\\
\mathcal{M}^{(1)}_{m,\ell\ell'} &= \int d^2r\,\frac{\Omega_{\mathrm{d}}^*\,\Omega_m\,g_{\mathrm{d}}^*(\mathbf{r})\,g_m(\mathbf{r})}{\Delta_{\mathrm{da}}(\mathbf{r})}\,
\phi_{\ell}^*(\mathbf{r})\,\phi_{\ell'}(\mathbf{r})\, ,\\
\mathcal{M}^{(2)}_{mm',\ell\ell'} &= \int d^2r\,\frac{\Omega_m^*\,\Omega_{m'}\,g_m^*(\mathbf{r})\,g_{m'}(\mathbf{r})}{\Delta_{\mathrm{da}}(\mathbf{r})}\,
\phi_{\ell}^*(\mathbf{r})\,\phi_{\ell'}(\mathbf{r})\, .
\end{align}

One can remove the residual one-body ``dipole'' term proportional to $M^{(0)}_{\ell\ell'}$ by introducing a second auxiliary drive of equal Rabi strength but opposite detuning. In the dressed-state (Autler--Townes) picture this produces two AC-Stark shifts of equal magnitude and opposite sign, which exactly cancel the disordered potential~\cite{cavity_proposal}. The details of this cancellation are given in Appendix \ref{sec:autler-townes}.
By tuning $\Omega_{\mathrm{d}} \gg \Omega_m$, one can also suppress the last term $\sim\mathcal{M}^{(2)}_{mm',\ell\ell'}$, so that the interaction $\sim\mathcal{M}^{(1)}_{m,\ell\ell'}$ dominates.
This term sets the desired linear Yukawa coupling between photons and fermions. 

This effective Hamiltonian maps to the YSYK model of Eq.~\eqref{eq:hamiltonian_explicit} by restricting to the $N$ lowest trap levels, 
choosing all detunings $\Delta_m = \omega_0$ (and including the zero-point energy), and relabeling $l\rightarrow i$ and $m \rightarrow k$. 
We then identify the Yukawa coupling of our model as
\begin{equation}
\label{Eq:three-leg2}
    \frac{g_{ij,k}}{\sqrt{2\omega_0 MN }} = \frac{\mathcal{M}^{(1)}_{k,ij}+\left( \mathcal{M}^{(1)}_{k,ji} \right)^*}2.
\end{equation}
Recalling from the main text that the trap involved is orders of magnitude below the other energy scales participating in the Hamiltonian, we neglect the single-particle energies $\epsilon_l$. With these definitions and the above simplifications, the effective Hamiltonian in Eq.~\eqref{eq:Hmicroscopic} reproduces the YSYK model of Eq.~\eqref{eq:hamiltonian_explicit}.

\subsection{Two-tones Cancellation of the Disordered Dipole Term}
\label{sec:autler-townes}

The adiabatic elimination of the excited state, which leads to Eq.~\eqref{Eq:H_mb_AE_fermion}, or equivalently Eq.~\eqref{Eq:H_AE_fermion_collected}, 
produces a disordered dipole potential,
\begin{equation}
\label{eq:Hdip}
H_{\mathrm{dip}}=\int d^{d}\mathbf{r}\;\frac{\lvert \Omega_{\mathrm{d}}\,g_{\mathrm{d}}(\mathbf{r})\rvert^{2}}{\Delta_{\mathrm{da}}(\mathbf{r})}\;
\psi^{\dagger}(\mathbf{r})\psi(\mathbf{r}) .
\end{equation}
This potential, if not compensated, may lead to undesired localization of the ground-state atoms.  

Fortunately, it can be compensated straightforwardly, exploiting the way the disordered detuning $\Delta_{\mathrm{da}}(\mathbf{r})$ is generated~\cite{cavity_proposal}. 
Denote by $\omega_\mathrm{a}^{(0)}$ the energy of the atomic excited state $\ket{\mathrm{e}}$ (relative to the ground state) in absence of the speckle pattern. 
The speckle beam couples $\ket{\mathrm{e}}$ to an auxiliary state $\ket{\mathrm{aux}}$, shifting the excited-state energy due to the AC-Stark effect to $\omega_\mathrm{a}(\mathbf{r})=\omega_\mathrm{a}^{(0)}+\delta_\mathrm{a}(\mathbf{r})$. 
By measuring frequencies relative to the classical drive, this amounts to the disordered detuning $\Delta_{\mathrm{da}}(\mathbf{r})=\omega_\mathrm{a}^{(0)}-\omega_\mathrm{d}+\delta_\mathrm{a}(\mathbf{r})$, which introduces the desired randomness into the model as discussed above. 
However, in the same way, the coupling due to the speckle beam induces an AC-Stark shift on the energy of $\ket{\mathrm{aux}}$ that is perfectly anticorrelated with the one of $\ket{\mathrm{e}}$, $\omega_\mathrm{aux}^{(0)}\rightarrow \omega_\mathrm{aux}^{(0)}-\delta_\mathrm{a}(\mathbf{r})$. 

One can thus introduce a second, off-resonant drive with Rabi amplitude $\Omega_{\mathrm{d}'}$, profile $g_{\mathrm{d}'}(\mathbf{r})$, and 
at frequency $\omega_{\mathrm{d}'}$ 
detuned by $\Delta^{\mathrm{d}'}_{\mathrm{aux}}(\mathbf{r})$ from the $g\!\leftrightarrow\!\mathrm{aux}$ transition. Eliminating $\psi_{\mathrm{aux}}$ yields an additional AC-Stark contribution,
\begin{equation}
\label{eq:shifttwodrive}
H'_{\mathrm{dip}}=\int d^{d}\mathbf{r}\;\frac{\lvert \Omega_{\mathrm{d}'}\,g_{\mathrm{d}'}(\mathbf{r})\rvert^{2}}{\Delta^{\mathrm{d}'}_{\mathrm{aux}}(\mathbf{r})}\;
\psi^{\dagger}(\mathbf{r})\psi(\mathbf{r}) .
\end{equation}
Thanks to the anti-correlation of the energy shift induced by the speckle pattern, by choosing $\omega_{d'}-\omega_\mathrm{aux}^{(0)}=\omega_\mathrm{a}^{(0)}-\omega_\mathrm{d}$ the detunings satisfy pointwise
\begin{equation}
\Delta_{\mathrm{da}}(\mathbf{r})=-\Delta^{\mathrm{d}'}_{\mathrm{aux}}(\mathbf{r}) . 
\end{equation}
By further matching the spatial profile and the intensity,
\begin{equation}
g_{\mathrm{d}'}(\mathbf{r})=g_{\mathrm{d}}(\mathbf{r}),\qquad \lvert \Omega_{\mathrm{d}'}\rvert=\lvert \Omega_{\mathrm{d}}\rvert ,
\end{equation}
the disordered potential on the ground-state atoms in Eq.~\eqref{eq:shifttwodrive} exactly cancels the one in Eq.~\eqref{eq:Hdip} deriving from $\ket{\mathrm{e}}$, i.e., $H_{\mathrm{dip}}+H'_{\mathrm{dip}}=0$. 

\subsection{Dressed decay channels $\tilde\Gamma$ and $\tilde\kappa$}\label{App:dressed-channels}

In this section, we account for the main dissipation channels in the experimental proposal of Sec.~\ref{sec:experiment}. The discussion closely follows the one in Ref.~\cite{cavity_proposal}. The time evolution of an operator $O$ is described in the adjoint–Lindblad picture~\cite{OpenSystems_Petruccione_book} as
\begin{equation}
\dot O \;=\; i[H_{\mathrm{mb}},O]\;+\;\sum_\ell L_\ell^\dagger O L_\ell - \tfrac12\{L^\dagger_\ell L_\ell,\,O\}
,
\label{eq:adjoint-lindblad}
\end{equation}
where the jump operators are $L(\mathbf r) = \sqrt{\Gamma} \psi_g^\dagger(\mathbf r)  \psi_e(\mathbf r) $ for spontaneous emission and $L_m = \kappa_m a_m$ for photon loss. In the following, we will denote $n_{g,e}(\mathbf r)=\psi_{g,e}^\dagger(\mathbf r)\psi_{g,e}(\mathbf r)$ the fermion state densities and $\sigma_{ge}(\mathbf r)=\psi_g^\dagger(\mathbf r)\psi_e(\mathbf r)$ the optical coherence.

Using Eq.~\eqref{eq:adjoint-lindblad}, the equation of motion for $\sigma_{ge}$ reads
\begin{equation}
    \dot{\sigma}_{ge}(\mathbf r)=
-\Big(\tfrac{\Gamma}{2}-i\Delta_{\mathrm{da}}(\mathbf r)\Big)\sigma_{ge}(\mathbf r) -\,i\Big[\Omega_{\mathrm{d}} g_{\mathrm{d}}(\mathbf r)+\tfrac12\sum_m \Omega_m g_m(\mathbf r)a_m\Big]n_g(\mathbf r),
\label{eq:psi-e-eom}
\end{equation}
where we took $n_e(\mathbf r) \sim 0$ in the dispersive regime $|\Delta_{\mathrm{da}}|\gg|\Omega_{\mathrm{d}}|,|\Omega_m|$. The same regime allows for an adiabatic elimination of $\sigma_{ge}(\mathbf r)$
, which yields
\begin{equation}
\sigma_{ge}(\mathbf r)=
\frac{\Omega_{\mathrm{d}}\,g_{\mathrm{d}}(\mathbf r)+\tfrac12\sum_m \Omega_m\,g_m(\mathbf r)\,a_m}{\Delta_{\mathrm{da}}(\mathbf r)+i\,\Gamma/2}\;n_g(\mathbf r).
\label{eq:slaved-coherence}
\end{equation}
In the regime $|\Omega_m/\Omega_{\mathrm{d}}|\ll1$, we can retain the drive–induced component of Eq.~\eqref{eq:slaved-coherence} only, and one finds the effective jump operator
\begin{equation}
L_\Gamma^{\mathrm{eff}} (\mathbf r)
=
\sqrt{\Gamma}\;\frac{\Omega_{\mathrm{d}}\,g_{\mathrm{d}}(\mathbf r)}{\Delta_{\mathrm{da}}(\mathbf r)+i\,\Gamma/2}\;n_g(\mathbf r).
\label{eq:L-gamma-eff}
\end{equation}
The effective decay rate magnitude $\tilde\Gamma$ for a single atom localized at position $\mathbf r$ is given by the absolute value squared of the prefactor in the expression above:
\begin{equation}
\tilde\Gamma(\mathbf r)
=\Gamma\;\frac{|\Omega_{\mathrm{d}}\,g_{\mathrm{d}}(\mathbf r)|^2}
{\Delta_{\mathrm{da}}(\mathbf r)^2+(\Gamma/2)^2}.
\label{eq:Gamma-tilde-local}
\end{equation}

In the regime given by Eq.~\eqref{eq:SYK4_scales}, relevant for approaching SYK$_4$, we can also eliminate the cavity mode. Using Eq.~\eqref{eq:adjoint-lindblad}, the time evolution of the cavity mode satisfies
\begin{equation}
\dot a_m
=
-\Big(\tfrac{\kappa}{2}+i\Delta_{\mathrm{cd}}\Big)a_m
-\,i\int d^2\mathbf r\;\tfrac12\,\Omega_m\,g_m(\mathbf r)\,\sigma_{ge}(\mathbf r).
\label{eq:cavity-eom}
\end{equation}
Inserting the drive–induced part of Eq.~\eqref{eq:slaved-coherence} and setting $\dot a_m\simeq 0$ gives
\begin{equation}
a_m
\simeq
\frac{-\,i}{\kappa/2 - i\Delta_{\mathrm{cd}}}
\int d^2\mathbf r\;
\frac{\tfrac12\,\Omega_m\Omega_{\mathrm{d}}\,g_m(\mathbf r)g_{\mathrm{d}}(\mathbf r)}{\Delta_{\mathrm{da}}(\mathbf r)+i\,\Gamma/2}\;n_g(\mathbf r).
\label{eq:am-slaved}
\end{equation}
Substituting Eq.~\eqref{eq:am-slaved} into~\eqref{eq:adjoint-lindblad} produces an effective jump operator
\begin{equation}
L_\kappa^{\mathrm{eff}}
=
\int d^2\mathbf r\;
\sqrt{\kappa}\;
\frac{\tfrac12\,\Omega_m\Omega_{\mathrm{d}}\,g_m(\mathbf r)g_{\mathrm{d}}(\mathbf r)}{\big(\kappa/2 - i\Delta_{\mathrm{cd}}\big)\big(\Delta_{\mathrm{da}}(\mathbf r)+i\,\Gamma/2\big)}\;n_g(\mathbf r),
\label{eq:L-kappa-eff}
\end{equation}
with an effective decay rate
\begin{equation}
\tilde\kappa(\mathbf r)
=
\kappa\;
\frac{\tfrac14\,|\Omega_m\Omega_{\mathrm{d}}|^2\,|g_m(\mathbf r)g_{\mathrm{d}}(\mathbf r)|^2}
{\big(\Delta_{\mathrm{cd}}^2+(\kappa/2)^2\big)\big(\Delta_{\mathrm{da}}(\mathbf r)^2+(\Gamma/2)^2\big)}.
\label{eq:kappa-tilde-local}
\end{equation}

The effective jumps in Eqs.~\eqref{eq:L-gamma-eff} and~\eqref{eq:L-kappa-eff} are proportional to $n(\mathbf r)$ and therefore act as dephasing within the ground manifold.

For completeness, we also derive the interaction scale used in Eq.~\eqref{J_dissipative} of the main text in presence of dissipation. 
Using the expressions in Eqs.~\eqref{eq:slaved-coherence} and \eqref{eq:am-slaved}, we can derive the effective Hamiltonian in presence of dissipation:
\begin{equation}
\begin{aligned}
H_{\mathrm{eff}}^{(2)}
=\;\frac12 \sum_m \frac{1}{\Delta_{\mathrm{cd}}+i\kappa/2}
\int\! d^2r\, d^2r'\;
\frac{\Omega_m\Omega_{\mathrm{d}}\,g_m(\mathbf{r})g_{\mathrm{d}}(\mathbf{r})}{\Delta_{\mathrm{da}}(\mathbf{r})+i\Gamma/2}\frac{\Omega_m\Omega_{\mathrm{d}}\,g_m(\mathbf{r}')g_{\mathrm{d}}(\mathbf{r}')}{\Delta_{\mathrm{da}}(\mathbf{r}')+i\Gamma/2}\;
n_g(\mathbf{r})\,n_g(\mathbf{r}') \;+\; \mathrm{h.c.}\,.
\end{aligned}
\tag{J10}
\end{equation}

We can again expand the ground-state density in the single-particle basis as in Appendix~\ref{app:light_matter_Hamiltonian} and define the equivalent of the tensor in Eq.~\eqref{eq:linear_int} in presence of dissipation:
\begin{equation}
\tilde{\mathcal{M}}^{(1)}_{m,ij}
=\int d^2r\;
\frac{\Omega_{\mathrm{d}}^*\Omega_m\,g_{\mathrm{d}}(\mathbf{r})g_m(\mathbf{r})}{\Delta_{\mathrm{da}}(\mathbf{r})+i\Gamma/2}\,
\varphi_i^*(\mathbf{r})\,\varphi_j(\mathbf{r}) .
\end{equation}
Collecting quartic terms and comparing with Eq.~\eqref{eq:Heff_4fint} gives
\begin{equation}
H_{\mathrm{eff}}^{(4)}
= -\frac12 \sum_{i,i',j,j'}
\Biggl[
\sum_m
\frac{\tilde{\mathcal{M}}^{(1)\,*}_{m,ii'}\,\tilde{\mathcal{M}}^{(1)}_{m,jj'}}{\Delta_{\mathrm{cd}}+i\kappa/2}
\Biggr]\,
c_i^\dagger c_{i'}\,c_j^\dagger c_{j'}  +~\mathrm{h.c.}\, ,
\tag{J11}
\end{equation}
and we can identify the SYK$_4$ couplings in presence of dissipation as
\begin{equation}
J_{ii';jj'} \;=\; \sum_m
\frac{\tilde{\mathcal{M}}^{(1)\,*}_{m,ii'}\,\tilde{\mathcal{M}}^{(1)}_{m,jj'}}{\Delta_{\mathrm{cd}}+i\kappa/2}.
\tag{J12}
\end{equation}
In the lossless limit, $\kappa,\Gamma\to0$, this reduces to Eq.~\eqref{Eq:four-leg}.
Under the same assumptions used in Eqs.~\eqref{eq:slaved-coherence}–\eqref{eq:am-slaved}, the typical interaction scale reads
\begin{equation}
J \;\simeq\;
\frac{|\Omega_{\mathrm{d}}|^2\,|\Omega_m|^2}{|\Delta_{\mathrm{da}}|^2}\,
\frac{1}{\bigl|\Delta_{\mathrm{cd}}+i\kappa/2\bigr|},
\tag{J13}
\end{equation}
which is the expression in Eq.~\eqref{J_dissipative} of the main text.

\end{document}